\documentclass[11pt]{article}

\usepackage{thesis}

\title{Superspace Formulation of
$N=4$ Super Yang-Mills Theory\\ with a Central Charge}%
\author{{\large\scshape Jun Saito}\\[1ex]
 \itshape Division of Physics, Graduate School of Science,
 Hokkaido University}
\date{\today}

\begin{document}
%%%%%%%%
%\maketitle
%\hfill\raisebox{35ex}[0ex][-35ex]{EPHOU-05-006}
\begin{center}
\hfill {EPHOU-05-006}\\
\hfill {hep-th/0512226}\\
\hfill December, 2005\\[10ex]
{\LARGE%
Superspace Formulation of $N=4$ Super Yang-Mills Theory\\
with a Central Charge}\\[5ex]
{\large\scshape Jun Saito}\\[2ex]
{\itshape Department of Physics, Graduate School of Science,
 Hokkaido University}\\[5ex]
\end{center}

\begin{center}
\renewcommand{\thefootnote}{\fnsymbol{footnote}}%
(\scshape\large Master's Thesis)%
\footnote{This article is a corrected version of the author's
master's thesis submitted to Graduate School of Science,
Hokkaido University in March 2005 and published
in \cite{JS}.}
\renewcommand{\thefootnote}{\arabic{footnote}}%
\setcounter{footnote}{0}
\end{center}
\allowdisplaybreaks

\begin{abstract}
A superspace formulation using superconnections and supercurvatures
is specifically constructed for $N=4$ extended super Yang-Mills theory
with a central charge in four dimensions, first proposed
by Sohnius, Stelle and West long ago.
We find that the constraints, almost uniquely derived from
the possible spin structure of the multiplet, can be algebraically
solved which results in an off-shell supersymmetric
formulation of the theory on the superspace.
\end{abstract}

\tableofcontents

\section{Introduction}
Theory of elementary particle is widely studied for the construction
of a unified theory which could explain every simple aspect
of nature including matters and interactions. We know that such a
theory can be realized to some extent as a quantum field theory (QFT)
based on the gauge principle. In fact, so called the standard theory of
particle physics has been constructed as a quantum field theory
with gauge invariance (Yang-Mills theory).
There are many successful examples where description of nature in the
standard theory agrees with experimental data with a considerably
surprising accuracy.

Though phenomenologically successful, the standard theory can not be
interpreted as a sufficient unified theory of particle physics.
This is clearly seen by noting the following fact.
A consistent and unified scheme which can treat the two fundamental
statistics of particles, i.e.\ bosonic and fermionic statistics,
should be essentially implemented into a unified theory.
In fact, fundamental matters in particle physics---quarks and leptons---%
are all described as fermions, while fundamental interactions---%
photons (which mediate electromagnetic interaction),
$W$ and $Z$ bosons (weak), gluons (strong) and gravitons (gravity)---%
are all described as bosons. So naively we expect that any theory
which unifies these fundamental matters and interactions should naturally
treat bosons and fermions in a unified way. The standard theory
dose not contain such a unified scheme.

Supersymmetry (SUSY) serves one systematic, somewhat realistic and 
almost unique possibility which treats bosons and fermions as unified
objects consistent with relativistic spacetime structure.
In a supersymmetric theory, bosons and fermions are always introduced as
a pair, and forms a multiplet over which several kinds of transformations,
called supertransformations, are defined. In this sense, the bosons and
fermions are treated in a unified manner and are called superpartners.
The theory is said to be supersymmetric when constructed to be symmetric
under the supertransformations of the superpartners.
Since a supertransformation converts a boson into a fermion, or vice
versa, it has to be generated by a fermionic charge, called a supercharge.
Such transformations, or charges,
which must be consistent with relativistic symmetry, namely
the Poincar\'{e} symmetry, are uniquely prescribed
by a graded Lie algebra, which is called the supersymmetry algebra
(Coleman-Mandula no-go theorem, Haag-\L opuszanski-Sohnius theorem).

Phenomenologically, only $N=1$ supersymmetry%
\footnote{In the following,
we denote the number of sets of conserved supercharges by $N$.},
in which a set of supercharges
is conserved, may seem to be realistic. In fact, the minimal supersymmetric
standard model (MSSM) with $N=1$ supersymmetry can be interpreted as a
successful phenomenological model from, for instance, the following reasons.
Firstly, MSSM unifies the coupling constants of the gauge
group $\GR{SU}(3)\times\GR{SU}(2)\times\GR{U}(1)$ in the model
as the meeting of the renormalization group flows of the each
coupling constant at the GUT scale ($\sim 10^{16}\ \MR{GeV}$), implying
that the gauge group itself is unified in the scale. This unification
of the coupling constants dose not occur in the usual standard model
without $N=1$ supersymmetry.
Secondly, so called the problem of naturalness, or the problem of hierarchy
of mass, can be solved in MSSM by the cancellation
of some bad divergences of self-energies due to the high degrees
of freedom of supersymmetry.

From the viewpoint of the unification, however, so called extended
supersymmetry, or $N\geqslant 2$ supersymmetry,
where several sets of supercharges are conserved,
may give more natural schema in many cases, as in the following examples.
First, those extended degrees of supersymmetry in four dimensions
can be naturally
interpreted as dimensionally reduced degrees of freedom of some
higher dimensional supersymmetric theories, including superstring
theories and supergravity theories in ten or eleven dimensions.
Such kind of naive correspondence of extended supersymmetry and
higher dimensional supersymmetry may play more explicitly a role in some
cases as in the AdS/CFT correspondence,
where the type IIA superstring (or supergravity) on $\MR{AdS}_5\times S^5$ is
naturally related to $N=4$ $\GR{U}(N)$ super Yang-Mills (SYM) theory
in four dimensions as a low energy effective theory.
Second, in the context of duality, which has been one of key notions
in recent theoretical works, extended supersymmetry
naturally appears and gives many fruitful topics, both
from physical and mathematical point of view.
Third, and as the fact which gives most directly the motivation of this
article, supersymmetry
on a lattice in four dimensions should be naturally and intrinsically
interpreted as (twisted) $N=4$ extended supersymmetry as is proposed
in recent works%
~\cite{D'Adda-Kanamori-Kawamoto-Nagata-1,D'Adda-Kanamori-Kawamoto-Nagata-2}.
In that paper, so called
twisted supersymmetry, an exotic version of extended supersymmetry
which may be related with some topological theories or some BRST quantized
gauge theories%
~\cite{Witten-1,Witten-2,Baulieu-Singer,Brooks-Montano-Sonnenschein,%
Labastida-Pernici,Birmingham-Rakowski-Thompson},
is introduced on a lattice, based on the facts that supercharges conserved
in a twisted supersymmetric theory are naturally associated with
the simplex structure of the lattice.
Introduced on the lattice, internal degrees of freedom of extended
supersymmetry may be interpreted as flavor degrees of freedom through the
Dirac-K\"{a}hler mechanism. Species doublers on the lattice,
notorious obstacles
which appear by introducing chiral fermions on a lattice, may
also be identified these internal or flavor degrees.
Thus there arises the possibility
that extended $N=4$ supersymmetry on a lattice can also give a unified
solution to the problem of chiral fermions on the lattice.

We should briefly remark here why we need to introduce a lattice theory.
Quantum field theory deals with inevitably infinite degrees of freedom
of the fields. This infinite degrees of freedom comes from that spacetime
in quantum field theory is assumed to be a four-dimensional continuum.
Continuous spacetime has infinitely many degrees
of freedom in an infinitely small area, or, in its momentum
representation, in an infinitely large momentum region over which
a divergent integration will be produced.
Those divergences are thus essential and unavoidable in quantum field
theories, so that they have to be estimated systematically as
finite quantities to make a rigorous calculation. Such a technical scheme
to evaluate those divergences is called a regularization.
Introducing a lattice structure into spacetime is considered to be one
of the most natural and important regularizations among others especially
by the following reasons. First, theory on a lattice is theoretically
important from the viewpoint of the unification. In fact, there is an approach
to formulate gravity as well as the other three interactions and matters on
a (random) lattice in a unified manner, and we know such a formulation
is successful in two dimensions.
Though no such theory has been completely formulated in
four dimensions, we expect the approach to a unified theory on a
four-dimensional random lattice could also be a successful scheme
for the unification.
Second, a theory on a lattice may be analyzed numerically to make a
realistic and practical computation.
As is well known this is the case in lattice QCD, which shows
the prominence of the lattice regularization especially
in a non-perturbative region where quantitative analysis can hardly be done
in other approaches.

Thus to construct a supersymmetric theory
on a lattice may seem to
serve a candidate of a realistic unification including gravity.

We should especially note here that
the formulation for introducing supersymmetry on a lattice in the
paper%
~\cite{D'Adda-Kanamori-Kawamoto-Nagata-1,D'Adda-Kanamori-Kawamoto-Nagata-2}
inevitably needs the
full twisted $N=4$ off-shell supersymmetry (i.e.,\ supersymmetry
respected exactly without using classical solutions) in a continuous
four-dimensional spacetime.
This aspect should be compared with the fact that in other theories with
twisted supersymmetry only a part of supercharges, especially
the scalar charges,
is taken into account so that off-shell formulation in such
theories is merely corresponding to only a part of supercharges,
not corresponding to the full supercharges.

One may thus consider to construct an off-shell formulation
of a full $N=4$ supersymmetric, especially super Yang-Mills, theory.
However, the $N=4$ super Yang-Mills theory with the full internal
symmetry $\GR{SU}(4)$ in four dimensions has been formulated only on-shell%
~\cite{Sohnius-1}.
Instead, there is one known off-shell formulation of $N=4$ super
Yang-Mills theory with the internal symmetry $\GR{USp}(4)$
~\cite{Sohnius-Stelle-West-1,Sohnius-Stelle-West-2}.
This model contains essentially a central charge to prohibit
higher spin components~\cite{Fayet-1}.
Since an off-shell formulation of a supersymmetric theory is
constructed most clearly and systematically by superspace formulation,
these off-shell $\GR{USp}(4)$ model should be naturally formulated
on a superspace. Here we have to emphasize that still
no superspace formulation of $N=4$ super Yang-Mills theory in four
dimensions is known.

Motivated by the facts above, I have attempted in this work to construct
a superspace formulation, mentioned briefly in~%
\cite{Sohnius-Stelle-West-2},
of the $\GR{USp}(4)$ super Yang-Mills theory in
four dimensions.
For the task, I applied a formulation using superconnections
and supercurvatures with the manifest gauge covariance.
I set up constraints on the superfields in the formulation
by noting the contents of multiplets as well as their spins
in the $\GR{USp}(4)$ model and solve the constraints to derive
the supertransformations of the contents. There remains some
difficulties which should be resolved by the future works.

This article is organized as follows. In section~\ref{sec-SUSY}
we review some foundations of extended supersymmetry algebra in four
dimensions, particularly one with central charges.
In section~\ref{sec-twistedSUSY} we discuss twisted supersymmetries
in four dimensions since those ideas serve an essential
background to this article.
In section~\ref{sec-USp-model} we present the construction
of the superspace formulation of the $\GR{USp}(4)$ model,
which includes the main results of this article.
Finally we conclude in section~\ref{sec-conclusion}, and
mention some possible ideas to complete our superspace formulation.

\section{$N$-Extended Supersymmetry in Four Dimensions}
\label{sec-SUSY}
In this section, we briefly review some general aspects of
the $N$-extended superalgebra in four dimensions and its irreducible
representations~\cite{Wess-Bagger,Gates-Grisaru-Rocek-Siegel,Sohnius-3}.
Especially we consider supersymmetry with
central charges. Notational and technical details are listed
in the appendix.

\subsection{Superalgebra}
$N$-extended superalgebra (super Poincar\'{e} algebra)
in four dimensions is prescribed by the following
(anti-)commutation relations:
\begin{xalignat}{2}
 \{Q_{i\alpha},Q_{j\beta}\}
    &=C_{\alpha\beta}Z_{ij},
 &
 \{\OLL{Q}^i{}_{\Dalpha},\OLL{Q}^j{}_{\Dbeta}\}
    &=C_{\Dalpha\Dbeta}\OLL{Z}^{ij},\label{BeginOfAlgebra}\\
 \{Q_{i\alpha},\OLL{Q}^j{}_{\Dbeta}\}
    &=2\delta_i{}^j(\sigma^\mu)_{\alpha\Dbeta}P_\mu,\label{QQbar}
 &
 [P_\mu,P_\nu]
    &=0,\\
 [Q_{i\alpha},P_\mu]
    &=0,
 &
 [\OLL{Q}^i{}_{\Dalpha},P_\mu]
    &=0,\\
 [J_{\mu\nu},Q_{i\alpha}]
    &=\frac{i}{2}(\sigma_{\mu\nu})_{\alpha}{}^{\beta}Q_{i\beta},
 &
 [J_{\mu\nu},\OLL{Q}^{i\Dalpha}]
    &=\frac{i}{2}(\bar{\sigma}_{\mu\nu})^{\Dalpha}{}_{\Dbeta}
           \OLL{Q}^{i\Dbeta},\\
 [J_{\mu\nu},P_\rho]
    &=i(\eta_{\mu\rho}P_\nu-\eta_{\nu\rho}P_\mu),
 &
 [J_{\mu\nu},J_{\rho\sigma}]
    &=i\begin{aligned}[t]
         & (\eta_{\nu\rho}J_{\mu\sigma}-\eta_{\nu\sigma}J_{\mu\rho}\\
         & -\eta_{\mu\rho}J_{\nu\sigma}+\eta_{\mu\sigma}J_{\nu\rho}),\\
       \end{aligned}\\
 [R^a,Q_{i\alpha}]
    &=(X^a)_i{}^jQ_{j\alpha},
 &
 [R^a,\OLL{Q}^i{}_{\Dalpha}]
    &=-(X^a)_j{}^i\OLL{Q}^j{}_{\Dalpha},\\
 [R^a,P_\mu]
    &=0,
 &
 [R^a,J_{\mu\nu}]
    &=0,\\
 [Z_{ij},\mbox{any}]
    &=0,
 &
 [\OLL{Z}^{ij},\mbox{any}]
    &=0,\label{EndOfAlgebra}
\end{xalignat}
where the supercharges $Q_{i\alpha},\ \OLL{Q}^i{}_{\Dalpha}$,
related by Hermitian conjugation as
\begin{equation}
 \OLL{Q}^i{}_{\Dalpha}=(Q_{i\alpha})^\dagger,
\end{equation}
are represented as Weyl spinors w.r.t.\ Lorentz transformation
$\GR{SO}(1,3)\cong\GR{SL}(2,\mathbb{C})$
in Minkowski spacetime,
and as the fundamental $\bs{N}$ and $\OLL{\bs{N}}$ representations
w.r.t.\ the $R$-symmetry (internal symmetry)%
\footnote{
In the following, we concentrate only on the $R$-symmetry
$\GR{SU}(N)$ for simplicity. The extension to the case
$\GR{U}(N)=\GR{U}(1)\times\GR{SU}(N)$, or to more general cases,
is trivial. Note, however,
if $N=4$ we cannot adopt the symmetry $\GR{U}(4)$ for the
super Yang-Mills multiplet in four dimensions because of the CPT theorem.}
$\GR{SU}(N)$, i.e.\ 
\begin{equation}
 Q_{i\alpha}\in(\bs{2},\bs{0},\bs{N}),\quad
 \OLL{Q}^i{}_{\Dalpha}\in(\bs{0},\bs{2},\OLL{\bs{N}}),
 \qquad\text{under\ \ }
 \GR{SL}(2,\mathbb{C})\times\GR{SU}(N);
\end{equation}
$Z_{ij}$ and $\OLL{Z}^{ij}$ are central charges which satisfy the
relations%
\footnote{In order for the central charges to be consistently
introduced in the superalgebra, the $R$-symmetry group
should be restricted to some extent as is seen below.}
\begin{equation}
 Z_{ij}+Z_{ji}=0,\quad
 \OLL{Z}^{ij}+\OLL{Z}^{ji}=0,
\end{equation}
i.e.\ 
\begin{equation}
 Z_{ij}\in\bs{\frac{N(N-1)}{2}},\quad
 \OLL{Z}^{ij}\in\OLL{\bs{\frac{N(N-1)}{2}}}\qquad
 \text{under\ \ }\GR{SU}(N),
\end{equation}
and
\begin{equation}
 \OLL{Z}^{ij}=(Z_{ij})^*,\quad \OLL{Z}+Z^\dagger=0;
\end{equation}
$\eta_{\mu\nu}$ is the flat spacetime metric
\begin{equation}
 \eta_{\mu\nu}=(+,-,-,-)\quad(\text{Minkowski}),
\end{equation}
with its contragradient $\eta^{\mu\nu}$ such that
\begin{equation}
 \eta^{\mu\rho}\eta_{\rho\nu}=\eta^\mu{}_\nu:=\delta^\mu{}_\nu;
\end{equation}
$P_\mu$ is the four momentum and $J_{\mu\nu}$ is the Lorentz
generator
\begin{equation}
 P_{\mu}\in (\bs{2},\bs{2}),\quad
 J_{\mu\nu}\in(\bs{3},\bs{0})\oplus(\bs{0},\bs{3})\qquad
 \text{under\ \ }
\GR{SL}(2,\mathbb{C}),
\end{equation}
with the convention
\begin{equation}
 \exp\left(-\frac{i}{2}\omega^{\mu\nu}J_{\mu\nu}\right)
  \in\GR{SO}(1,3);
\end{equation}
$R^a$ is the generator of the $R$-symmetry (internal symmetry)
$\GR{SU}(N)$, i.e.\ 
\begin{equation}
 R^a\in\MF{su}(N)
\end{equation}
with the convention
\begin{equation}
 \exp\left(i\sum_a t^a R^a\right)\in\GR{SU}(N),
\end{equation}
and $(X^a)_i{}^j$ is its adjoint representation
\begin{equation}
 (X^a)_i{}^j\in(\bs{N^2-1})\qquad
 \text{under\ \ }\GR{SU}(N),
\end{equation}
with
\begin{equation}
 (X^a{}^*)^i{}_j:=((X^a)_i{}^j)^*=(X^a)_j{}^i\quad(\text{Hermitian});
\end{equation}
$C_{\alpha\beta}$ and $C_{\Dalpha\Dbeta}$ are the
$\GR{SL}(2,\mathbb{C})$ invariant tensors with relations
\begin{equation}
 C_{\alpha\beta}+C_{\beta\alpha}=0,\quad
 C_{\Dalpha\Dbeta}+C_{\Dbeta\Dalpha}=0,
\end{equation}
i.e.\ 
\begin{equation}
 C_{\alpha\beta}\in(\bs{1},\bs{0}),\quad
 C_{\Dalpha\Dbeta}\in(\bs{0},\bs{1})\qquad
 \text{under\ \ }
 \GR{SL}(2,\mathbb{C}),
\end{equation}
and
\begin{equation}
 C_{\Dalpha\Dbeta}=(C_{\alpha\beta})^*,
\end{equation}
while also $C^{\alpha\beta}$ and $C^{\Dalpha\Dbeta}$ are defined as
\begin{equation}
 C_{\alpha\gamma}C^{\beta\gamma}=\delta_\alpha{}^\beta,\quad
 C_{\Dalpha\Dgamma}C^{\Dbeta\Dgamma}=\delta_{\Dalpha}{}^{\Dbeta};
\end{equation}
$(\sigma^\mu)_{\alpha\Dbeta}$ and $(\bar{\sigma}^\mu)^{\Dalpha\beta}$
are defined by the Pauli matrices $\tau^i$ as
\begin{equation}
 \sigma^\mu=(\bs{1},\tau^i),\quad \bar{\sigma}^\mu=(\bs{1},-\tau^i)
\end{equation}
in Minkowski spacetime,
and interpreted as
\begin{equation}
 (\sigma^\mu)_{\alpha\Dbeta}\in(\bs{2},\bs{2}),\quad
 (\bar{\sigma}^\mu)^{\Dalpha\beta}\in(\bs{2},\bs{2})\qquad
 \text{under\ \ }
 \GR{SL}(2,\mathbb{C}),
\end{equation}
with relations
\begin{equation}
 (\sigma^\mu\bar{\sigma}^\nu+\sigma^\nu\bar{\sigma}^\mu)_\alpha{}^\beta
     =2\eta^{\mu\nu}\delta_\alpha{}^\beta,\quad
 (\bar{\sigma}^\mu\sigma^\nu+\bar{\sigma}^\nu\sigma^\mu)^{\Dalpha}{}_{\Dbeta}
     =2\eta^{\mu\nu}\delta^{\Dalpha}{}_{\Dbeta};
\end{equation}
and $\sigma_{\mu\nu}$ and $\bar{\sigma}_{\mu\nu}$ are defined by
\begin{equation}
 (\sigma^{\mu\nu})_\alpha{}^\beta
  :=\frac{1}{2}(\sigma^\mu\bar{\sigma}^\nu
      -\sigma^\nu\bar{\sigma}^\mu)_\alpha{}^\beta,\quad
 (\bar{\sigma}^{\mu\nu})^{\Dalpha}{}_{\Dbeta}
  :=\frac{1}{2}(\bar{\sigma}^\mu\sigma^\nu
      -\bar{\sigma}^\nu\sigma^\mu)^{\Dalpha}{}_{\Dbeta},
\end{equation}
which are interpreted as
\begin{equation}
 \sigma^{\mu\nu}\in(\bs{3},\bs{0}),\quad
 \bar{\sigma}^{\mu\nu}\in(\bs{0},\bs{3})\qquad
 \text{under\ \ }
 \GR{SL}(2,\mathbb{C}).
\end{equation}

Let us quickly recall how central charges are consistently implemented
into the superalgebra above. For this we consider the case where the
algebra has more general internal symmetry which includes the $R$-symmetry
$\AL{su}(N)$. The Coleman-Mandula theorem restricts such an internal
symmetry to be generated by Lorentz invariant and compact Lie algebra
$\MC{A}$. More precisely, we can take as
$\MC{A}=\AL{su}(N)\oplus\MC{A}_1\oplus\MC{A}_2$ with
$\MC{A}_1$ and $\MC{A}_2$ being invariant semi-simple and Abelian
subalgebras, respectively.
Accordingly let us denote the irreducible representations
of this internal symmetry algebra by
$Q_{I\alpha},\ \OLL{Q}^{J}{}_{\Dbeta}$ with
$I=(i,\MF{i}),\ J=(j,\MF{j})$, and let the Hermitian generators of
this algebra be $B^l$ including $R^a$. Assume these generators
satisfy the relations
\begin{xalignat}{2}
 [B^l,Q_{I\alpha}]
  &=(S^l)_I{}^J Q_{J\alpha},
 &
 [B^l,\OLL{Q}^I{}_{\Dalpha}]
  &=-(S^l)_J{}^I \OLL{Q}^J{}_{\Dalpha},\\
 [B^l,P_\mu]
  &=0,
 &
 [B^l,J_{\mu\nu}]
  &=0. %,\\
% [B^l,B^m]
%  &=i\sum_n f^{lmn}B^n.
% &
%  &
\end{xalignat}
Since $\{Q_{I\alpha},Q_{J\beta}\}$
is symmetric under the exchange of $(I,\alpha)\LR (J,\beta)$, it should
have most generally the form
\begin{equation}
 \{Q_{I\alpha},Q_{J\beta}\}
   =C_{\alpha\beta}Z_{IJ}
    +\frac{1}{2}(\sigma^{\mu\nu})_{\alpha\beta}(Y_{\mu\nu})_{IJ},
\end{equation}
where $Z_{IJ}$ is antisymmetric under $I\LR J$
while $(Y_{\mu\nu})_{IJ}$ is symmetric, and
$(\sigma^{\mu\nu})_{\alpha\beta}
 :=(\sigma^{\mu\nu})_\alpha{}^\gamma C_{\gamma\beta}$
is symmetric under $\alpha\LR\beta$.
According to the Coleman-Mandula theorem, we have to set
\begin{equation}
 Z_{IJ}=\sum_l (a^l)_{IJ}B^l \in\MC{A},\qquad
 (Y_{\mu\nu})_{IJ}=Y_{IJ}J_{\mu\nu},
\end{equation}
Then the Jacobi identity, firstly, w.r.t.\ 
$P_\mu,\ Q_{I\alpha},\ Q_{J\beta}$ leads to that $Y_{IJ}=0$,
secondly, the identity w.r.t.\ 
$B^l,\ Q_{I\alpha},\ Q_{J\beta}$ shows that linear combinations
$Z_{IJ}$ generates an invariant subalgebra $\MC{Z}$ of $\MC{A}$,
and, thirdly, the identity w.r.t.\ 
$Q_{I\alpha},\ Q_{J\beta},\ \OLL{Q}^{K}{}_{\Dgamma}$
and the fact that $B^l$ is Hermitian prove that
$\MC{Z}$ is Abelian. Thus we find that $\MC{Z}=\MC{A}_2$ which is the
center of the algebra $\MC{A}$, namely,
\begin{equation}
 [Z_{IJ},B^l]=0,\qquad
 \{Q_{I\alpha},Q_{J\beta}\}=C_{\alpha\beta}Z_{IJ}.
\end{equation}
Therefore a nonzero central charge exists if and only if
the internal symmetry algebra has the invariant Abelian subalgebra,
and, of course, one has to work on supersymmetry for $N\geqslant 2$.
In what follows, we only consider the $R$-symmetry and the symmetry generated
by the center $\MC{Z}$ as nontrivial internal symmetries of our system,
namely the case $\MC{A}=\AL{su}(N)\oplus\MC{Z}$.
Since $\MC{Z}$ is Abelian, its action is a trivial scalar multiplication
so that its adjoint representation is trivial.
Thus we can label the quantities
simply by indices $I=i,\ J=j$ and treats only $R^a$ as nontrivial
generators of the internal symmetry, which in turn leads to the superalgebra
listed at the beginning of this section.

\subsection{Central Charges}
In the preceding section, we have seen central charges can be interpreted
as generators of Abelian internal symmetry. However, one can not
introduce a nonzero central charge yet consistently to an arbitrary
structure of the $R$-symmetry. In fact, the Jacobi identity w.r.t.\ 
$R^a,\ Q_{i\alpha},\ Q_{j\beta}$ together with that $[R^a,Z_{ij}]=0$
leads to
\begin{equation}
\label{automorphism}
 (X^a)_i{}^k Z_{kj}+Z_{ik}(X^a)_j{}^k=0,
\end{equation}
or equivalently, since $X^a$ is Hermitian and
$\DS Z_{ij}=\sum_{B^l\in\MC{A}}(a^l)_{ij}B^l$,
\begin{equation}
\label{intertwine}
 (X^a)_i{}^k (a^l)_{kj}=-(a^l)_{ik}(X^a{}^\ast)^k{}_j.
\end{equation}
If $(a^l)_{ij}$ is nondegenerate, i.e.\ if the matrix $((a^l)_{ij})$
has its inverse, the above condition is more directly written as
\begin{equation}
 (X^a)_i{}^j=(a^l)_{ik}(X^a{}^\ast)^k{}_l (a^l)^{lj},
\end{equation}
where $(a^l)^{ij}:=(((a^l)^T)^{-1})^{ij}$.
These equations show that each coefficient $(a^l)_{ij}$ relates each
representation of $R$-symmetry generator $X^a$ to its conjugate representation
$X^a{}^\ast$. Central charges can exist if and only if such an intertwiner
$a^l$ does exist. In other words, the representation of the $R$-symmetry algebra,
or at least some invariant subalgebra, has to be (pseudo) real,
namely, the $R$-symmetry algebra has to be automorphic, for a
central charge to exist.
 
If we persist to require the full $\GR{SU}(N)$ as the $R$-symmetry,
such intertwiner $a^l$ does not exist except the case%
\footnote{
For $N=2$, $\GR{SU}(2)\cong\GR{USp}(2)$
contains such an intertwiner; the $\GR{SU}(2)$ invariant tensor
$\varepsilon_{ij}$ plays the role.}
$N=2$, since
for $N\geqslant 3$ the fundamental representation $\bs{N}$
of $\GR{SU}(N)$ is not real, i.e.\ $\bs{N}$ and $\OL{\bs{N}}$
are inequivalent%
\footnote{See appendix \ref{sec-unitary-fund}.}.
These facts can also be seen easily by a direct computation;
let \eRef{intertwine} hold for all generators $X^a\ (a=1,\cdots,N^2-1)$
in $\AL{su}(N)$, multiply the both side by $(X^a)_m{}^n$,
take summation w.r.t.\ $a=1,\cdots,N^2-1$ by using the completeness
relation \eRef{complete-su}
in $\AL{su}(N)$,
and then take contraction w.r.t.\ indices $n$ and $j$ to give
\begin{equation}
 0=(N^2-N-2)(a^l)_{im}=(N-2)(N+1)(a^l)_{im},
\end{equation}
which shows that nonzero intertwiner $a^l$ exists only if $N=2$.

Thus we have to break the full $R$-symmetry group $\GR{SU}(N)$ into
some automorphic subgroup of $\GR{SU}(N)$ in order to introduce
a nonzero central charge. If $N$ is even, say $N=2n$,
the unitary symplectic group%
\footnote{See appendix~\ref{sec-symplectic}.}
$\GR{USp}(2n)\subset\GR{SU}(2n)$
can be taken as a candidate, for, $\GR{USp}(2n)$ contains
the invariant tensor $\Omega_{ij}$ such that
\begin{equation}
 (X^a)_i{}^j=\Omega_{ik}(X^a{}^\ast)^k{}_l\Omega^{lj},\quad
 \Omega^{ij}:=((\Omega^T)^{-1})^{ij}.
\end{equation}
Later we consider the automorphic subgroup
$\GR{USp}(4)\subset\GR{SU}(4)$ for $N=4$ super Yang-Mills theory with a
central charge, which is the main subject of this article.
Another nontrivial example could be served by the spinor representation
of $\GR{SO}(N)\subset\GR{SU}(N)$ for some appropriate $N$,
where the charge conjugation matrix, %and its Hodge dual,
or equivalently the $B$-conjugation matrix%
\footnote{See appendix~\ref{sec-Clifford}.}, can be used
as invariant intertwiners.

Central charges can be interpreted as some sort of additional masses.
This can be seen by the following observations.
Firstly, by diagonalizing the on-shell superalgebra w.r.t.\ the central
charges as will be seen in the next section,
we can show the BPS bound relation
\begin{equation}
\label{BPS}
 2m\geqslant z^l,
\end{equation}
where $m$ is the mass of the on-shell algebra and $z_a$ are
the eigenvalues of the central charges. This relation allows
one to interpret $z_a$, i.e.\  the central charges, as some kind of masses.
Secondly, if we consider a superalgebra in higher dimensions and
then take a dimensional reduction by, for instance, the Kaluza-Klein
compactification of spacetime,
we could obtain a superalgebra in four dimensions with
a central charge which originates from the momentum corresponding to the
compactified direction. In such case, the central charge can be
clearly regarded as a mass of an extra dimension.
Thirdly, by an appropriate spontaneous breaking of a gauge
symmetry in some extended supersymmetric gauge theories,
a central charge emerges as a nonzero vev (of a moduli)
both algebraically and field theoretically~\cite{Fayet-1}.
In this formulation, origins of some part of central charges are gauge bosons
in higher dimensions and as an effect are related somewhat directly
to the momenta in the higher dimensions. In fact, as an on-shell relation
one can show that $\DS P^\mu P_\mu=\sum_l (z^l)^2$, which is
identical to the BPS condition.

\subsection{Representations}
Let us now examine the irreducible one particle representations
of the superalgebra eqs.~\Ref{BeginOfAlgebra}--\Ref{EndOfAlgebra}.

\subsubsection{Massive Multiplet without Central Charge}
First we consider the superalgebra with no central charge.
For massive multiplets with mass $m>0$ i.e.\ $P^2=m^2$,
we can take the rest frame
in which $P^\mu=(m,\bs{0})$. Then relevant anticommutation relations are
written as
\begin{gather}
 \{Q_{i\alpha},\OLL{Q}^j{}_{\Dbeta}\}=2m\delta_{\alpha\Dbeta}\delta_i{}^j,\\
 \{Q_{i\alpha},Q_{j\beta}\}=0,\qquad
 \{\OLL{Q}^i{}_{\Dalpha},\OLL{Q}^j{}_{\Dbeta}\}=0.
\end{gather}
Then defining
\begin{equation}
 a_{i\alpha}:=\frac{1}{\sqrt{2m}}Q_{i\alpha},\quad
 (a_{i\alpha})^\dagger:=\frac{1}{\sqrt{2m}}\OLL{Q}^i{}_{\Dalpha},
\end{equation}
we find that
\begin{gather}
 \{a_{i\alpha},(a_{j\beta})^\dagger\}=\delta_\alpha{}^\beta \delta_i{}^j,\\
 \{a_{i\alpha},a_{j\beta}\}=0,\qquad
 \{(a_{i\alpha})^\dagger,(a_{j\beta})^\dagger\}=0.
\end{gather}
Hence we obtain $2N$ sets of fermionic creation-annihilation operators
which constructs the irreducible representation with total of
$2^{2N}(2j+1)$ states on a Clifford vacuum with spin $j$.
The maximum spin for such representations is $N/2+j$.

\subsubsection{Massless Multiplet without Central Charge}
\label{sec-massless}
For massless multiplets, we can take a light-cone frame where
$P^2=0$, so for instance $P^\mu=(P^0,0,0,-P^0),\ P^0>0$.
Defining
\begin{gather}
 a_i:=\frac{1}{2\sqrt{P^0}}Q_{1i},\quad
 (a_i)^\dagger:=\frac{1}{2\sqrt{P^0}}\OLL{Q}^i{}_{\dot{1}},\\
 b_i:=\frac{1}{2\sqrt{P^0}}Q_{2i},\quad
 (b_i)^\dagger:=\frac{1}{2\sqrt{P^0}}\OLL{Q}^i{}_{\dot{2}},
\end{gather}
we find that
\begin{gather}
 \{a_i,(a_j)^\dagger\}=\delta_i{}^j,\\
 \{a_i,a_j\}=0,\qquad
 \{(a_i)^\dagger,(a_j)^\dagger\}=0,
\end{gather}
while
\begin{gather}
 \{b_i,(b_j)^\dagger\}=0,\quad
 \{b_i,b_j\}=0,\quad\{(b_i)^\dagger,(b_j)^\dagger\}=0,\\
 \{a_i,b_j\}=0,\quad\{(a_i)^\dagger,(b_j)^\dagger\}=0,\quad
 \{a_i,(b_j)^\dagger\}=0,\quad\{(a_i)^\dagger,b_j\}=0.
\end{gather}
Hence we obtain only $N$ sets of fermionic creation-annihilation operators,
with $b_i,\ (b_i)^\dagger$ merely represented as $0$ on
a Hilbert representation space.
These operators thus create total of $2^N$ states%
\footnote{
We work on a CPT invariant field theory so that we require
the CPT-conjugate of a multiplet is also included in the whole multiplet.
Thus unless the multiplet composed of the $2^{N}$ states
created purely by the supercharges
is CPT-self-conjugate, it needs its CPT-conjugate to be
accompanied so that the whole multiplet consists of
$2^{N+1}$ states.}
on a Clifford vacuum with helicity $\lambda$.
The maximum helicity for such representations is clearly
$N/2+\lambda$.

\subsubsection{Multiplet with Nonzero Central Charges}
\label{sec-centralcharge}
Finally let us consider the multiplets on which central charges take
nonzero values. We take a massive rest frame, by the reason
which will be clear shortly, with momentum $P^2=m^2,\ P^\mu=(m,\bs{0})$.
The relevant anticommutators are
\begin{gather}
 \{Q_{i\alpha},\OLL{Q}^j{}_{\Dbeta}\}=2m\delta_{\alpha\Dbeta}\delta_i{}^j,\\
 \{Q_{i\alpha},Q_{j\beta}\}=C_{\alpha\beta}Z_{ij},\qquad
 \{\OLL{Q}^i{}_{\Dalpha},\OLL{Q}^j{}_{\Dbeta}\}
    =C_{\Dalpha\Dbeta}(Z^\ast)^{ij}.
\end{gather}
Since $[Z_{ij},\text{any}]=[(Z^\ast)^{ij},\text{any}]=0$,
the central charges can be simultaneously diagonalized.
Further, because $Z_{ij}$ is antisymmetric,
we can take, by a suitable orthogonal (thus unitary)
transformation $(U_i{}^j)$
w.r.t.\ the internal indices, a basis in which
the central charges take the standard form as in
\begin{gather}
 Z_{ij}=U_{i}{}^k (Z^{\MR{std}})_{kl} U_j{}^l,\\
 (Z^{\MR{std}})_{ij}:=iC\otimes Z^{\MR{d}}\quad(N\text{ even})\qquad
 (Z^{\MR{std}})_{ij}:=
 \left(\begin{array}{cc}
         iC\otimes Z^{\MR{d}} & 0 \\
            0        & 0
       \end{array}\right)        \quad(N\text{ odd}),
\end{gather}
where $Z^{\MR{d}}=\diag(z^1,\cdots,z^{\MR{rank}(Z)})$ with
$z^l\geqslant 0$ being the eigenvalues of $(Z_{ij})$ and
the tensor product
$(A_i{}^k)\otimes(B_j{}^l)=(A_i{}^k B_j{}^l)$ is represented by
the dictionary order
\begin{equation}
 (i,j)=(1,1),\cdots,(n,1),\cdots,(1,n),\cdots,(n,n),
\end{equation}
and similarly for $(k,l)$.
Then by pursuing the transformation
\begin{equation}
 \tilde{Q}_{i\alpha}=(U^{-1})_i{}^j Q_{j\alpha},\quad
 (\tilde{Q}_{i\alpha})^\dagger
   =\tilde{\OLL{Q}}^i{}_{\Dalpha}
   =U_j{}^i \OLL{Q}^j{}_{\Dalpha},
\end{equation}
and by labeling as $i=(am)\ (m=1,\cdots,[N/2])$
just corresponding to the standard form $Z^{\MR{std}}$,
we find that
\begin{gather}
 \{\tilde{Q}_{am\alpha},(\tilde{Q}_{bn\beta})^\dagger\}
  =2m\delta_\alpha{}^\beta\delta_a{}^b\delta_m{}^n,\\
 \{\tilde{Q}_{am\alpha},\tilde{Q}_{bn\beta}\}
  =iC_{\alpha\beta}C_{ab}(Z^{\MR{d}})_{mn},\qquad
 \{(\tilde{Q}_{am\alpha})^\dagger,(\tilde{Q}_{bn\beta})^\dagger\}
  =iC^{\alpha\beta}C^{ab}(Z^{\MR{d}})^{mn}.
\end{gather}
Note here that for the case of $N$ odd we generalize the diagonal
matrix $Z^{\MR{d}}$ as
\begin{equation}
 Z^{\MR{d}}=\diag(z^1,\cdots,z^{\rank{Z}},0,\cdots,0).
\end{equation}
Hence letting
\begin{gather}
 a_{m\alpha}:=\frac{1}{\sqrt{2}}\left(
              \tilde{Q}_{1m\alpha}
               +iC_{\alpha\gamma}(\tilde{Q}_2{}^m{}_\gamma)^\dagger
               \right),\quad
 (a_{m\alpha})^\dagger
            :=\frac{1}{\sqrt{2}}\left(
              (\tilde{Q}_{1m\alpha})^\dagger
               -iC^{\alpha\gamma}\tilde{Q}_2{}^m{}_\gamma\right),\\
 b_{m\alpha}:=\frac{1}{\sqrt{2}}\left(
              \tilde{Q}_{1m\alpha}
               -iC_{\alpha\gamma}(\tilde{Q}_2{}^m{}_\gamma)^\dagger
               \right),\quad
 (b_{m\alpha})^\dagger
            :=\frac{1}{\sqrt{2}}\left(
              (\tilde{Q}_{1m\alpha})^\dagger
               +iC^{\alpha\gamma}\tilde{Q}_2{}^m{}_\gamma\right),
\end{gather}
we obtain
\begin{gather}
 \{a_{m\alpha},(a_{n\beta})^\dagger\}
     =\delta_{\alpha\beta}(2m\delta_{mn}+(Z^{\MR{d}})_{mn}),\quad
 \{b_{m\alpha},(b_{n\beta})^\dagger\}
     =\delta_{\alpha\beta}(2m\delta_{mn}-(Z^{\MR{d}})_{mn}),\\
 \{a_{m\alpha},a_{n\beta}\}=0,\quad
 \{b_{m\alpha},b_{n\beta}\}=0,\quad
 \{(a_{m\alpha})^\dagger,(a_{n\beta})^\dagger\}=0,\quad
 \{(b_{m\alpha})^\dagger,(b_{n\beta})^\dagger\}=0,\\
 \{a_{m\alpha},b_{n\beta}\}=0,\quad
 \{(a_{m\alpha})^\dagger,(b_{n\beta})^\dagger\}=0,\quad
 \{a_{m\alpha},(b_{n\beta})^\dagger\}=0,\quad
 \{(a_{m\alpha})^\dagger,b_{n\beta}\}=0.
\end{gather}
The left hand side of the first two equations is positive definite,
so is the right, which leads to
\begin{equation}
 2m\geqslant z^l,\quad\text{for}\ l=1,\cdots,\rank{Z}.
\end{equation}
This is the BPS bound condition \eRef{BPS}. Especially
massless states can not have nonzero central charge, this is why
we consider the massive representations here.

If central charges satisfy the condition
\begin{equation}
 z^i=2m\quad\text{for}\ i=1,\cdots,r\leqslant\rank{Z},
\end{equation}
the corresponding operators $b_i,\ (b_i)^\dagger$ vanish.
Then we have total of $2(N-r)$ sets of creation-annihilation
operators. The irreducible representations constructed by
such operators on a Clifford vacuum with spin $j$ (or helicity $\lambda$)
have total of $2^{2(N-r)}$ states and the maximum spin (or helicity)
is $(N-r)/2+j$ (or $(N-r)+\lambda$). If $r=\rank(Z)=N/2$ (for $N$ even)
the representations are the same as those in the massless case.
Note in particular for a system with some central charges
the maximum spin (helicity) can be consistently reduced.

\section{Twisted Supersymmetry in Four Dimensions}
\label{sec-twistedSUSY}
As a variant of the notion of supersymmetry, one
could consider ``twisted'' supersymmetry.
Since, as noted above, this subject serves one of the main
motivations to this article, let us here examine the technical
aspect of twisted supersymmetry.

\subsection{A Historical Review}
Twisted supersymmetry was first proposed by Witten%
~\cite{Witten-1,Witten-2}
in the context of topological field theory (TFT). %
Topological field theories are independent on metrics of the base
manifolds on which those theories are defined so that are only
sensitive to topological structures of the manifolds.
Witten showed that, firstly, so called the Donaldson invariants on a four
manifold~\cite{Donaldson} can be computed within the framework of
a quantum field theory,
which we call the Donaldson-Witten theory,
and, secondly, the Donaldson-Witten theory can be obtained by twisting
$N=2$ super Yang-Mills theory. Shortly after the Witten's work,
that the twisted supersymmetry could be
interpreted as a kind of BRST symmetry so that the Donaldson-Witten
theory is derived by a BRST quantization of topological
Yang-Mills theory in four dimensions%
~\cite{Baulieu-Singer,Brooks-Montano-Sonnenschein,Labastida-Pernici,%
Birmingham-Rakowski-Thompson}.
These facts implies that as twisted supersymmetric theories one might
construct some topological models. Actually, for instance,
topological matters in two~\cite{Eguchi-Yang,Labastida-Llatas}
and four~\cite{Alvarez-Labastida}
dimensions were shown to be
constructed as twisted $N=2$ supersymmetric theories in two and four
dimensions, respectively. These observations were further
generalized to twisting of $N=4$ supersymmetric theories.
Yamron introduced~\cite{Yamron} two inequivalent types of twisting
$N=4$ super Yang-Mills theories, and pointed out that one more type of
twisting would be possible. Then Vafa and Witten constructed%
~\cite{Vafa-Witten} a topological theory as a twisted Yang-Mills theory
with twisting procedure equivalent to one of two introduced by Yamron.
Yamron's third type of twist was later
considered by Marcus~\cite{Marcus} and Blau and Thompson%
~\cite{Blau-Thompson}. These works were analyzed in detail from
the viewpoint of the Mathai-Quillen formalism in \cite{Labastida-Lozano}.

We also remark that one alternative twisting of supersymmetry was
introduced in%
~\cite{Kawamoto-Tsukioka,Kato-Kawamoto-Uchida,Kato-Kawamoto-Miyake,%
Kato-Miyake}.
There twisted supersymmetry is strongly motivated by
the Dirac-K\"{a}hler mechanism, and in fact the twisting process
can be understood
to be essentially the same as the Dirac-K\"ahler fermion formulation.
We thus call this twist the Dirac-K\"{a}hler twist.
Here we emphasize that the Dirac-K\"{a}hler twist, in four dimensions,
is equivalent
to the Yamron's third type as we will see later and, in two dimensions, to the
two-dimensional twist mentioned above,
although the motivations by which these twisting procedures
are considered might be quite different.

In the following, we briefly review how these kinds of twisted
supersymmetry in four dimensions can be defined,
and list the corresponding twisted superalgebras.

\subsection{General  Twisting Procedure}
Twist of supersymmetry can be understood to be the process
of identifying (some part of) the internal symmetry ($R$-symmetry)
with (some part of) the spacetime
symmetry (Lorentz symmetry) and then defining a new representation
of spacetime symmetry for the original supercharges.
Such a process may include breaking of the internal symmetry group
into some subgroups, one of which is isomorphic to (a subgroup of) the
Lorentz group, and then taking the diagonal sum%
\footnote{More precisely, the diagonal sum of representations of
the algebras corresponding to the internal symmetry and the spacetime
symmetry. The corresponding representation space is the tensor
product of the original representation spaces.} of those
of the isomorphic subgroups.

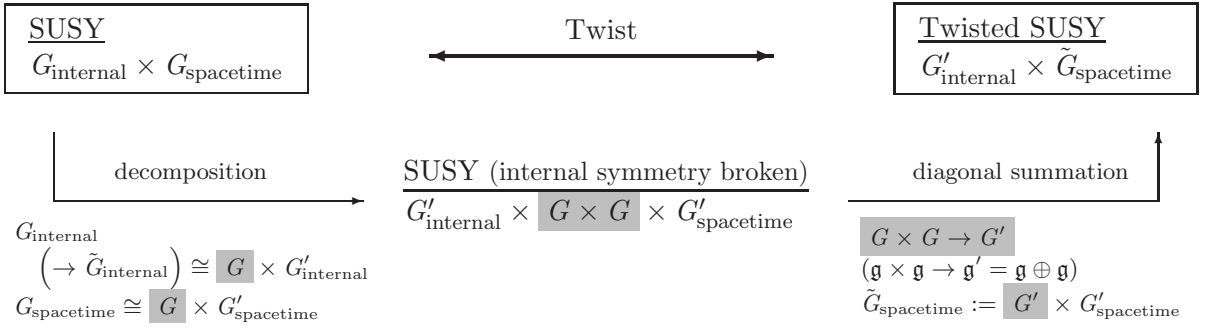
\begin{figure}
\begin{center}
 \setlength{\unitlength}{1.3pt}
 \fbox{\begin{tabular}{l}
    $\UL{\text{SUSY}}$ \\
    $\GR{G}_{\text{internal}}\times\GR{G}_{\text{spacetime}}$
 \end{tabular}}
 \hfill
 \begin{picture}(100,10)
    \allinethickness{1pt}
    \put(50,0){\vector(-1,0){50}}
    \put(50,0){\vector(1,0){50}}
    \put(50,5){\makebox(0,0)[b]{Twist}}
 \end{picture}
 \hfill
 \fbox{\begin{tabular}{l}
    $\UL{\text{Twisted SUSY}}$ \\
    $\GR{G}'_{\text{internal}}\times\tilde{\GR{G}}_{\text{spacetime}}$
 \end{tabular}}
 \vspace{2em} \\
 \mbox{\rule[-40\unitlength]{0mm}{0mm}%
 \begin{picture}(100,10)(0,50)
%    \allinethickness{1pt}
    \footnotesize
    \put(10,70){\line(0,-1){20}}
    \put(10,50){\vector(1,0){90}}
    \put(50,55){\makebox(0,0)[b]{decomposition}}
    \put(50,45){\makebox(0,0)[t]{%
     \begin{tabular}{l}
      $\GR{G}_{\MR{internal}}$ \\
      \(
        \quad\left(\RA\tilde{\GR{G}}_{\MR{internal}}\right)
        \cong\colorbox{LightGray}{$\GR{G}$}\times\GR{G}'_{\MR{internal}}
      \)                       \\
      \(
        \GR{G}_{\MR{spacetime}}
        \cong\colorbox{LightGray}{$\GR{G}$}\times\GR{G}'_{\MR{spacetime}}
      \)
     \end{tabular}}}
%    \put(0,10){\circle{1}}
%    \put(100,60){\circle{1}}
 \end{picture}}
 \hfil
 \begin{tabular}{l}
    $\UL{\text{SUSY \small(internal symmetry broken)}}$ \\
    \(
      \GR{G}'_{\text{internal}}\times
      \colorbox{LightGray}{$\GR{G}\times\GR{G}$}\times
      \GR{G}'_{\text{spacetime}}
    \)
 \end{tabular}
 \hfil
 \mbox{\rule[-40\unitlength]{0mm}{0mm}%
 \begin{picture}(100,10)(0,50)
%    \allinethickness{1pt}
    \footnotesize
    \put(90,50){\vector(0,1){20}}
    \put(0,50){\line(1,0){90}}
    \put(50,55){\makebox(0,0)[b]{diagonal summation}}
    \put(50,45){\makebox(0,0)[t]{%
     \begin{tabular}{l}
      \colorbox{LightGray}{$\GR{G}\times\GR{G}\RA\GR{G}'$} \\
      $(\AL{g}\times\AL{g}\RA\AL{g}'=\AL{g}\oplus\AL{g})$  \\
      \(
       \tilde{\GR{G}}_{\text{spacetime}}
       :=\colorbox{LightGray}{$\GR{G}'$}\times\GR{G}'_{\text{spacetime}}
      \)
     \end{tabular}}}
%    \put(0,10){\circle{1}}
%    \put(100,60){\circle{1}}
 \end{picture}}
\end{center}
\caption{Schematic description of twisting process}
\label{fig-twist}
\end{figure}

To be more specific, let us consider a supersymmetric theory with
internal symmetry group $G_{\text{internal}}$ and spacetime
symmetry group $G_{\text{spacetime}}$. For an $N$-extended
supersymmetric theory in four dimensions, these symmetry groups
should be taken as $\GR{SU}(N)$ (or $\GR{U}(N)$) and
$\GR{P}(1,3)$ (Poincar\'{e} group), respectively.
We assume that the spacetime symmetry group can be decomposed
irreducibly as
$\GR{G}_{\text{spacetime}}\cong \GR{G}\times\GR{G}'_{\text{spacetime}}$.
A twisting procedure for this system can be taken if there exists
a subgroup $\GR{G}'$ of $\GR{G}_{\text{internal}}$ such that
$\GR{G}_{\text{internal}}\cong\GR{G}'\times\GR{G}'_{\text{internal}}$
and also $\GR{G}'\cong\GR{G}$ (isomorphic).
However, such decomposion of the internal symmetry group
may not be possible in general unless we break the group into
its appropriate subgroup. Here, to illustrate the twisting process,
we first break the internal symmetry as
$\GR{G}_{\text{internal}}\RA \tilde{\GR{G}}_{\text{internal}}$,
and then assume $\tilde{\GR{G}}_{\text{internal}}$ can be decomposed as
\(
 \tilde{\GR{G}}_{\text{internal}}
  \cong\GR{G}'\times\GR{G}'_{\text{internal}}.
\)
Since $\GR{G}'\cong \GR{G}$, we denote the former simply by $\GR{G}$
from now on.
We therefore have a supersymmetric theory with associated symmetries
\(
 \GR{G}'_{\text{internal}}\times\GR{G}\times\GR{G}
 \times\GR{G}'_{\text{spacetime}}.
\)
Now we are at the position of twisting of this system.
The process is described schematically in Figure~\ref{fig-twist}
and consists of steps below:
\begin{computation}
Identification of two representations of $\GR{G}$.
To do this, we have to take two equivalent irreducible representations
$(\rho_{\MR{I}},V_{\MR{I}})$ and $(\rho_{\MR{S}},V_{\MR{S}})$
of the two $\GR{G}$'s, respectively.
\end{computation}

\begin{computation}
Taking the tensor product representation of the two
irreducible representations of $\GR{G}$.
More precisely---%
the original symmetry $\GR{G}\times\GR{G}$ should be represented
by $(\rho\times\rho,V\otimes V)$, where since
$(\rho_{\MR{I}},V_{\MR{I}})\cong(\rho_{\MR{S}},V_{\MR{S}})$
we simply denote them by $(\rho,V)$.
Then replace this by the tensor representation
$(\rho\otimes\rho,V\otimes V)$.
We also denote $\GR{G}\times\GR{G}$ simply by $\GR{G}'$
and understand that it is represented by the tensor product representation
as above. Note here that $\GR{G}'$ is nothing but
$\GR{G}$; we just rename it just for the readability.
Corresponding algebra now becomes
$\AL{g}'=\AL{g}\oplus\AL{g}$, which should be represented by
$(d\rho\otimes\bs{1}+\bs{1}\otimes d\rho,V\otimes V)$, or simply by
$(d\rho+d\rho,V\otimes V)$---.
\end{computation}

\begin{computation}
Decomposion of the representation into irreducible representations.
This step is equivalent to the Clebsch-Gordan decomposition of the
tensor product representation $(\rho\otimes\rho,V\otimes V)$, or
$(d\rho+d\rho,V\otimes V)$.
\end{computation}

\begin{computation}
Redefinition of the spacetime symmetry by
$\tilde{\GR{G}}_{\text{spacetime}}:= \GR{G}'\times\GR{G}'_{\text{spacetime}}$.
Since $\GR{G}'\equiv\GR{G}$, this definition certainly restore
the spacetime symmetry group though its representation is
changed from the original one.
\end{computation}

Through all these steps, we now obtain a system which has
the twisted supersymmetry with associated symmetries
$\GR{G}'_{\text{internal}}\times\tilde{\GR{G}}_{\text{spacetime}}$.
One of the main outcome of such twisting process
is that one can extract explicitly the scalar part of supercharges
of the original supersymmetry as the singlet representation
in the Clebsch-Gordan decomposition in the step (iii) above.
In the context of TFT,
such scalar supercharge can be interpreted as a topological
charge which assures the topological nature of those theories,
for, the scalar charge can be introduced on an arbitrary manifold
(because it is a scalar), and the charge is nilpotent
(if there is no central charge).
We review some specific examples of twisting process
of extended supersymmetries in four dimensions.

\subsection{Twisted $N=2$ Supersymmetry}
First consider the twist of $N=2$ supersymmetry in four-dimensional
Euclidean spacetime,
with $R$-symmetry group
$\GR{U}(2)\cong\GR{SU}(2)_{\MR{I}}\times\GR{U}(1)$
and Lorentz symmetry group
$\GR{SO}(4)\cong\GR{SU}(2)_{\MR{L}}\times\GR{SU}(2)_{\MR{R}}$.
The labels $\MR{I},\ \MR{L},\ \MR{R}$ specify the representations.
We denote the $R$-symmetry generators by $R_{ij}$
and the Lorentz generators by $J_{\alpha\beta},\ J_{\Dalpha\Dbeta}$,
where
\begin{equation}
 J_{\alpha\beta}:=\frac{1}{4}(\sigma^{\mu\nu})_{\alpha\beta}J_{\mu\nu},\quad
 J_{\Dalpha\Dbeta}
                :=\frac{1}{4}(\bsigma^{\mu\nu})_{\Dalpha\Dbeta}J_{\mu\nu}
\end{equation}
correspond to the decomposition above.
The superalgebra is given as \Ref{BeginOfAlgebra}--\Ref{EndOfAlgebra},
namely
\begin{xalignat}{2}
 \{Q_{i\alpha},Q_{j\beta}\}
  &=C_{ij}C_{\alpha\beta}Z,
 &
 \{\OLL{Q}^i{}_{\Dalpha},\OLL{Q}^j{}_{\Dbeta}\}
  &=C^{ij}C_{\Dalpha\Dbeta}Z,\\
 \{Q_{i\alpha},\OLL{Q}^j{}_{\Dbeta}\}
  &=\delta_i{}^j P_{\alpha\Dbeta},
 &
 [P_{\alpha\Dbeta},P_{\gamma\Ddelta}]
  &=0,\\
 [Q_{i\alpha},P_{\gamma\Ddelta}]
  &=0,
 &
 [\OLL{Q}^i{}_{\Dalpha},P_{\gamma\Ddelta}]
  &=0,\\
 [J_{\alpha\beta},Q_{i\gamma}]
  &=\frac{i}{2}C_{\gamma(\alpha}Q_{i\beta)},
 &
 [J_{\alpha\beta},\OLL{Q}^i{}_{\Dgamma}]
  &=0,\\
 [J_{\Dalpha\Dbeta},Q_{i\gamma}]
  &=0,
 &
 [J_{\Dalpha\Dbeta},\OLL{Q}^i{}_{\Dgamma}]
  &=\frac{i}{2}C_{\gamma(\Dalpha}\OLL{Q}^i{}_{\Dbeta)},\\
 [J_{\alpha\beta},P_{\gamma\Ddelta}]
  &=\frac{i}{2}C_{\gamma(\alpha}P_{\beta)\Ddelta},
 &
 [J_{\Dalpha\Dbeta},P_{\gamma\Ddelta}]
  &=\frac{i}{2}C_{\Ddelta(\Dalpha}P_{\gamma\Dbeta)},\\
 [J_{\alpha\beta},J_{\gamma\delta}]
  &=\frac{i}{2}C_{(\alpha|(\gamma}J_{\delta)|\beta)},
 &
 [J_{\Dalpha\Dbeta},J_{\Dgamma\Ddelta}]
  &=\frac{i}{2}C_{(\Dalpha|(\Dgamma}J_{\Ddelta)|\Dbeta)},\\
 [J_{\alpha\beta},J_{\Dgamma\Ddelta}]
  &=0,
 &
 [Z,\text{any}]
  &=0,\\
 [R_{ij},Q_{k\gamma}]
  &=\frac{i}{2}C_{k(i}Q_{j)\gamma},
 &
 [R_{ij},\OLL{Q}_{k\gamma}]
  &=\frac{i}{2}C_{k(i}\OLL{Q}_{j)\gamma},\\
 [R_{ij},R_{kl}]
  &=\frac{i}{2}C_{(i|(k}R_{l)|j)},
 &
 [R_{ij},P_{\alpha\Dbeta}]
  &=0,\\
 [R_{ij},J_{\alpha\beta}]
  &=0,
 &
 [R_{ij},J_{\Dalpha\Dbeta}]
  &=0,
\end{xalignat}
where $P_{\alpha\Dbeta}:=(\sigma^\mu)_{\alpha\Dbeta}P_\mu$.
Supercharges $Q_{i\alpha}$ and $\OLL{Q}^i{}_\Dalpha$ have
the $\GR{U}(1)$ charges%
\footnote{More precisely,
we only have $\GR{U}(1)/Z_4$ symmetry unless the central charge vanishes.}
$+1$ and $-1$, respectively, and the other generators have charges 0.
Notice here that we have taken the representation of the $R$-symmetry
to be equivalent to the representations of $\GR{SU}(2)_{\MR{L,R}}$
of the Lorentz symmetry. Indices $i,\ j$ can be raised or lowered
by the $\GR{SU}(2)$ invariant tensors $C_{ij},\ C^{ij}$.

Then according to the procedure above, we first identify
$\GR{SU}(2)_{\MR{I}}$ as either of $\GR{SU}(2)_{\MR{L}}$
or $\GR{SU}(2)_{\MR{R}}$, and after that, we take the tensor
product representation. In this case, these process should be described
as the identification of indices $i$ as either of $\alpha$
or $\Dalpha$, and then as taking the diagonal summation
\begin{equation}
 J'_{\alpha\beta}:=J_{\alpha\beta}+R_{\alpha\beta},\quad\text{or}\quad
 J'_{\Dalpha\Dbeta}:=J_{\Dalpha\Dbeta}+R_{\Dalpha\Dbeta}.
\end{equation}
We represent these twisting prescription simply as
\begin{equation}
 \begin{array}{ccc}
   \bs{2} & \RA & (\bs{2},\bs{1}) \\
     i    & \RA &  \alpha
 \end{array},
 \quad\text{or}\quad
 \begin{array}{ccc}
   \bs{2} & \RA & (\bs{1},\bs{2}) \\
     i    & \RA &  \Dalpha
 \end{array}.
\end{equation}
In what follows, we only treats the former diagonal summation
since these two are obviously equivalent, so that the tensor product
representation can be expressed as
$Q_{\alpha\beta},\ \OLL{Q}_{\alpha\Dbeta}$.
Next, we take the Clebsch-Gordan decomposition
\begin{equation}
 \begin{array}{ccccc}
  \bs{2}\otimes\bs{2} & =  & \bs{1} &  \oplus & \bs{3} \\
  Q_{\alpha\beta}     &\RA &  Q_{[\alpha\beta]}, &    &  Q_{(\alpha\beta)},
 \end{array}
\end{equation}
and, for simplicity, we denote as
\begin{equation}
 Q:=\frac{1}{2}C^{\alpha\beta}Q_{[\alpha\beta]},\quad
 H_{\alpha\beta}:=Q_{(\alpha\beta)},\quad
 G_{\alpha\Dbeta}:=\OLL{Q}_{\alpha\Dbeta}.
\end{equation}
The remaining task is to define a new representation of Lorentz symmetry.
This can be realized, as will be clear shortly, just by
considering the symmetry to be generated by
$J'_{\alpha\beta}$ and $J_{\Dalpha\Dbeta}$.

Thus we complete a twisting of the $N=2$ supersymmetry.
The resultant twisted $N=2$ superalgebra is readily derived from the
original superalgebra and the definitions of the twisted supercharges
above. Renaming $J'_{\alpha\beta}$ simply as $J_{\alpha\beta}$,
we obtain
\begin{xalignat}{2}
 \{Q,Q\}
  &=Z,
 &
 \{H_{\alpha\beta},H_{\gamma\delta}\}
  &=C_{(\alpha|(\gamma}C_{\delta)|\beta)}Z,\\
 \{G_{\alpha\Dbeta},G_{\gamma\Ddelta}\}
  &=C_{\alpha\beta}C_{\Dbeta\Ddelta}Z,
 &
 \{Q,H_{\alpha\beta}\}
  &=0,\\
 \{Q,G_{\alpha\Dbeta}\}
  &=P_{\alpha\Dbeta},
 &
 \{H_{\alpha\beta},G_{\gamma\Ddelta}\}
  &=C_{\alpha(\gamma}P_{\beta)\Ddelta},\\
 [Q,P_{\gamma\Ddelta}]
  &=0,
 &
 [H_{\alpha\beta},P_{\gamma\Ddelta}]
  &=0,\\
 [G_{\alpha\Dbeta},P_{\gamma\Ddelta}]
  &=0,
 &
 [P_{\alpha\Dbeta},P_{\gamma\Ddelta}]
  &=0,\\
 [J_{\alpha\beta},Q]
  &=0,
 &
 [J_{\Dalpha\Dbeta},Q]
  &=0,\\
 [J_{\alpha\beta},H_{\gamma\delta}]
  &=\frac{i}{2}C_{(\alpha|(\gamma}H_{\delta)|\beta)},
 &
 [J_{\Dalpha\Dbeta},H_{\gamma\delta}]
  &=0,\\
 [J_{\alpha\beta},G_{\gamma\Ddelta}]
  &=\frac{i}{2}C_{(\alpha|\gamma}G_{\beta)\Ddelta},
 &
 [J_{\Dalpha\Dbeta},G_{\gamma\Ddelta}]
  &=\frac{i}{2}C_{(\Dalpha|\Ddelta}G_{\gamma\Dbeta)},\\
 [J_{\alpha\beta},P_{\gamma\Ddelta}]
  &=\frac{i}{2}C_{(\alpha|\gamma}P_{\beta)\Ddelta},
 &
 [J_{\Dalpha\Dbeta},P_{\gamma\Ddelta}]
  &=\frac{i}{2}C_{(\Dalpha|\Ddelta}P_{\gamma\Dbeta)},\\
 [J_{\alpha\beta},J_{\gamma\delta}]
  &=\frac{i}{2}C_{(\alpha|(\gamma}J_{\delta)|\beta)},
 &
 [J_{\Dalpha\Dbeta},J_{\Dgamma\Ddelta}]
  &=\frac{i}{2}C_{(\Dalpha|(\Dgamma}J_{\Ddelta)|\Dbeta)},\\
 [J_{\alpha\beta},J_{\Dgamma\Ddelta}]
  &=0,
 &
 [Z,\text{any}]
  &=0.
\end{xalignat}
Note here that $J_{\alpha\beta}$ and $J_{\Dalpha\Dbeta}$ can
actually interpreted as the generators of a new Lorentz group.
The twisted charges
$Q,\ H_{\alpha\beta},\ G_{\alpha\Dbeta}$ transform under this
new Lorentz symmetry as if they are a scalar, a (anti-)selfdual tensor,
and a vector, as implied by their indices, and they have
the $\GR{U}(1)$ charges $+1,\ +1,\ -1$, respectively.

We could have taken, as mentioned above, another diagonal summation
based on the identification $i\RA\Dalpha$. The result is almost
the same as obtained above, except that the role of selfduality
and anti-selfduality should be exchanged. Thus we find that
there is essentially one twist for $N=2$ supersymmetry
in four dimensions.

Since we have obtain the twisted superalgebra, a twisted
supersymmetric field theory can be constructed based on the algebra.
However, we can find such a theory by directly twisting
a theory with the ordinary $N=2$ supersymmetry. As an example,
consider the $N=2$ super Yang-Mills theory. It contains
a gauge vector $A_{\alpha\Dbeta}$, Weyl spinors
$\psi_{i\alpha},\ \OLL{\psi}^i{}_\Dalpha$, complex scalars
$\phi,\ \OLL{\phi}$, and auxiliary fields $G_{ij}$. The twist above
can be applied to this theory and leads to the twisted field contents
\begin{align}
 A_{\alpha\Dbeta} &\RA A_{\alpha\Dbeta},\\
 \lambda_{i\beta} &\RA \lambda_{\alpha\beta} \RA \eta,\ \chi_{\alpha\beta},\\
 \Blambda_{i\Dbeta} &\RA \psi_{\alpha\Dbeta},\\
 \phi             &\RA \phi,\\
 \OLL{\phi}       &\RA \rho,\\
 G_{ij}           &\RA G_{\alpha\beta},
\end{align}
where $\eta=C^{\alpha\beta}\psi_{\alpha\beta}$,
$\chi_{\alpha\beta}=\psi_{(\alpha\beta)}$, etc.
The action and supertransformations can also be twisted based on
these (re-)definitions of fields and supercharges.

\subsection{Twisted $N=4$ Supersymmetries}
We now review the survey of possible twists of $N=4$ supersymmetry
in Euclidean four-dimensional spacetime.

\paragraph{Breaking of the Internal Symmetries}
According to the general procedure for twisting explained above,
we have to first take a representation of a subgroup of the $R$-symmetry
which can be identified as a representation of (a subgroup of)
the Lorentz symmetry. In Euclidean four-dimensional spacetime,
the Lorentz symmetry should be given by
$\GR{SO}(4)\cong\GR{SU}(2)_{\MR{L}}\times\GR{SU}(2)_{\MR{R}}$
as before. For an $N=4$ supersymmetric theory,
the full $R$-symmetry should be $\GR{SU}(4)$. Since $\GR{SU}(4)$
has no subgroup which can factorize directly as in
$\GR{SU}(4)\cong \GR{G}\times\GR{G}'$, we have to break the $\GR{SU}(4)$
into some suitable subgroup which is then factorized by
$\GR{SU}(2)$ or $\GR{SU}(2)\times\GR{SU}(2)\cong\GR{SO}(4)$
corresponding to the factorization of the Lorentz symmetry.

Actually, such nontrivial decomposition of $\GR{SU}(4)$ is determined uniquely
once we choose an $\GR{SU}(2)$ subgroup to be factorized,
since remaining factors must be the (maximal) subgroup which
makes this chosen $\GR{SU}(2)$ factor to be an invariant subgroup.
We have essentially two possibilities:
\begin{align}
 &\left\{
 \begin{aligned}
   \GR{SU}(4)&\RA \GR{SU}(2)\times\GR{SU}(2),\\
   \AL{su}(4)&\RA \AL{su}(2)\oplus\AL{su}(2)
                = \left\{\sqrt{1/2}\left(
                          X_{\MR{s,a}}^{12}+X_{\MR{s,a}}^{34}
                                   \right),\ 
                    \sqrt{1/2}X^1
                   +\sqrt{1/6}\left(\sqrt{2}X^3-X^2\right)\right\}\\
             &\phantom{\RA \AL{su}(2)\oplus\AL{su}(2)=-XX}
                  \oplus
                  \left\{\sqrt{1/2}\left(
                          X_{\MR{s,a}}^{13}+X_{\MR{s,a}}^{24}
                                   \right),\ 
                    \sqrt{1/3}\left(\sqrt{2}X^2+X^3\right)\right\},
 \end{aligned}\right.\\
 &\left\{
 \begin{aligned}
   \GR{SU}(4)&\RA \GR{SU}(2)\times\GR{SU}(2)\times\GR{U}(1),\\
   \AL{su}(4)&\RA \AL{su}(2)\oplus\AL{su}(2)\oplus\tilde{\AL{u}}(1)
                = \left\{X_{\MR{s,a}}^{12},\ X^1\right\}\oplus
                  \left\{X_{\MR{s,a}}^{34},\ 
                    \sqrt{1/3}\left(\sqrt{2}X^3-X^2\right)\right\}\\
             &\phantom{\RA \AL{su}(2)\oplus\AL{su}(2)=-XX}
                  \oplus\left\{\sqrt{1/3}\left(\sqrt{2}X^2+X^3\right)\right\},
 \end{aligned}\right.
\end{align}
where the $\GR{SU}(N)$ generators $X$, with $N=4$, are defined by
\Ref{SU-gen-1} and \Ref{SU-gen-2}.
These decompositions are further restricted by specifying representations
of each $\GR{SU}(2)$ factor.
In order to carry out twisting, the former should
be assigned as
\begin{equation}
 \GR{SU}(2)_{\MR{L}}\times\GR{SU}(2)_{\MR{I}},\quad
 \text{or equivalently},\quad
 \GR{SU}(2)_{\MR{R}}\times\GR{SU}(2)_{\MR{I}},
\end{equation}
which corresponds to
\begin{equation}
 \left\{\begin{aligned}
   \bs{4} &\RA (\bs{2},\bs{1})\oplus(\bs{2},\bs{1}),\\
     i    &\RA  a\alpha,
 \end{aligned}\right. \quad\text{or equivalently},\quad
 \left\{\begin{aligned}
   \bs{4} &\RA (\bs{1},\bs{2})\oplus(\bs{1},\bs{2}),\\
     i    &\RA  a\Dalpha,
 \end{aligned}\right.
\end{equation}
while the latter should be assigned as
\begin{gather}
 \GR{SU}(2)_{\MR{L}}\times\GR{SU}(2)_{\MR{I}}\times\GR{U},\quad
 \text{or equivalently},\quad
 \GR{SU}(2)_{\MR{R}}\times\GR{SU}(2)_{\MR{I}}\times\GR{U},
\end{gather}
or
\begin{equation}
 \GR{SU}(2)_{\MR{L}}\times\GR{SU}(2)_{\MR{R}}\times\GR{U},
\end{equation}
which corresponds to
\begin{gather}
 \left\{\begin{aligned}
   \bs{4} &\RA (\bs{2},\bs{1})\oplus(\bs{1},\bs{1})\oplus(\bs{1},\bs{1}),\\
     i    &\RA  (\alpha, a)\equiv a\oplus \alpha,
 \end{aligned}\right.\quad\text{or equivalently},\quad
 \left\{\begin{aligned}
   \bs{4} &\RA (\bs{1},\bs{2})\oplus(\bs{1},\bs{1})\oplus(\bs{1},\bs{1}),\\
     i    &\RA  (\Dalpha, a)\equiv \Dalpha\oplus a,
 \end{aligned}\right.
\end{gather}
or
\begin{equation}
 \left\{\begin{aligned}
   \bs{4} &\RA (\bs{2},\bs{1})\oplus(\bs{1},\bs{2}),\\
     i    &\RA  (\alpha,\Dalpha)\equiv \alpha\oplus\Dalpha.
 \end{aligned}\right.
\end{equation}
These three (out of five) decompositions of representations
provide inequivalent possible twisting of $N=4$ supersymmetry
\cite{Yamron,Labastida-Lozano}.
Let us now examine how supercharges are represented after each of these
twisting processes.

\paragraph{Vafa-Witten Twist}
This is the twist for the decomposition \cite{Yamron,Vafa-Witten}
\begin{equation}
 \GR{SU}(4)\RA \GR{SU}(2)_{\MR{L}}\times\GR{SU}(2)_{\MR{I}},\quad
 \left\{\begin{aligned}
   \bs{4} &\RA (\bs{2},\bs{1})\oplus(\bs{2},\bs{1}),\\
     i    &\RA  a\alpha.
 \end{aligned}\right.
\end{equation}
Diagonal summation is taken w.r.t.\ this $\GR{SU}(2)_{\MR{L}}$
and the one in the Lorentz group. The resultant twisted theory
thus has the symmetry group
$\GR{SU}(2)_{\MR{I}}\times\GR{SU}(2)_{\MR{L}'}\times\GR{SU}(2)_{\MR{R}}$.
Supercharges are twisted as
\begin{equation}
 \left\{\begin{aligned}
    Q_{i\beta}&\RA Q_{a\,\alpha,\beta}
               \RA Q_{a},\ Q_{a\alpha\beta},\\
    \OLL{Q}^i{}_{\Dbeta} &\RA Q^a_{\alpha\Dbeta},
 \end{aligned}\right.
\end{equation}
while the on-shell $N=4$ super Yang-Mills multiplet is twisted as
\begin{align}
 A_{\alpha\Dbeta}&\RA A_{\alpha\Dbeta},\\
 \lambda_{i\beta}&\RA \lambda_{a\,\alpha,\beta}
                  \RA \eta_a,\ \chi_{a\,\alpha\beta},\\
 \OLL{\lambda}^i{}_{\Dbeta}
                 &\RA \psi^a{}_{\alpha\Dbeta},\\
 \phi_{ij}       &\RA \phi_{a\alpha,b\beta}
                  \RA \varphi_{ab},\ G_{\alpha\beta},
\end{align}
where $_{ab}$ and $_{\alpha\beta}$ denote symmetric labeling
w.r.t.\ $a\LR b$ and $\alpha\LR\beta$, respectively.
Note here that $\phi_{ij}+\phi_{ji}=0$ so that
$\phi_{a\alpha,b\beta}$ must be antisymmetric w.r.t.\ $a\LR b$
($\alpha\LR\beta$) and at the same time be symmetric
w.r.t.\ $\alpha\LR\beta$ ($a\LR b$, respectively).

\paragraph{Half Twist}
This corresponds to \cite{Yamron}
\begin{equation}
 \GR{SU}(4)\RA \GR{SU}(2)_{\MR{L}}\times\GR{SU}(2)_{\MR{I}}
               \times\GR{U}(1),\quad
 \left\{\begin{aligned}
   \bs{4} &\RA (\bs{2},\bs{1})\oplus(\bs{1},\bs{1})\oplus(\bs{1},\bs{1}),\\
     i    &\RA  (a,\alpha)\equiv a\oplus\alpha.
 \end{aligned}\right.
\end{equation}
Diagonal summation is taken for this $\GR{SU}(2)_{\MR{L}}$,
so the resultant symmetry group is
\(
 \GR{SU}(2)_{\MR{I}}\times\GR{U}(1)\times
 \GR{SU}(2)_{\MR{L}'}\times\GR{SU}(2)_{\MR{R}}.
\)
Twisted supercharges are then
\begin{equation}
 \left\{\begin{aligned}
    Q_{i\beta}&\RA Q_{a\oplus\alpha,\beta}
              \RA Q^{(-1)}_{a\beta},\ Q^{(+1)},\ Q^{(+1)}_{\alpha\beta},\\
    \OLL{Q}^i{}_{\Dbeta}&\RA Q^{a\oplus\alpha}{}_{\Dbeta}
               \RA Q^{(+1)}_{a\Dbeta},\ Q^{(-1)}_{\alpha\Dbeta},
 \end{aligned}\right.
\end{equation}
where we have denoted the $\GR{U}(1)$ charges%
\footnote{These charges can be assigned essentially
as the eigenvalues of $\sqrt{2}X^2+X^3$.}
in the superscripts.
The $N=4$ on-shell SYM multiplet is twisted as
\begin{align}
 A_{\alpha\Dbeta} &\RA A^{(0)}_{\alpha\beta},\\
 \lambda_{i\beta}&\RA \lambda_{a\oplus\alpha\,\beta}
                  \RA \lambda^{(+1)}_{a\beta},\
                      \eta^{(-1)},\ \chi^{(-1)}_{\alpha\beta},\\
 \OLL{\lambda}^i{}_{\Dbeta}
                  &\RA \OLL{\lambda}^{a\oplus\alpha}{}_{\Dbeta}
                   \RA \psi^{(+1)}_{\alpha\Dbeta},\
                       \zeta^{(-1)\,a}_{\Dbeta},\\
 \phi_{ij}        &\RA \phi_{a\oplus\alpha,b\oplus\beta}
                   \RA B^{(-2)},\ C^{(+2)},\ 
                       G^{(0)}_{a\alpha}(=\phi_{a\beta},\ \phi_{b\alpha}).
\end{align}

\paragraph{Amphicheiral Twist}
The last one is for \cite{Yamron,Marcus,Blau-Thompson}
\begin{equation}
 \GR{SU}(4)\RA \GR{SU}(2)_{\MR{L}}\times\GR{SU}(2)_{\MR{R}}
               \times\GR{U}(1),\quad
 \left\{\begin{aligned}
   \bs{4} &\RA (\bs{2},\bs{1})\oplus(\bs{1},\bs{2}),\\
     i    &\RA  (\alpha,\Dbeta)\equiv \alpha\oplus\Dbeta.
 \end{aligned}\right.
\end{equation}
The twisted supercharges are
\begin{equation}
 \left\{\begin{aligned}
    Q_{i\gamma}&\RA Q_{\alpha\oplus\Dbeta,\gamma}
              \RA Q^{(+1)},\ Q^{(-1)}_{\gamma\Dbeta},
                  \ Q^{(+1)}_{\alpha\gamma},\\
    \OLL{Q}^i{}_{\Dgamma}&\RA \OLL{Q}^{\alpha\oplus\Dbeta}{}_\Dgamma
                \RA \tilde{Q}^{(+1)},\ \tilde{Q}^{(-1)}_{\alpha\Dgamma},
                    \ \tilde{Q}^{(+1)}_{\Dbeta\Dgamma}.
 \end{aligned}\right.
\end{equation}
Thus we find that such theory contains two scalar supercharges
with the same $\GR{U}(1)$ charges.
The $N=4$ SYM multiplet is now twisted as
\begin{align}
 A_{\alpha\Dbeta} &\RA A^{(0)}_{\alpha\Dbeta},\\
 \lambda_{i\gamma}&\RA \lambda_{\alpha\oplus\Dbeta\,\gamma}
                   \RA \tilde{\psi}^{(+1)}_{\gamma\Dbeta},\ 
                       \eta^{(-1)},\ \chi^{(-1)}_{\alpha\gamma},\\
 \OLL{\lambda}^i{}_{\Dgamma}
                  &\RA \OLL{\lambda}^{\alpha\oplus\Dbeta}{}_\Dgamma
                   \RA \psi^{(+1)}_{\alpha\Dgamma},\ 
                       \tilde{\eta}^{(-1)},\ 
                       \tilde{\chi}^{(-1)}_{\Dbeta\Dgamma},\\
 \phi_{ij}        &\RA \phi_{\alpha\oplus\Dbeta,\gamma\oplus\Ddelta}
                   \RA B^{(-2)},\ C^{(+2)},\ V^{(0)}_{\alpha\Dbeta}.
\end{align}

\subsection{Dirac-K\"{a}hler Twist}
Let us now consider one more type of twist which we call
the Dirac-K\"{a}hler twist (or the DK twist for short)%
~\cite{Kawamoto-Tsukioka,Kato-Kawamoto-Uchida,Kato-Kawamoto-Miyake,%
Kato-Miyake}.
As noted above, there are essentially three different types of
twist of $N=4$ supersymmetry in four dimensions.
Thus, the DK twist is expected to be equivalent to one of
the three. In fact, the twist, in four dimensions, will be recognized
as the amphicheiral twist listed at the end of the preceding section.

In principle, the DK twist can be defined in any dimensions.
Here again we consider Euclidean spacetime.
In Euclidean spacetime with dimension $D$, spacetime symmetry group should be
$\GR{SO}(D)\cong\GR{Spin}(D)$, where $\GR{Spin}(D)$ is the double
(and indeed the universal) cover of $\GR{SO}(D)$.
As for $R$-symmetry
group $\GR{G}_{\MR{I}}$, the DK twist restricts it to be
$\GR{Spin}(D)$, or, more generally, to be a group including
$\GR{Spin}(D)$ as a subgroup. The DK twist
is the process of taking the diagonal sum of these two $\GR{Spin}(D)$,
one is the Lorentz group and the other is the $R$-symmetry group,
and then identifying the resultant $\GR{Spin}(D)$ as a new Lorentz
symmetry group.

To be specific, consider supercharges $Q_{A\alpha}$ and
$\OLL{Q}^{A\alpha}$ in $D$-dimensional Euclidean spacetime
with $R$-symmetry $G_\MR{I}=\GR{Spin}(D)$.
For simplicity, we work on the case where $D$ is even. Then
$A$ and $\alpha$ are labels of $2^{D/2}$-component spinors.
The Dirac conjugate is defined as%
\footnote{See appendix~\ref{sec-Clifford} for these conventions.}
\begin{equation}
 \OLL{Q}^{A\alpha}=\sum_{B\beta}(Q_{B\beta})^\dagger
                    (\Gamma^0)_B{}^A(\Gamma^0)_\beta{}^\alpha.
\end{equation}
We also assume that the $\GR{Spin}(D)$-Majorana
(or $\GR{Spin}(D)$-Majorana-Weyl, if possible) conditions are imposed on the
supercharges, as in
\begin{equation}
 Q_{A\alpha}=(Q_\MR{c})_{A\alpha}:=C_{AB}C_{\alpha\beta}(\OLL{Q}^T)^{B\beta},
\end{equation}
where the charge conjugation matrices are such that
\begin{equation}
 C^T=\VE' C,\quad C^\dagger C=1,\quad
 C\gamma^\mu C^{-1}=\eta'(\gamma^\mu)^T,\quad
 C\gamma^i C^{-1}=\eta'(\gamma^i)^T,\quad
 \VE',\ \eta'=\pm 1.
\end{equation}
The DK twist now reads
\begin{equation}
 \GR{G}_{\MR{I}}\RA \GR{Spin}(D),\quad
 \left\{\begin{aligned}
   \bs{2^{D/2}} &\RA  \bs{2^{D/2}},\\
     A          &\RA  \alpha,\\
  Q_{A\beta}    &\RA Q_{\alpha\beta}.
 \end{aligned}\right.
\end{equation}
Using \eRef{completeness-Cliff}, we obtain a Clebsch-Gordan
decomposition as
\begin{equation}
\label{DKtwist}
 Q_{\alpha\beta}=\sum_{p=0}^D \frac{1}{p!}
                  Q_{\mu_1\cdots\mu_p}
		  (\gamma^{\mu_1\cdots\mu_p}C^{-1})_{\alpha\beta},\quad
 Q_{\mu_1\cdots\mu_p}
   := (-1)^{p(p-1)/2}\frac{1}{2^{D/2}}
      (C\gamma_{\mu_1\cdots\mu_p})^{\alpha\beta} Q_{\alpha\beta}.
\end{equation}
We call $Q_{\mu_1\cdots\mu_p}$ the DK twisted supercharges.
Note that they contain particularly the scalar charge
$\DS Q:=\frac{1}{2^{D/2}}C^{\alpha\beta}Q_{\alpha\beta}$
for $p=0$ as well as
the pseudo scalar charge
$\tilde{Q}:=\frac{1}{2^{D/2}}(C\Gamma^5)^{\alpha\beta}Q_{\alpha\beta}$
for $p=D$.
Note also that by taking the linear combinations
\begin{align}
 Q_{\mu_1\cdots\mu_p}^{\pm}
 :&=\frac{1}{2}\left(
    Q_{\mu_1\cdots\mu_p}\pm
    \frac{1}{(D-p)!}(-1)^{D/4+p(p+1)/2}
      \VE_{\mu_1\cdots\mu_p}{}^{\mu_{p+1}\cdots\mu_D}
      Q_{\mu_{p+1}\cdots\mu_D}\right) \notag\\
  &=\frac{1}{2^{D/2}}(-1)^{p(p-1)/2}
    \left(C\gamma_{\mu_1\cdots\mu_p}\frac{1}{2}\left(1\pm\Gamma^5\right)
    \right)^{\alpha\beta}Q_{\alpha\beta},
\end{align}
we can explicitly denote the chiralities of the supercharges.

The corresponding superalgebra should be
\begin{gather}
 \{Q_{A\alpha},Q_{B\beta}\}
  =2C^{-1}_{AB}(\gamma^\mu C^{-1})_{\alpha\beta}P_\mu
    +\Gamma_{\alpha\beta}Z_{AB},\\
 [J_{\mu\nu},Q_{A\alpha}]
  =\frac{i}{2}(\gamma_{\mu\nu})_\alpha{}^\beta Q_{A\beta},\qquad
 [R_{ij},Q_{A\alpha}]
  =\frac{i}{2}(\gamma_{ij})_A{}^B Q_{B\alpha},
\end{gather}
where, just for simplicity, we have chosen $\eta'=1$
so that $C^{-1}_{AB}$ and $(\gamma_\mu C^{-1})_{\alpha\beta}$ should
become both antisymmetric or both symmetric, and
$\Gamma_{\alpha\beta}$ is restricted to be $C^{-1}_{\alpha\beta}$ or
$(\Gamma^5 C^{-1})_{\alpha\beta}$ or both by the Coleman-Mandula theorem.
The DK twisted algebra can be readily obtained using
\eRef{orthonormal} and the definition of the twisted supercharges.
For example, in a simple case
$\Gamma_{\alpha\beta}Z_{AB}=C_{\alpha\beta}C_{AB}Z$, we have
\begin{align}
 \{Q_{\mu_1\cdots\mu_p},Q_{\nu_1\cdots\nu_{p-1}}\}
 &=\frac{2}{2^{D/2}}\eta_{\mu_1\cdots\mu_p,\nu_1\cdots\nu_p}P^{\nu_p},\quad
 \{Q_{\mu_1\cdots\mu_p},Q_{\nu_1\cdots\nu_p}\}
 =\frac{1}{2^{D/2}}\eta_{\mu_1\cdots\mu_p,\nu_1\cdots\nu_p}Z,\\
 [J_{\mu\nu},Q_{\rho_1\cdots\rho_p}]
 &=\frac{i}{2}\VE'(-1)^{p(p-1)/2}\Bigl(
    Q_{\mu\nu\rho_1\cdots\rho_p}
    +\sum_{i=1}^p\eta_{[\mu|\rho_i}
         Q_{\rho_1\cdots\nu]\cdots\rho_p} \notag\\
 &\phantom{\sgn(C^{-1}\gamma_{\rho_1\cdots\rho_p})++}
    +\sum_{1\leqslant i<j\leqslant p}(-1)^{i+j}
      (\eta_{[\mu|\rho_i}\eta_{\nu]\rho_j})
      Q_{\rho_1\cdots\breve{\rho_i}\cdots\breve{\rho_j}\cdots\rho_p}
                                 \Bigr),\\
 [R_{\mu\nu},Q_{\rho_1\cdots\rho_p}]
 &=\frac{i}{2}\VE'(-1)^{p(p-1)/2}\Bigl(
    -Q_{\mu\nu\rho_1\cdots\rho_p}
    +\sum_{i=1}^p\eta_{[\mu|\rho_i}
         Q_{\rho_1\cdots\nu]\cdots\rho_p} \notag\\
 &\phantom{\sgn(C^{-1}\gamma_{\rho_1\cdots\rho_p})++}
    -\sum_{1\leqslant i<j\leqslant p}(-1)^{i+j}
      (\eta_{[\mu|\rho_i}\eta_{\nu]\rho_j})
      Q_{\rho_1\cdots\breve{\rho_i}\cdots\breve{\rho_j}\cdots\rho_p}
                                 \Bigr).
\end{align}
It is then crucial that by taking the diagonal summation
$J'_{\mu\nu}:=J_{\mu\nu}+R_{\mu\nu}$, we have
\begin{equation}
 [J'_{\mu\nu},Q_{\rho_1\cdots\rho_p}]
 =i\VE'(-1)^{p(p-1)/2}
    \sum_{i=1}^p\eta_{[\mu|\rho_i}
         Q_{\rho_1\cdots\nu]\cdots\rho_p},
\end{equation}
which can be identified as the ordinary Lorentz transformations
of a tensor quantity
 so that, in particular,
$[J'_{\mu\nu},Q]=0$.

Finally let us take a closer look at the DK twist of $N=4$ supersymmetry
in Euclidean four-dimensional spacetime. The $R$-symmetry
has to be $\GR{Spin}(4)\cong\GR{SO}(4)$. Twisted supercharges
are defined as in \Ref{DKtwist}. Here we redefine them, changing some
trivial constant factors including sings, as
\begin{alignat}{3}
 Q&:=C^{\alpha\beta}Q_{\alpha\beta},
 &\quad
 Q_\mu&:=(C\gamma_\mu)^{\alpha\beta}Q_{\alpha\beta},
 &\quad
 Q_{\mu\nu}
  &:=(C\gamma_{\mu\nu})^{\alpha\beta}
        Q_{\alpha\beta},\\
 \tilde{Q}&:=(C\gamma^5)^{\alpha\beta}
        Q_{\alpha\beta},
 &\quad
 \tilde{Q}_\mu
  &:=(C\gamma^5\gamma_\mu)^{\alpha\beta}
        Q_{\alpha\beta},
 &\quad
 \tilde{Q}_{\mu\nu}
  &:=(C\gamma^5\gamma_{\mu\nu})^{\alpha\beta}
        Q_{\alpha\beta}.
\end{alignat}
Now take specific representations
\begin{equation}
 \gamma^\mu:=\left(\begin{array}{cc}
               0       &   \sigma^\mu \\
           \bsigma^\mu &   0
             \end{array}\right),\quad
 \sigma^\mu=(\bs{1},i\tau^i),\quad
 \bsigma^\mu=(\bs{1},-i\tau^i),
\end{equation}
so that
\begin{gather}
 C:=i\gamma^1\gamma^3
           =\left(\begin{array}{cc}
               -\tau^2 &   0 \\
                  0 &   -\tau^2
             \end{array}\right),\quad
 \gamma^5:=\gamma^0 \gamma^1 \gamma^2 \gamma^3
           =\left(\begin{array}{cc}
                  1 &  0 \\
                  0 & -1
             \end{array}\right).
\end{gather}
Then in terms of the Weyl basis
\begin{equation}
 \psi_\alpha\RA
   \left(\begin{array}{c}
      \psi_\alpha \\
      \OLL{\psi}^\Dalpha
   \end{array}\right),\quad
 Q_{\alpha\beta}\RA\left(\begin{array}{cc}
  Q_{\alpha\beta}   &  Q_{\alpha}{}^{\Dbeta} \\
  Q^{\Dalpha}{}_{\beta}  &  Q^{\Dalpha\Dbeta}
 \end{array}\right),
\end{equation}
we find that
\begin{xalignat}{2}
 Q&=C_{(2)}^{\alpha\beta}Q_{\alpha\beta}
             +C^{(2)}_{\Dalpha\Dbeta}Q^{\Dalpha\Dbeta},
 &
 \tilde{Q}
  &=C_{(2)}^{\alpha\beta}Q_{\alpha\beta}
             -C^{(2)}_{\Dalpha\Dbeta}Q^{\Dalpha\Dbeta},\\
 Q_\mu
  &=-(\sigma_\mu)_{\alpha\Dbeta}Q^{\alpha\Dbeta}
             +(\bsigma_\mu)^{\Dalpha\beta}Q_{\Dalpha\beta},
 &
 \tilde{Q}_\mu
  &=(\sigma_\mu)_{\alpha\Dbeta}Q^{\alpha\Dbeta}
             +(\bsigma_\mu)^{\Dalpha\beta}Q_{\Dalpha\beta},\\
 Q_{\mu\nu}
  &=(\sigma_{\mu\nu})^{\alpha\beta}Q_{\alpha\beta}
             +(\bsigma_{\mu\nu})_{\Dalpha\Dbeta}Q^{\Dalpha\Dbeta},
 &
 \tilde{Q}_{\mu\nu}
  &=(\sigma_{\mu\nu})^{\alpha\beta}Q_{\alpha\beta}
             -(\bsigma_{\mu\nu})_{\Dalpha\Dbeta}Q^{\Dalpha\Dbeta}.
\end{xalignat}
Thus it is clear that they are essentially the same as
supercharges in the amphicheiral twist above, for
\begin{xalignat}{2}
 Q_{[\alpha\beta]}&=\frac{1}{2}(Q+\tilde{Q})C^{(2)}_{\alpha\beta},
  &\quad
 Q^{[\Dalpha\Dbeta]}&=\frac{1}{2}(Q-\tilde{Q})C_{(2)}^{\Dalpha\Dbeta},\\
 Q_{\alpha\Dbeta}&=\frac{1}{4}(Q_\mu+\tilde{Q}_\mu)(\sigma^\mu)_{\alpha\Dbeta},
  &\quad
 Q^{\Dalpha\beta}&=-\frac{1}{4}(Q_\mu-\tilde{Q}_\mu)
           (\bsigma^\mu)^{\Dalpha\beta}, \\
 Q_{(\alpha\beta)}&=\frac{1}{4}(Q_{\mu\nu}+\tilde{Q}_{\mu\nu})
           (\sigma^{\mu\nu})_{\alpha\beta},
  &\quad
 Q^{(\Dalpha\Dbeta)}&=\frac{1}{4}(Q_{\mu\nu}-\tilde{Q}_{\mu\nu})
           (\bsigma^{\mu\nu})^{\Dalpha\Dbeta},
\end{xalignat}
and the $\GR{U}(1)$ charges in the amphicheiral twist
can be identified as products of the chiralities
w.r.t.\ to the two spinor indices, one originates
from those for the $R$-symmetry and the other from
those for the Lorentz symmetry.
We therefore understand that the Dirac-K\"{a}hler twist
for $N=4$ supersymmetry in four dimensions is equivalent
to the amphicheiral twist, though the former is applicable
in any dimensions.

\section{Superspace Formulation of $N=D=4$ SYM with a Central Charge}
\label{sec-USp-model}
In this section, a superspace formulation using superconnections and
supercurvatures is applied to $N=4$ super Yang-Mills theory
with a central charge in four dimensions.
There is no known superspace formulation for
$N=4$ supersymmetric theories with manifest supersymmetry.
Thus in particular the $N=4$ model, whose off-shell formulation
is known~\cite{Sohnius-Stelle-West-1,Sohnius-Stelle-West-2}, is
carefully surveyed to see how difficulties occur in the extension
to its manifest supersymmetric superspace formulation.

\subsection{A Historical Review}
Super Yang-Mills theory (SYM) is defined to be a supersymmetric
gauge theory which contains a vector gauge boson with spin $1$
and has no field contents with higher spins.
From the theoretical point of view, an off-shell formulation
of such theories is quite important. Among others, superspace
formulations, especially that using superconnections and
supercurvatures on the superspace, should serve the most natural
and powerful scheme to construct the off-shell super
Yang-Mills theories.

Formulations of super Yang-Mills theories in four dimensions
were developed long ago with $N=1$%
~\cite{Wess-Zumino-1,Ferrara-Zumino,Salam-Strathdee},
for $N=2$~\cite{Ferrara-Zumino,Salam-Strathdee,Fayet-2}
and for (on-shell) $N=4$~\cite{Gliozzi-Scherk-Olive} supersymmetries.
Counterparts in two dimensions were also found as well%
~\cite{Ferrara,DiVecchia-Ferrara}.
Later, systematic construction of supersymmetric gauge theories
on superspace with superconnections and supercurvatures
was proposed for both Abelian
and non-Abelian gauge groups with $N=1$ supersymmetries in four
dimensions~\cite{Wess-Zumino-2,Wess}.
Then it was generalized to superspace formulation
for four-dimensional $N=2$ super Yang-Mills theory~\cite{Grimm-Sohnius-Wess}
and also applied to many works, %including two-dimensional case,
%~\cite{Hamada-Takao}
including, for example, the $N=2$ twisted
superspace formulation
with twisted superconnections and supercurvatures
of some topological field theories in four and two dimensions%
~\cite{Alvarez-Labastida,Kato-Kawamoto-Miyake}.
However further generalization of such formulations simply to
$N=4$ super Yang-Mills theory based on the internal $\GR{SU}(4)$
symmetry%
\footnote{
We call this theory the $\GR{SU}(4)$ model.}
has failed unless it breaks to
the $N=2$ super Yang-Mills or merely results in an on-shell
$N=4$ formulation~\cite{Sohnius-1}.
In fact, no superspace formulation for the $\GR{SU}(4)$ SYM
which contains fields with spins only less than 1 has
been constructed.
Even an off-shell formulation for the $\GR{SU}(4)$ model which includes
the usual on-shell field contents, namely,
1 vector, 4 Majorana spinors and 6 real scalars%
~\cite{Brink-Schwarz-Scherk}, as well as complete lists of
an auxiliary fields, has not been constructed.
However, an $N=4$ supersymmetric gauge theory which
contains higher spin fields as well as higher derivatives in an action
has been formulated on twisted superspace in four dimensions%
~\cite{Kato-Kawamoto-Miyake}. This is the only known example
of $N=4$ superspace formulation with manifest supersymmetry.

One of the essential difficulties to construct $N=4$ super Yang-Mills
theory on superspace comes from the structure of the field contents
of the multiplets in $N=4$ supersymmetry.
As we have seen in section \ref{sec-massless},
massless representations, including vector multiplet for super
Yang-Mills theory, of $N$-extended superalgebra has maximum
helicity (spin) $\geqslant N/2$.
Thus, an $N=4$ supersymmetric theory
may contain the maximum helicity $\geqslant 4/2=2$.
Especially the simple generalization of%
~\cite{Grimm-Sohnius-Wess} to the $\GR{SU}(4)$ SYM
on superspace is based on the off-shell
Clifford vacuum with helicity $0$, so that it must contain
fields with helicity $>1$ (up to $2$) which are not allowed
in the SYM multiplet. Then one may consider if one can
consistently prohibit for such higher spin fields to emerge
by imposing some suitable constraints. However no such procedure
has succeeded~\cite{Sohnius-1} as noted above.
This is one of the main reason
one cannot complete superspace formulation of SYM with
the usual field contents.

In \cite{Sohnius-Stelle-West-1,Sohnius-Stelle-West-2}, it was shown
that
$N=4$ super Yang-Mills with another kind of $N=4$ SYM multiplet
which includes a central charge in the superalgebra based on
the internal $\GR{USp}(4)$ symmetry
can be formulated off-shell both for Abelian and non-Abelian
gauge groups%
\footnote{We call this theory the $\GR{USp}(4)$ model.}.
The basic idea is to introduce a central charge
into the $N=4$ super Yang-Mills theory in order to reduce
the possible maximum spin of the filed contents so that
the theory contains spins less than 1. We have seen the
mechanism how central charges reduce the on-shell spins (helicities).
Such a mechanism to reduce the spins successfully works
in their formulations.
However, they did not construct the corresponding superspace
formulation in completely closed in four dimensions.

There also proposed in \cite{Sohnius-Stelle-West-1,Sohnius-Stelle-West-2}
off-shell, though not on superspace, $N=2$
super Yang-Mills theory with a central charge.
Corresponding and applied superspace formulations were later developed
with a $N=1$ vector superfield~\cite{Milewski}
and, by gauging the central charge%
~\cite{Claus-deWit-Faux-Kleijn-Siebelink-Termonia,Gaida},
with $N=2$ superconnections and supercurvatures%
~\cite{Hindawi-Ovrut-Waldram} and
with the harmonic superspace formalism%
~\cite{Dragon-Ivanov-Kuzenko-Sokatchev-Theis}, to be contrasted with
the $N=2$ super Yang-Mills theory with ungaged central charge%
~\cite{Sohnius-2}.

On the other hand, the $N=4$ super Yang-Mills theory ($\GR{USp}(4)$
model) has not been constructed explicitly on superspace, though
briefly mentioned in~\cite{Sohnius-Stelle-West-2}.
In the following, we will thus attempt to formulate
the $\GR{USp}(4)$ SYM with a central charge.
Since it is formulated off-shell, it seems possible
to develop the corresponding superspace formulation.

\subsection{The $\GR{USp}(4)$ Superalgebra}
Let us now go on to the analysis of the $\GR{USp}(4)$ model.
First recall that, as noted in section~\ref{sec-SUSY}, in order
to introduce a central charge, we have to break the full internal
symmetry $\GR{SU}(4)$ into some automorphic subgroup.
As such subgroup, we can adopt $\GR{USp}(4)$, and
just one real, or Hermitian, central charge can be implemented
into the superalgebra.

The $\GR{USp}(4)$ superalgebra is given as
eqs.~\Ref{BeginOfAlgebra}--\Ref{EndOfAlgebra} with its $R$-symmetry
group replaced by $\GR{USp}(4)$, namely%
\footnote{
For notations on $\GR{USp}(4)$ and $\AL{usp}(4)$, see
appendix~\ref{sec-symplectic}.},
\begin{xalignat}{2}
 \{Q_{i\alpha},Q_{j\beta}\}
    &=C_{\alpha\beta}\Omega_{ij}Z,
 &
 \{\OLL{Q}_i{}_{\Dalpha},\OLL{Q}_j{}_{\Dbeta}\}
    &=C_{\Dalpha\Dbeta}\Omega_{ij}Z, \label{QQ-symp}\\
 \{Q_{i\alpha},\OLL{Q}_{j\Dbeta}\}
    &=2\Omega_{ij}(\sigma^\mu)_{\alpha\Dbeta}P_\mu,
 &
 [P_\mu,P_\nu]
    &=0,\\
 [Q_{i\alpha},P_\mu]
    &=0,
 &
 [\OLL{Q}_{i\Dalpha},P_\mu]
    &=0,\\
 [J_{\mu\nu},Q_{i\alpha}]
    &=\frac{i}{2}(\sigma_{\mu\nu})_{\alpha}{}^{\beta}Q_{i\beta},
 &
 [J_{\mu\nu},\OLL{Q}_i{}^{\Dalpha}]
    &=\frac{i}{2}(\bar{\sigma}_{\mu\nu})^{\Dalpha}{}_{\Dbeta}
           \OLL{Q}_i{}^{\Dbeta},\\
 [J_{\mu\nu},P_\rho]
    &=i(\eta_{\nu\rho}P_\mu-\eta_{\mu\rho}P_\nu),
 &
 [J_{\mu\nu},J_{\rho\sigma}]
    &=i\begin{aligned}[t]
         & (\eta_{\nu\rho}J_{\mu\sigma}-\eta_{\nu\sigma}J_{\mu\rho}\\
         & -\eta_{\mu\rho}J_{\nu\sigma}+\eta_{\mu\sigma}J_{\nu\rho}),\\
       \end{aligned}\\
 [R^a,Q_{i\alpha}]
    &=(X^a)_i{}^jQ_{j\alpha},
 &
 [R^a,\OLL{Q}_{i\Dalpha}]
    &=(X^a)_i{}^j\OLL{Q}_{j\Dalpha},\\
 [R^a,P_\mu]
    &=0,
 &
 [R^a,J_{\mu\nu}]
    &=0,\\
 [Z,\mbox{any}]
    &=0,
 &  &
\end{xalignat}
where supercharges $Q_{i\alpha}$ and $\OLL{Q}^i{}_{\Dalpha}$,
related as
\begin{equation}
 \OLL{Q}^i{}_{\Dalpha}:=(Q_{i\alpha})^\dagger,
\end{equation}
transform as Weyl spinors w.r.t.\ Lorentz transformations
and as fundamental representations $\bs{4}$ and $\OL{\bs{4}}$, respectively,
w.r.t.\ the internal $\GR{USp}(4)$. These two internal symmetry
representations are, however, equivalent and related so that
indices are raised or lowered freely by the $\GR{USp}(4)$ invariant
metric $\Omega^{ij}$ and $\Omega_{ij}$ as in
\begin{equation}
 Q^i=\Omega^{ij}Q_j,\quad
 Q_i=Q^j\Omega_{ji},\qquad
 \Omega_{ij}+\Omega_{ji}=0,\quad
 \Omega^{ik}\Omega_{jk}=\delta_j{}^i.
\end{equation}
Thus we do not need to distinguish strictly the contravariant
and covariant indices for $\GR{USp}(4)$.
We have denoted the $\GR{USp}(4)$ generators by $R^a\in\AL{usp}(4)$ and
its representations by $X^a$ which satisfy that
\begin{equation}
 (X^a)_{ij}=(X^a)_{ji},\quad
 (X^a)_{ij}:=(X^a)_i{}^k\Omega_{kj}.
\end{equation}
Through this section, we work on the theory in four-dimensional Minkowski
spacetime, with $\eta_{\mu\nu}=(+,-,-,-)$.
Note in passing that the central charge is in fact Hermitian
\begin{equation}
 Z^\dagger=Z
\end{equation}
as it has to be so since the Hermitian conjugate of one of the equation in
\Ref{QQ-symp} converts to the other.

\subsection{The $\GR{USp}(4)$ SYM}
The $\GR{USp}(4)$ super Yang-Mills theory is given as%
~\cite{Sohnius-Stelle-West-1,Sohnius-Stelle-West-2}%
\begin{align}
 S&=\tr\int d^4 x\,\Bigl(
       -\frac{1}{4}F_{\mu\nu}F^{\mu\nu}
       -\frac{1}{2}V_\mu V^\mu
       +\frac{1}{2}D_\mu \phi_{ij}D^\mu \phi^{ij}
       +\frac{1}{2}H_{ij}H^{ij} \notag \\
  &\phantom{\tr\int d^4 x\,FF}
       -\frac{i}{4}\OLL{\lambda}^i\stackrel{\LR}{\Slash{D}}\lambda_i
       -\OLL{\lambda}^i[\lambda^j,\phi_{ij}]
       +\frac{1}{4}[\phi_{ij},\phi_{kl}][\phi^{ij},\phi^{kl}]\Bigr),
%       +\frac{1}{4}[\phi_{ik},\phi^k{}_j][\phi^i_l,\phi^{lj}]\Bigr),
       \label{action}
\end{align}
where the field contents of the multiplet are arranged as
\begin{align*}
 A_{\mu}    &\ \text{:\ \ vector, $4-1$ (gauge d.o.f.) $=3$ components,}\\
 \lambda_i  &\ \text{:\ \ $\GR{USp}(4)$-Majorana spinors,
                       $4\times 4=16$ components,}\\
 \phi_{ij}  &\ \text{:\ \ scalars, $6$ (antisymm.\ w.r.t.\ $i,j$)
                              $-1$ ($\Omega$-traceless) $=5$ components,}\\
 V_\mu      &\ \text{:\ \ pseudo vector, $4-1$ (gauge d.o.f.)
                              $=3$ components,}\\
 H_{ij}     &\ \text{:\ \ auxiliary scalars, $5$ components as in the scalar,}
\end{align*}
so that the d.o.f.\ of bosons and fermions are the same and are 16.
The $\GR{USp}(4)$-Majorana condition%
\footnote{See appendix~\ref{sec-Clifford} for the
Majorana spinor representations.} is given as
\begin{equation}
 \lambda_{i\alpha}=\OLL{\lambda}^{j\beta}(C^{-1})_{\beta\alpha}\Omega_{ji},
\end{equation}
where $C^{-1}$ is the charge conjugation matrix.

Supertransformations of these component fields are given as
\begin{align}
 \delta A_\mu &= i\OLL{\zeta}^{i}(\gamma_\mu)
                   \lambda_{i}, \label{delta-A} \\
 \delta \phi_{ij}
              &=-i\left(\OLL{\zeta}_i\lambda_j-\OLL{\zeta}_j\lambda_i
                +\frac{1}{2}\Omega_{ij}\OLL{\zeta}{}^k\lambda_k\right),\\
 \delta \lambda_i
              &= -\frac{1}{2}\gamma_{\mu\nu}F^{\mu\nu}\zeta_i
                 +2\gamma^\mu D_\mu \phi_i{}^j\zeta_j
                 +\gamma^5\gamma^\mu V_\mu\zeta_i   %\notag\\
%              &\phantom{=\ }
                 +2\gamma^5 H_i{}^j\zeta_j
                 -2i[\phi_{ik},\phi^{kj}]\zeta_j,\\
 \delta H_{ij}&=i(\OLL{\zeta}_i\gamma^5\gamma^\mu D_\mu\lambda_j
                -\OLL{\zeta}_j\gamma^5\gamma^\mu D_\mu\lambda_i
                +\frac{1}{2}\Omega_{ij}
                 \OLL{\zeta}^k\gamma^5\gamma^\mu D_\mu\lambda_k)\notag \\
              &\phantom{=\ }
                -2(\OLL{\zeta}_i\gamma^5[\lambda^l,\phi_{jl}]
                   -\OLL{\zeta}_j\gamma^5[\lambda^l,\phi_{il}]
                   +\frac{1}{2}\Omega_{ij}\OLL{\zeta}^k\gamma^5
                               [\lambda^l,\phi_{kl}])
                   +\OLL{\zeta}^k\gamma^5[\lambda_k,\phi_{ij}]\\
 \delta V_\mu &=i\OLL{\zeta}^i\gamma^5\gamma_{\mu\nu}
                          D^\nu \lambda_i
                +2\OLL{\zeta}^i\gamma^5\gamma_\mu
                         [\lambda^j,\phi_{ij}], \label{delta-V}
\end{align}
and transformations generated by the central charge are
\begin{align}
 \delta_z A_\mu    &=\omega V_\mu,\\
 \delta_z \phi_{ij}&=-\omega H_{ij},\\
 \delta_z \lambda_i&=-\omega(\gamma^5\gamma^\mu D_\mu \lambda_i
                       -2i\gamma^5[\lambda^j,\phi_{ij}]),\\
 \delta_z H_{ij}   &=\omega\left(-D^\mu D_\mu \phi_{ij}
              +i(\frac{1}{4}\Omega_{ij}\{\OLL{\lambda}_k,\lambda^k\})
               -\left[[\phi_{kl},\phi^{kl}],\phi_{ij}\right]\right),\\
 \delta_z V_\mu    &=\omega\left(
                       D_\nu F^\nu{}_\mu
                      -\frac{1}{2}\{\OLL{\lambda},\gamma_\mu \lambda\}
                      +i[\phi^{ij},D_\mu \phi_{ij}]\right),
\end{align}
where we have omitted the spinor indices.
These supertransformations close off-shell, up to field dependent
gauge transformations.
The action \Ref{action} is invariant under these supertransformations
as long as the additional constraint
\begin{equation}
\label{additional}
 0=D^\mu V_\mu+ \frac{1}{2}\{\OLL{\lambda}^i,\gamma^5\lambda_i\}
              -i[\phi^{ij},H_{ij}]
\end{equation}
is satisfies.

\subsection{The $\GR{USp}(4)$ superspace}
The $\GR{USp}(4)$ superspace is represented by the coordinates
$(x^\mu,\theta^{i\alpha},\OLL{\theta}^{i\Dalpha},z)$ which are the conjugate
to the generators $(P_\mu,Q_{i\alpha},\OLL{Q}_{i\Dalpha},Z)$.
In other words, the superspace parameterizes, as local coordinate system,
the manifold composed of the $\GR{USp}(4)$ superalgebra.
In this viewpoint, these generators of the superalgebra are represented
as the adjoint representation on the manifold itself, and are
expressed by the superspace coordinates as in
\begin{align}
 P_{\mu}
  &=i\del_\mu,\\
 Q_{i\alpha}
  &=\frac{\del}{\del \theta^{i\alpha}}
         -i(\sigma^\mu)_{\alpha\Dbeta}\OLL{\theta}_{i}{}^{\Dbeta}\del_\mu
         -\frac{i}{2}\theta_{i\alpha}\del_z,\\
 \OLL{Q}_{i\Dalpha}
  &=\frac{\del}{\del \OLL{\theta}^{i\Dalpha}}
         -i\theta_i{}^{\beta}(\sigma^\mu)_{\beta\Dalpha}\del_\mu
         -\frac{i}{2}\OLL{\theta}_{i\Dalpha}\del_z,\\
 Z
  &=i\del_z,
\end{align}
which satisfy that
\begin{xalignat}{2}
 \{Q_{i\alpha},Q_{j\beta}\}
  &=-C_{\alpha\beta}\Omega_{ij}Z,
 &
 \{\OLL{Q}_{i\Dalpha},\OLL{Q}_{j\Dbeta}\}
  &=-C_{\Dalpha\Dbeta}\Omega_{ij}Z,\\
 \{Q_{i\alpha},\OLL{Q}_{j\Dbeta}\}
  &=-2\Omega_{ij}(\sigma^\mu)_{\alpha\Dbeta}P_\mu,
 &
 [P_\mu,P_\nu]
  &=0,\\
 [Q_{i\alpha},P_\mu]
  &=0,
 &
 [\OLL{Q}_{i\Dalpha},P_\mu]
  &=0,\\
 [Z,\text{any}]
  &=0.
 &
  &
\end{xalignat}

Supercovariant derivatives, which all (anti-)commute with
generators above are then given, on the superspace, as
\begin{align}
 D_{\mu}
  &=\del_\mu,\\
 D_{i\alpha}
  &=\frac{\del}{\del \theta^{i\alpha}}
         +i(\sigma^\mu)_{\alpha\Dbeta}\OLL{\theta}_{i}{}^{\Dbeta}\del_\mu
         +\frac{i}{2}\theta_{i\alpha}\del_z,\\
 \OLL{D}_{i\Dalpha}
  &=\frac{\del}{\del \OLL{\theta}^{i\Dalpha}}
         +i\theta_i{}^{\beta}(\sigma^\mu)_{\beta\Dalpha}\del_\mu
         +\frac{i}{2}\OLL{\theta}_{i\Dalpha}\del_z,\\
 D_z
  &=\del_z,
\end{align}
which satisfy the following algebra:
\begin{xalignat}{2}
 \{D_{i\alpha},D_{j\beta}\}
  &=iC_{\alpha\beta}\Omega_{ij}D_z,
 &
 \{\OLL{D}_{i\Dalpha},\OLL{D}_{j\Dbeta}\}
  &=iC_{\Dalpha\Dbeta}\Omega_{ij}D_z,\\
 \{D_{i\alpha},\OLL{D}_{j\Dbeta}\}
  &=2i\Omega_{ij}(\sigma^\mu)_{\alpha\Dbeta}D_\mu,
 &
 [D_\mu,D_\nu]
  &=0,\\
 [D_{i\alpha},D_{\mu}]
  &=0,
 &
 [\OLL{D}_{i\Dalpha},D_\mu]
  &=0,\\
 [D_z,\text{any}]
  &=0,
 & & \\
 \{Q_{i\alpha},\text{any\ }D\}
  &=0,
 &
 \{\OLL{Q}_{i\Dalpha},\text{any\ }D\}
  &=0,\\
 [P_\mu,\text{any\ }D]
  &=0,
 &
 [Z,\text{any\ }D]
  &=0,
\end{xalignat}
with the last four equations assuring the supercovariance of these
derivatives.

\subsection{Supersymmetric Gauge Theories on the $\GR{USp}(4)$ Superspace}
Let us now formulate a supersymmetric gauge theory on the
$\GR{USp}(4)$ superspace constructed in the preceding section.
We follow the formulation using superconnections and supercurvatures
defined on the superspace. This is technically the natural
generalization of the formulation of ordinary gauge theory,
i.e.\ Yang-Mills theory, using gauge connections and curvatures
on the ordinary spacetime. We emphasize that this formulation has a great
advantage among others since there the structure of the superalgebra
can be manifestly implemented.

\subsubsection{Superconnections}
Supercovariant derivatives $D_I=(D_\mu,D_{i\alpha},\OLL{D}_{i\Dalpha},D_z)$
are further gauge covariantized by adding the superconnections
$\Gamma_I$ to define the gauge-supercovariant derivatives
\begin{equation}
 \nabla_I:=D_I-i\Gamma_I,
\end{equation}
with gauge transformations
\begin{equation}
\label{gaugetransf-superconn}
 \nabla'_I=e^K\nabla e^{-K},\quad\text{or}\quad
 \delta_K\nabla_I=[\nabla_I,K],\quad
 \delta_K\Gamma_I=i[\nabla_I,K],
\end{equation}
where $K$ is an anti-Hermitian gauge parameter superfield.

Here we note that
by some suitable gauge transformations
the lowest components
w.r.t.\ $\theta,\ \OLL{\theta}$ of the fermionic superconnections
$\Gamma_{i\alpha}$ and $\OLL{\Gamma}_{i\Dalpha}$ can be
algebraically gauged away. On the other hand,
the whole components of one of the five bosonic superconnections
$\Gamma_\mu$ and $\Gamma_z$
can also be gauged away as in the usual gauge theory.
In fact, take the anti-Hermitian
gauge parameter superfield $K$ as, say,
\begin{equation}
 K=\kappa(x,z)+\theta^{i\alpha}\left(i\Gamma_{i\alpha}
                                -i[\kappa(x,z),\Gamma_{i\alpha}]\right)
              +\OLL{\theta}^{i\Dalpha}\left(i\OLL{\Gamma}_{i\Dalpha}
                              -i[\kappa(x,z),\OLL{\Gamma}_{i\Dalpha}]\right)
              +K'(x,z,\theta,\OLL{\theta}),
\end{equation}
where $K'$, which is also anti-Hermitian,
contains no linear and zeroth order terms w.r.t.\ 
$\theta$ or $\OLL{\theta}$ and the lowest parameter pure-imaginary
field $\kappa$ is determined as such that
\begin{equation}
 \del_z\kappa(x,z)=i[\Gamma_z(x,\theta=0,\OLL{\theta}=0,z),\kappa(x,z)]
  +i\Gamma_z(x,\theta=0,\OLL{\theta}=0,z).
\end{equation}
Since this condition on $\kappa$ is a first order differential equation,
it has certainly a solution. Then it is easy to see
\begin{equation}
 \Gamma_{i\alpha}'|=\Gamma_{i\alpha}+\delta\Gamma_{i\alpha}\bigr|=0,\quad
 \OLL{\Gamma}_{i\Dalpha}'|
                =\OLL{\Gamma}_{i\Dalpha}+\delta\OLL{\Gamma}_{i\Dalpha}\bigr|=0,
 \quad
 \Gamma'_z|=\Gamma_z+\delta\Gamma_z\bigr|=0,
\end{equation}
where for any function on the superspace,
\begin{equation}
 B|:=B(x,\theta=0,\OLL{\theta}=0,z).
\end{equation}
We call such special gauge the Wess-Zumino gauge;
namely in the Wess-Zumino gauge, we have
\begin{equation}
 \Gamma_{i\alpha}|=
 \OLL{\Gamma}_{i\Dalpha}|
 \Gamma_z|=0.
\end{equation}
In the following computations, we usually take this gauge.

\subsubsection{Supercurvatures}
Supercurvatures $F_{IJ}$ are then defined as in
\begin{equation}
\label{curvature}
 [\nabla_I,\nabla_J]_\pm=-iF_{IJ},
\end{equation}
where the left hand side means anticommutators
if both $\nabla_I$ and $\nabla_J$ are fermionic and denoted
by the subscription $+$ while, otherwise, it means
commutators and labeled by the subscription $-$.
Gauge transformations on supercurvatures are read off from those
on gauge-supercovariant derivatives as
\begin{align}
 \delta_K F_{IJ}
   &=i[\delta_K \nabla_I,\nabla_J]_\pm+i[\nabla_I,\delta_K \nabla_J]_\pm
                =i[[\nabla_I,K],\nabla_J]_\pm+i[\nabla_I,[\nabla_J,K]]_\pm \\
   &=i[[\nabla_I,\nabla_J]_\pm,K]=[F_{IJ},K].
\end{align}
Thus supercurvatures are gauge covariant. They should therefore
be the ingredients for constructing the super Yang-Mills theory.
Also supercurvatures are supercovariant, i.e.\ they are superfields
which transform correctly under a supertransformation.
This fact can also be shown easily from the supercovariance of the
superconnections.

Note for example
\begin{equation}
 \{\nabla_{i\alpha},\OLL{\nabla}_{j\Dbeta}\}_\pm
  =2\Omega_{ij}(\sigma^\mu)_{\alpha\Dbeta}\nabla_\mu
   -i\{D_{i\alpha},\OLL{\Gamma}_{j\Dbeta}\}
   -i\{\OLL{D}_{j\Dbeta},\Gamma_{i\alpha}\}
   -\{\Gamma_{i\alpha},\OLL{\Gamma}_{j\Dbeta}\}
   +2i\Omega_{ij}(\sigma^\mu)_{\alpha\Dbeta}\Gamma_\mu.
\end{equation}
The first term in the right hand side of this equation
contains a gauge-supercovariant derivative itself. In literature,
such terms are sometimes called supertorsions and distinguished from
supercurvatures. It is of course merely the difference of terminology,
and here, we call, including such terms, the right hand side
of \eRef{curvature} supercurvatures.

\subsubsection{Constraints on Supercurvatures}
In an ordinary Yang-Mills theory, once a gauge covariant curvature
are defined, one can construct the Lagrangian (action) as
$\MC{L}\sim -\tr F_{\mu\nu}F^{\mu\nu}$. In supersymmetric theories,
however, superfields have large degrees of freedom so highly
reducible, containing possibly many extra component fields.
Thus to formulate the correct theory one would like to construct,
one has to impose some appropriate constraints on superfields
and restricts them covariantly to the symmetry of the system.
In our case we need some suitable gauge and supercovariant
constraints on the supercurvatures we defined above.

To obtain appropriate constraints, let us start by writing
the supercurvatures constructed from two fermionic derivatives as
\begin{align}
 \{\nabla_{i\alpha},\nabla_{j\beta}\}
  & =i\Omega_{ij}C_{\alpha\beta}\nabla_z
    -iF_{ij\alpha\beta},\label{nablanabla} \\
 \{\nabla_{i\alpha},\OLL{\nabla}_{j\Dbeta}\}
  & =i\Omega_{ij}(\sigma^\mu)_{\alpha\Dbeta}\nabla_\mu
    -iF_{ij\alpha\Dbeta},
\end{align}
where%
\footnote{We have rescale the gauge-supercovariant derivatives
as
\[
 \nabla_\alpha\RA\sqrt{2}\nabla_\alpha,\quad
 \OLL{\nabla}_{\Dalpha}\RA\sqrt{2}\OLL{\nabla}_{\Dalpha},\quad
 \nabla_z\RA 2\nabla_z
\]
just for a conventional reason.}
\begin{equation}
 F_{ij\alpha\beta}=F_{ji\beta\alpha}
\end{equation}
since the left hand side of \eRef{nablanabla}
is symmetric under the exchange $(i,\alpha)\LR(j,\beta)$.
Then it can be decomposed into symmetric and antisymmetric parts as in
\begin{equation}
F_{ij\alpha\beta}
  =W^{\MR{a}}_{ij\alpha\beta}+W^{\MR{s}}_{ij\alpha\beta},
\end{equation}
where
\begin{gather}
 W^{\MR{a}}_{ij\alpha\beta}+W^{\MR{a}}_{ji\alpha\beta}
 = W^{\MR{a}}_{ij\alpha\beta}+W^{\MR{a}}_{ij\beta\alpha}=0,\\
 W^{\MR{s}}_{ij\alpha\beta}=W^{\MR{s}}_{ji\alpha\beta}
 = W^{\MR{s}}_{ij\beta\alpha}.
\end{gather}
The antisymmetric part is trivially written as
\begin{equation}
 W^{\MR{a}}_{ij\alpha\beta}
   =C_{\alpha\beta}W^{\MR{a}}_{ij},\quad
 W^{\MR{a}}_{ij}+W^{\MR{a}}_{ji}=0.
\end{equation}
Then further we can decompose the antisymmetric factor into the parts
which is
proportional to the invariant metric $\Omega_{ij}$ and which
does not contain $\Omega_{ij}$ as in
\begin{equation}
 W^{\MR{a}}_{ij}= W_{ij}+\Omega_{ij}W,\quad
 \text{ where}\qquad \Omega^{ij}W_{ij}=0 \qquad(\text{$\Omega$-traceless}).
\end{equation}
Similarly, the symmetric part can be represented as
\begin{gather}
 W^{\MR{s}}_{ij\alpha\beta}
   =\frac{1}{2}\sum_a(\sigma_{\mu\nu})_{\alpha\beta}(X^a)_{ij}
                       W^{\mu\nu}{}^a,\\
 \text{where}\qquad (X^a)_{ij}\in\AL{usp}(4),
\end{gather}
since $(\sigma_{\mu\nu})_{\alpha\beta}=(\sigma_{\mu\nu})_{\beta\alpha}$
and $X_{ij}=X_{ji}$ for $(X_{ij})\in\AL{usp}(4)$.

\paragraph{Spins and Constraints}
We now concentrate on the spins contained in the supercurvatures.
Spins (helicities) are of course eigenvalues of the Lorentz
generators $J_{\mu\nu}$. Here we notice
\begin{equation}
 [S^\pm_i,S^\pm_j]=i\epsilon_{ijk}S^{\pm k},\quad
 [S^\pm_i,S^\mp_j]=0,\qquad
 S^\pm_i:=\frac{1}{2}\left(-\frac{1}{2}
             \epsilon_{ijk}J^{jk}\mp J^4{}_i\right)
         =-\frac{1}{2}\epsilon_{ijk}(J^\pm)^{jk},
\end{equation}
where $(J^\pm)_{\mu\nu}$ are the selfdual and anti-selfdual parts.
We thus recognize $S^\pm_i$ as the spin generators.
The third component of spins (helicities) are then
\begin{equation}
 J:=(S^+)^3+(S^-)^3=J^{12}.
%\frac{1}{2}\left(-J^{12}\mp J^4{}_3\right).
\end{equation}
Then using
\begin{equation}
 [J_{\mu\nu},Q_{i\alpha}]
  =\frac{i}{2}(\sigma_{\mu\nu})_{\alpha}{}^{\beta}Q_{i\beta},\quad
 [J_{\mu\nu},\OLL{Q}_{i\Dalpha}]
  =\frac{i}{2}(\bar{\sigma}_{\mu\nu})^{\Dbeta}{}^{\Dalpha}
           \OLL{Q}_{i\Dbeta},
\end{equation}
we can easily compute spins of the superfields $F_{ij\alpha\beta}$ and
$F_{ij\alpha\Dbeta}$, as in
\begin{gather}
 [J,F_{ij1\dot{1}}]=+F_{ij1\dot{1}},\qquad
 [J,F_{ij2\dot{2}}]=-F_{ij2\dot{2}},\\
 [J,F_{ij1\dot{2}}]=[J,F_{ij2\dot{1}}]=0.
\end{gather}
and
\begin{gather}
 [J,W^{\MR{s}}_{ij11}]=+W^{\MR{s}}_{ij11},\qquad
 [J,W^{\MR{s}}_{ij22}]=-W^{\MR{s}}_{ij22},\\
 [J,W^{\MR{s}}_{ij12}]=[J,W^{\MR{s}}_{ij21}]=0.
\end{gather}
Thus we find that $F_{ij\alpha\Dbeta}$ and $W^{\MR{s}}$ have
spin 1. As will be clear later, the multiplet in
our formulation is determined by successively multiplying
$\nabla_{\alpha},\ \OLL{\nabla}_{\Dalpha}$
as ``creation operators'' on supercurvatures as the basic superfields.
In other words, component fields are created as the form
\begin{equation}
 \nabla_{I_1}\cdots\nabla_{I_r}F,
\end{equation}
where $F$ is a nonzero supercurvature in our system.
Since $F_{ij\alpha\Dbeta}$ and $W^{\MR{s}}$ have spins 1
themselves, the created fields from such superfields can
obviously contain
spins higher than 1, which are not allowed in the ($\GR{USp}(4)$)
super Yang-Mills multiplet. We therefore drop such superfields:
\begin{equation}
 F_{ij\alpha\Dbeta}=W^{\MR{s}}=0.
\end{equation}

Hence we consider the constraints
\begin{align}
 \{\nabla_{i\alpha},\nabla_{j\beta}\}
  & =i\Omega_{ij}C_{\alpha\beta}\nabla_z
     -i\Omega_{ij}C_{\alpha\beta}W-iC_{\alpha\beta}W_{ij},\\
 \{\nabla_{i\alpha},\OLL{\nabla}_{j\Dbeta}\}
  & =i\Omega_{ij}(\sigma^\mu)_{\alpha\Dbeta}\nabla_\mu,
\end{align}
where $\Omega_{ij}C_{\alpha\beta}W$ at the right hand side
of the first equation can be absorbed into
{$\Omega_{ij}C_{\alpha\beta}\Gamma_z$ by a trivial redefinition
of the superconnection $\Gamma_z$.

\paragraph{Reality Condition}
Finally let us consider the reality of $W_{ij}$ (and, strictly, $W$,
in order to be absorbed in $\Gamma_z$ as noted above).
Since $\GR{USp}(4)$ is real in the sense of representations,
we can impose the condition,
consistently to the internal symmetry,
\begin{equation}
 (W_{ij})^\ast\equiv\OL{W}^{ij} = W^{ij},\quad
 W^\ast = W.
\end{equation}
where
\begin{equation}
 W^{ij}:=\Omega^{ik}\Omega^{jl}W_{kl}.
\end{equation}
Since we are looking for the irreducible (or minimal) multiplet,
we must impose all we can impose without breaking any nontrivial
structure of the constraints. More practically, the scalar
fields contained in the $\GR{USp}(4)$ model are real,
so that we do not need extra complex d.o.f.\ of the superfields
$W_{ij}$ (and $W$). We therefore adopt these reality
constraints. We emphasize here that in the earlier
attempt to construct the $\GR{SU}(4)$ model in
\cite{Sohnius-1}, the nontrivial complex structure of $\GR{SU}(4)$
prevented from imposing any such reality constraint
without forcing the superalgebra close only on-shell.

Thus our constraints in the
final form are
\begin{xalignat}{2}
 \{\nabla_{i\alpha},\nabla_{j\beta}\}
  & =i\Omega_{ij}C_{\alpha\beta}\nabla_z
     -iC_{\alpha\beta}W_{ij},
 &
 \{\OLL{\nabla}_{i\Dalpha},\OLL{\nabla}_{j\Dbeta}\}
  & =i\Omega_{ij}C_{\Dalpha\Dbeta}\nabla_z
     +iC_{\Dalpha\Dbeta}W_{ij}, \label{BeginOfConstraints}\\
 \{\nabla_{i\alpha},\OLL{\nabla}_{j\Dbeta}\}
  & =i\Omega_{ij}(\sigma^\mu)_{\alpha\Dbeta}\nabla_\mu,
 &
  & \\
 [\nabla_{i\alpha},\nabla_\mu]
  &=-iF_{i\alpha\mu},
 &
 [\OLL{\nabla}_{i\Dalpha},\nabla_\mu]
  &=+i\OLL{F}_{i\Dalpha\mu},\\
 [\nabla_{i\alpha},\nabla_z]
  &=-iG_{i\alpha},
 &
 [\OLL{\nabla}_{i\Dalpha},\nabla_z]
  &=+i\OLL{G}_{i\Dalpha},\\
 [\nabla_\mu,\nabla_z]
  &=-ig_\mu
 &
 [\nabla_\mu,\nabla_\nu]
  &=-iF_{\mu\nu}, \label{EndOfConstraints}
\end{xalignat}
where
\begin{equation}
 \Omega^{ij}W_{ij}=0,\quad (W_{ij})^*=W^{ij}.
\end{equation}

\subsection{Solving the Constraints}
Let us then move on to solving the constraints
\Ref{BeginOfConstraints}--\Ref{EndOfConstraints} derived above.

Specifically, we will solve them in the following way.
First, applying the constraints into the Bianchi identities
\begin{equation}
 [\nabla_A,[\nabla_B,\nabla_C\}\}
 \pm[\nabla_B,[\nabla_C,\nabla_A\}\}
 \pm[\nabla_C,[\nabla_A,\nabla_B\}\}=0,
\end{equation}
we obtain various relations among the supercurvatures
and their derivatives.
Then we will find each higher derivatives of the supercurvatures,
particularly of the superfield $W_{ij}$,
can be expressed only by some combinations of all the other supercurvatures,
including themselves, and their lower derivatives. In other words, we
can compute each higher derivatives of the supercurvatures
self-consistently using the other supercurvatures in our system.
Once each derivatives of supercurvatures are computed self-consistently,
componentwise expansions of the supercurvatures can be fulfilled,
so that the supercurvatures are determined completely
in a self-consistent manner. In this sense we say we solve the
constraints. We will then identify the set of
component fields, which are necessary and sufficient to compute
all the supercurvatures componentwisely,
as the independent fields in our system.

\subsubsection{Bianchi Identities}
We now go through with the process described above.
Since the computations are lengthy and cumbersome, here we only list
the results in the following%
\footnote{See appendix~\ref{sec-comp} for the detail of those computations.}.

\paragraph{Three Fermionic Derivatives}
First we list the results of
applying Bianchi identities for three fermionic derivatives,
say $(\nabla_\alpha,\nabla_\beta,\nabla_\gamma)$.
\begin{xalignat}{2}
 [\nabla_{j\alpha},W^j{}_i]
  &=5iG_{i\alpha},\label{BeginOfBianchi}
 &
 [\Bnabla_{j\Dalpha},W^j{}_i]
  &=5i\OLL{G}_{i\Dalpha},\\
 G_{i\alpha}
  &=-\frac{1}{4}(\sigma^\mu C)_\alpha{}^{\Dbeta}\OLL{F}_{i\Dbeta\mu},
    \label{GandF}
 &
 \OLL{G}_{i\Dalpha}
  &=-\frac{1}{4}(C\sigma^\mu)^\beta_\Dalpha F_{i\beta\mu},\\
 [\nabla_{i\alpha},W_{jk}]
  &=2i\Omega_{i[j}G_{k]\alpha}+i\Omega_{jk}G_{i\alpha},
 &
 [\Bnabla_{i\Dalpha},W_{jk}]
  &=2i\Omega_{i[j}\OLL{G}_{k]\Dalpha}+i\Omega_{jk}\OLL{G}_{i\Dalpha},\\
 F_{i\alpha\mu}
  &=(\bar{\sigma}_\mu C)^{\Dgamma}{}_\alpha\OLL{G}_{i\Dgamma},
 &
 \OLL{F}_{i\Dalpha}
  &=(C\bsigma_\mu)_\Dalpha{}^\beta G_{i\beta}.
    \label{FandG}
\end{xalignat}

\paragraph{Four Fermionic Derivatives}
\begin{align}
 \{\nabla_{i\alpha},G_{j\beta}\}
  &=-\frac{i}{4}\Omega_{ij}(\sigma^{\mu\nu})_{\alpha\beta}F_{\mu\nu}
    -\frac{1}{2}C_{\alpha\beta}[\nabla_z,W_{ij}]
    -\frac{1}{4}C_{\alpha\beta}[W_{ik},W^k{}_j],\\
 \{\Bnabla_{i\Dalpha},\OLL{G}_{j\Dbeta}\}
  &=-\frac{i}{4}\Omega_{ij}(\bsigma^{\mu\nu})_{\Dalpha\Dbeta}F_{\mu\nu}
    -\frac{1}{2}C_{\Dalpha\Dbeta}[\nabla_z,W_{ij}]
    +\frac{1}{4}C_{\Dalpha\Dbeta}[W_{ik},W^k{}_j],\\
 \{\OLL{\nabla}_{i\Dalpha},G_{j\beta}\}
  &=-\frac{1}{2}(\sigma^\mu)_{\beta\Dalpha}\Bigl(
       i\Omega_{ij}g_\mu-[\nabla_\mu,W_{ij}]\Bigr),\\
 \{\nabla_{i\alpha},\OLL{G}_{j\Dbeta}\}
  &=-\frac{1}{2}(\sigma^\mu)_{\alpha\Dbeta}\Bigl(
       i\Omega_{ij}g_\mu+[\nabla_\mu,W_{ij}]\Bigr).
\end{align}

\paragraph{Five Fermionic Derivatives}
\begin{align}
 [\nabla_{i\alpha},g_\mu]
  &=-(\sigma^{\mu\nu})_\alpha{}^\beta[\nabla^\nu,G_{i\beta}]
    +(\bar{\sigma}_\mu C)^{\Dbeta}{}_{\alpha}
      [\OLL{G}^j{}_{\Dbeta},W_{ij}],\\
 [\Bnabla_{i\Dalpha},g_\mu]
  &=-(\bsigma^{\mu\nu})^\Dbeta{}_\Dalpha[\nabla^\nu,\OLL{G}_{i\Dbeta}]
    +(C\bar{\sigma}_\mu)_\Dalpha{}^\beta
      [G^j{}_\beta,W_{ij}],\\
 [\nabla_z,G_{i\alpha}]
  &=-(\sigma^\mu C)_\alpha{}^{\Dalpha}[\nabla_\mu,\OLL{G}_{i\Dalpha}]
     +[G^j{}_\alpha,W_{ij}],\\
 [\nabla_z,\OLL{G}_{i\Dalpha}]
  &=-(C\sigma^\mu)^\beta{}_\Dalpha[\nabla_\mu,G_{i\beta}]
     -[\OLL{G}^j{}_\Dalpha,W_{ij}],\\
 [\nabla_{i\alpha},[\nabla_z,W_{jk}]]
  &=-i(\sigma^\mu C)_\alpha{}^{\Dalpha}\Bigl(
       \Omega_{jk}[\nabla_\mu,\OLL{G}_{i\Dalpha}]
        +2\Omega_{i[j}[\nabla_\mu,\OLL{G}_{k]\Dalpha}]\Bigr) \notag \\
  &\phantom{=+}
    +i\Bigl(\Omega_{jk}[G^l{}_\alpha,W_{il}]
        +2\Omega_{i[j}[G^l{}_\alpha,W_{k]l}]\Bigr) \notag \\
  &\phantom{=+}
    -i[G_{i\alpha},W_{jk}],\\
 [\Bnabla_{i\Dalpha},[\nabla_z,W_{jk}]]
  &=-i(C\sigma^\mu)^\beta{}_\Dalpha\Bigl(
       \Omega_{jk}[\nabla_\mu,G_{i\beta}]
        +2\Omega_{i[j}[\nabla_\mu,G_{k]\beta}]\Bigr) \notag \\
  &\phantom{=+}
    -i\Bigl(\Omega_{jk}[\OLL{G}^l{}_\Dalpha,W_{il}]
        +2\Omega_{i[j}[\OLL{G}^l{}_\Dalpha,W_{k]l}]\Bigr) \notag \\
  &\phantom{=+}
    +i[\OLL{G}_{i\Dalpha},W_{jk}].
\end{align}

\paragraph{Six Fermionic Derivatives}
\begin{align}
 [\nabla_z,[\nabla_z,W_{jk}]]
  &=[\nabla_\mu,[\nabla^\mu,W_{jk}]] \notag \\
  &\phantom{=+}
    +i(\Omega_{jk}\{G_{i\alpha},G^{i\alpha}\}
            -4\{G_{j\alpha},G_k{}^\alpha\}) \notag \\
  &\phantom{=+}
    -i(\Omega_{jk}\{\OLL{G}_{i\Dalpha},\OLL{G}^{i\Dalpha}\}
           -4\{\OLL{G}_{j\Dalpha},\OLL{G}_k{}^{\Dalpha}\}) \notag \\
  &\phantom{=+}
    +\frac{1}{4}[W_{[j|l},[W_{k]m},W^{ml}]],\\
 [\nabla_z,g_\mu]
  &=[\nabla^\nu,F_{\mu\nu}]
    -2(\bar{\sigma}^{\Dalpha\alpha})\{G_{i\alpha},\OLL{G}^i{}_{\Dalpha}\}
    +\frac{i}{4}[W^{ij},[\nabla_\mu,W_{ij}]],\\
 [\nabla^\mu,g_\mu]
  &=-\{G_{i\alpha},G^{i\alpha}\}
    -\{\OLL{G}_{i\Dalpha},\OLL{G}^{i\Dalpha}\}
    +\frac{i}{4}[W^{ij},[\nabla_z,W_{ij}]].\label{additional-2}
\end{align}

Note each first fermionic derivative of all the superfields in the theory
has been consistently computed in terms of only these superfields.
In this sense, these computations close in the set of superfields
at our disposal. Especially no additional (super)field can be
excited by applying fermionic derivatives in any way.
In other words, we can compute any higher derivatives of these superfields.
Such computations including further fermionic derivatives
are necessary to obtain an invariant action.

\subsubsection{Definition of Component Fields}
According to the computations above, we find the independent component
fields, or essentially the independent superfields,
which should correspond to the following degrees of freedom:
\begin{align}
 \phi_{ij}
   :&=W_{ij}|,\\
 \lambda_{i\alpha}
   :&=2iG_{i\alpha}|
     =\frac{2}{5}[\nabla_{j\alpha},W^j{}_i]\bigr|
     =-2[\nabla_{i\alpha},\nabla_z]\bigr|,\\
 \OLL{\lambda}_{i\Dalpha}
   :&=-2i\OLL{G}_{i\Dalpha}|
     =-\frac{2}{5}[\Bnabla_{j\Dalpha},W^j{}_i]\bigr|
     =-2[\Bnabla_{i\Dalpha},\nabla_z]\bigr|,\\
 H_{ij}
   :&=-i[\nabla_z,W_{ij}]|\\
    &=\frac{1}{10}\{\nabla_{[i\alpha},[\nabla_k{}^\alpha,W^k{}_{j]}]\}\bigr|
     =\frac{i}{2}\{\nabla_{[i\alpha},G_{j]}{}^\alpha\}\bigr|\\
    &=\frac{1}{10}\{\Bnabla_{[i\Dalpha},[\Bnabla_k{}^\Dalpha,W^k{}_{j]}]\}
      \bigr|
     =\frac{i}{2}\{\Bnabla_{[i\Dalpha},\OLL{G}_{j]}{}^\Dalpha\}\bigr|,\\
 V_\mu
   :&=ig_\mu|=-[\nabla_\mu,\nabla_z]\bigr|\\
    &=\frac{i}{20}(\bsigma_\mu)^{\Dbeta\beta}
      \{\nabla_{i\beta},[\Bnabla_{j\Dbeta},W^{ji}]\}\bigr|\\
    &=\frac{i}{20}(\bsigma_\mu)^{\Dbeta\beta}
      \{\Bnabla_{i\Dbeta},[\nabla_{j\beta},W^{ji}]\}\bigr|,\\
 A_\mu
   :&=i\nabla_\mu|,\\
 \biggl(\quad  
  F_{\mu\nu}
   :&=i[\nabla_\mu,\nabla_\nu]\bigr| \notag \\
    &=\frac{i}{8}\left(
      (\sigma_{\mu\nu})^{\alpha\beta}
       \{\nabla_{i\alpha},G^i{}_\beta\}
      +(\bsigma_{\mu\nu})^{\Dalpha\Dbeta}
       \{\Bnabla_{i\Dalpha},\OLL{G}^i{}_\Dbeta\}\right)\Bigr|\notag \\
    &=\frac{1}{40}\left(
      (\sigma_{\mu\nu})^{\alpha\beta}
       \{\nabla_{i\alpha},[\nabla_{j\beta},W^{ji}]\}
      +(\bsigma_{\mu\nu})^{\Dalpha\Dbeta}
       \{\Bnabla_{i\Dalpha},[\Bnabla_{j\Dbeta},W^{ji}]\}\right)\Bigr|
  \quad\biggr),
\end{align}
where $|$ denotes to take the lowest components.
Note these field contents are exactly the same as in the
$\GR{USp}(4)$ model in \cite{Sohnius-1}.
In particular, the highest spin is 1, successfully prohibiting
the higher spins.

\subsubsection{Supertransformations}
Let us now derive the supertransformations on these
component fields. Actually we find that these are the same
as in \eRef{delta-A}--\Ref{delta-V},
up to some trivial constant factors and the difference
of the spinor representations, as they should be.
Similarly the central charge transformations are also
rederived.

First we recall how those transformations are computed
in our formulation.
Let $B$ be a general superfield and $b:=B|$ its lowest component
field. Supertransformation of the component field $b$ is defined to be
\begin{equation}
 \delta b =\delta (B|)
         :=(\delta B)\Bigr|
          =[\xi^{i\alpha}Q_{i\alpha}
                +\OLL{\xi}^{i\Dalpha}\OLL{Q}_{i\Dalpha},B]\Bigr|
          =[\xi^{i\alpha}D_{i\alpha}
                +\OLL{\xi}^{i\Dalpha}\OLL{D}_{i\Dalpha},B]\Bigr|,
\end{equation}
where supercharges are represented as differential operators
on the superspace. The last equality holds since the computation
is for the lowest terms; the difference of
$Q_{i\alpha},\ \OLL{Q}_{i\Dalpha}$ and
$D_{i\alpha},\ \OLL{D}_{i\Dalpha}$ contains only linear terms w.r.t.\ 
$\theta,\ \OLL{\theta}$, so that it vanishes when taking
the lowest contributions (unless the superfield $B$ contains
fermionic derivatives, which is the case in our computations below.).
In the following computations, we always take the Wess-Zumino gauge,
where the fermionic superconnections
(and the central charge superconnection as well)
have no lowest terms. Then we find
\begin{equation}
  \delta b=\xi^{i\alpha}[D_{i\alpha},B]_\pm
             +\OLL{\xi}^{i\Dalpha}[\OLL{D}_{i\Dalpha},B]_\pm\Bigr|
          =\xi^{i\alpha}[\nabla_{i\alpha},B]_\pm
                +\OLL{\xi}^{i\Dalpha}[\OLL{\nabla}_{i\Dalpha},B]_\pm\Bigr|.
\end{equation}
Thus supertransformations of $b=B|$ can be readily computed
from the fermionic first derivatives of
superfields $[\nabla_{i\alpha},B]$ and $[\Bnabla_{i\Dalpha},B]$.
In our case, each independent component
field is defined as a lowest component of the superfield in the system
whose fermionic first derivatives have been computed completely
by Bianchi identities as shown in the preceding section.
We can therefore derive any supertransformations of each
component field in our formulation.
Similarly, the central charge transformations of the field
$b=B|$ are computed as
\begin{equation}
 \delta_z b =\delta_z (B|):=(\delta_z B) \Bigr|
   =[i\omega D_z,B]\Bigr|
   =i\omega[\nabla_z,B]\Bigr|\qquad
       \text{(due to the WZ gauge)}.
\end{equation}

We then list below the explicit results.
\allowdisplaybreaks
\begin{align}
 \delta \phi_{ij}
 &= \delta W_{ij}\Bigr|\notag\\
 &= \xi^{k\alpha}[\nabla_{k\alpha},W_{ij}]
   +\OLL{\xi}^{k\Dalpha}[\Bnabla_{k\Dalpha},W_{ij}]\Bigr|\notag\\
% &= \xi^{i\alpha}\Bigl(
%     i\Omega_{ij}G_{k\alpha}+2i\Omega_{k[i}G_{j]\alpha}\Bigr)
%   +\OLL{\xi}^{i\Dalpha}\Bigl(
%     i\Omega_{ij}\OLL{G}_{k\Dalpha}+2i\Omega_{k[i}\OLL{G}_{j]\Dalpha}\Bigr)
%    \Bigr|\notag\\
 &=\left(\xi_{[i}{}^\alpha\lambda_{j]\alpha}
      +\frac{1}{2}\Omega_{ij}\xi^{k\alpha}\lambda_{k\alpha}\right)
   -\left(\OLL{\xi}_{[i}{}^\Dalpha\OLL{\lambda}_{j]\Dalpha}
   +\frac{1}{2}\Omega_{ij}\OLL{\xi}^{k\Dalpha}\OLL{\lambda}_{k\Dalpha}\right),
  \label{BeginOfSupertransf}\\
 \delta \lambda_{i\alpha}
 &=2i\left(\xi^{k\beta}\{\nabla_{k\beta},G_{i\alpha}\}
           +\OLL{\xi}^{k\Dbeta}\{\Bnabla_{k\Dbeta},G_{i\alpha}\}\right)
   \Bigr|\notag\\
 &=\frac{1}{2}\xi_i{}^\beta(\sigma^{\mu\nu})_{\beta\alpha}F_{\mu\nu}\notag\\
 &\phantom{=\,}
   +i(\sigma^\mu)_{\alpha\Dbeta}\OLL{\xi}^{k\Dbeta}[\nabla_\mu,\phi_{ki}]
   -i(\sigma^\mu)_{\alpha\Dbeta}\OLL{\xi}_i{}^\Dbeta V_\mu
   +i\xi^k{}_\alpha H_{ki}
   -\frac{i}{2}\xi^k{}_\alpha[\phi_{kl},\phi^l{}_i],\\
 \delta \OLL{\lambda}_{i\Dalpha}
 &=-2i\left(\xi^{k\beta}\{\nabla_{k\beta},\OLL{G}_{i\Dalpha}\}
           +\OLL{\xi}^{k\Dbeta}\{\Bnabla_{k\Dbeta},\OLL{G}_{i\Dalpha}\}\right)
   \Bigr|\notag\\
 &=-\frac{1}{2}\OLL{\xi}_i{}^\Dbeta(\bsigma^{\mu\nu})_{\beta\alpha}F_{\mu\nu}
    \notag \\
 &\phantom{=\,}
   +i(\sigma^\mu)_{\beta\Dalpha}\xi^{k\beta}[\nabla_\mu,\phi_{ki}]
   +i(\sigma^\mu)_{\beta\Dalpha}\xi_i{}^\beta V_\mu
   +i\OLL{\xi}^k{}_\Dalpha H_{ki}
   -\frac{i}{2}\OLL{\xi}^k{}_\Dalpha[\phi_{kl},\phi^l{}_i],\\
 \delta H_{ij}
 &=-i\left(\xi^{k\beta}[\nabla_{k\beta},[\nabla_z,W_{ij}]]
          +\OLL{\xi}^{k\Dbeta}[\Bnabla_{k\Dbeta},[\nabla_z,W_{ij}]]\right)
    \Bigr|\notag \\
 &=-i\left(\xi_{[i}{}^\beta(\sigma^\mu)_{\beta\Dbeta}
            [\nabla_\mu,\OLL{\lambda}_{j]}{}^\Dbeta]
           +\frac{1}{2}\Omega_{ij}\xi^{k\beta}(\sigma_\mu)_{\beta\Dbeta}
            [\nabla_\mu,\OLL{\lambda}_k{}^\Dbeta]\right)\notag\\
 &\phantom{=\,}
    -i\left(\Bxi_{[i\Dbeta}(\bsigma^\mu)^{\Dbeta\beta}
            [\nabla_\mu,\lambda_{j]\beta}]
           +\frac{1}{2}\Omega_{ij}\Bxi^k{}_\Dbeta(\bsigma_\mu)^{\Dbeta\beta}
            [\nabla_\mu,\lambda_{k\beta}]\right)\notag\\
 &\phantom{=\,}
    -i\left(\xi_{[i}{}^\beta[\lambda^l{}_\beta,\phi_{j]l}]
            +\frac{1}{2}\Omega_{ij}\xi^{k\beta}
             [\lambda^l{}_\beta,\phi_{kl}]\right)\notag\\
 &\phantom{=\,}
    -i\left(\Bxi_{[i}{}^\Dbeta[\OLL{\lambda}^l{}_\Dbeta,\phi_{j]l}]
            +\frac{1}{2}\Omega_{ij}\Bxi^{k\Dbeta}
             [\OLL{\lambda}^l{}_\Dbeta,\phi_{kl}]\right)\notag\\
 &\phantom{=\,}
    +\frac{i}{2}\xi^{k\beta}[\lambda_{k\beta},\phi_{ij}]
    +\frac{i}{2}\Bxi^{k\Dbeta}[\OLL{\lambda}_{k\Dbeta},\phi_{ij}],\\
 \delta V_\mu
 &=i\xi^{i\alpha}[\nabla_{i\alpha},g_\mu]
   +i\Bxi^{i\Dalpha}[\Bnabla_{i\Dalpha},g_\mu]\Bigr|\notag\\
 &=-\frac{1}{2}\xi^{i\alpha}(\sigma_{\mu\nu})_\alpha{}^\beta
           [\nabla^\nu,\lambda_{i\beta}]
   +\frac{1}{2}\Bxi^i{}_\Dalpha(\bsigma_{\mu\nu})^\Dalpha{}_\Dbeta
           [\nabla^\nu,\Bnabla_i{}^\Dbeta]\notag\\
 &\phantom{=\,}
   +\frac{1}{2}\xi^{i\alpha}(\sigma_{\mu})_{\alpha\Dalpha}
           [\Blambda^{j\Dalpha},\phi_{ij}]
   +\frac{1}{2}\Bxi^i{}_\Dalpha(\bsigma_{\mu})^{\Dalpha\alpha}
           [\lambda^j{}_\alpha,\phi_{ij}],\\
 \delta A_\mu
 &=i\left(\xi^{i\alpha}[\nabla_{i\alpha},\nabla_\mu]
          +\Bxi^{i\Dalpha}[\Bnabla_{i\Dalpha},\nabla_\mu]\right)\Bigr|
     \notag\\
 &=-\frac{i}{2}\xi^{i\alpha}(\sigma_\mu)_{\alpha\Dalpha}\Blambda_i{}^\Dalpha
   +\frac{i}{2}\Bxi^i{}_\Dalpha(\bsigma_\mu)^{\Dalpha\alpha}\lambda_{i\alpha}.
   \label{EndOfSupertransf}
\end{align}
Similarly, the central charge transformations are give as follows:
\begin{align}
 \delta_z\phi_{ij}
 &=i\omega[\nabla_z,W_{ij}]\Bigr|\notag\\
 &=-\omega H_{ij},\\
 \delta_z\lambda_{i\alpha}
 &=-2\omega[\nabla_z,G_{i\alpha}]\Bigr|\notag\\
 &=i\omega\left(
   (\sigma^\mu)_{\alpha\Dalpha}[\nabla_\mu,\Blambda_i{}^\Dalpha]
    +[\lambda^j{}_\alpha,\phi_{ij}]\right),\\
 \delta_z\Blambda_{i\Dalpha}
 &=-2\omega[\nabla_z,G_{i\alpha}]\Bigr|\notag\\
 &=-i\omega\left(
   (C\bsigma^\mu)_\Dalpha{}^\alpha[\nabla_\mu,\lambda_{i\alpha}]
    +[\Blambda^j{}_\Dalpha,\phi_{ij}]\right),\\
 \delta_z H_{ij}
 &=\omega[\nabla_z,[\nabla_z,W_{ij}]]\Bigr|\notag\\
 &=\omega\Biggl(
   [\nabla_\mu,[\nabla^\mu,\phi_{ij}]]
   -\frac{1}{4}[\phi_{[i|k},[\phi_{j]l},\phi^{kl}]]\notag\\
 &\phantom{=-}
   -i\left(\frac{1}{4}\Omega_{ij}\{\lambda_{k\gamma},\lambda^{k\gamma}\}
           -\{\lambda_{i\gamma},\lambda_j{}^{\gamma}\}\right)\notag\\
 &\phantom{=--}
   +i\left(\frac{1}{4}\Omega_{ij}\{\Blambda_{k\Dgamma},\Blambda^{k\Dgamma}\}
           -\{\Blambda_{i\Dgamma},\Blambda_j{}^{\Dgamma}\}\right)\Biggr),\\
 \delta_z V_\mu
 &=-\omega[\nabla_z,g_\mu]\Bigr|\notag\\
 &=\omega\left(
   [\nabla_\nu,F^\nu{}_\mu]
   -\frac{1}{2}(\bsigma)^{\Dalpha\alpha}
    \{\Blambda_{i\Dalpha},\lambda^i{}_\alpha\}
   +\frac{1}{4}[\phi^{ij},[\nabla_\mu,\phi_{ij}]]\right),\\
 \delta_z A_\mu
 &=-\omega[\nabla_z,\nabla_\mu]\Bigr|\notag\\
 &=-\omega V_\mu.
\end{align}

\subsubsection{Off-Shell Closure}
Finally we show that the supertransformations derived above
close off-shell. Since the supertransformations are
essentially intrinsic to our superspace formulation and
derived automatically, the off-shell
closure of those supertransformations shows, in a sense,
the superspace has been completely and successfully constructed
for the $\GR{USp}(4)$ model.

Let $B$ be one of the superfields in our case
and $b:=B|$ be the lowest component as before.
Recall first the off-shell closure is generally represented as
\begin{align}
 (\delta_1\delta_2-\delta_2\delta_1)b
 &=-i\left(
  (\xi_1)_{i\alpha}(\xi_2)^{i\alpha}
   +(\Bxi_1)_{i\Dalpha}(\Bxi_2)^{i\Dalpha}\right)
  [\del_z,b] \notag \\
 &\phantom{=\ }
   -i\left(
  (\xi_1)_i{}^{\alpha}(\sigma^\mu)_{\alpha\Dalpha}(\Bxi_2)^{i\Dalpha}
  -(\xi_2)_i{}^{\alpha}(\sigma^\mu)_{\alpha\Dalpha}(\Bxi_1)^{i\Dalpha}\right)
  [\del_\mu,b].\label{closeness}
\end{align}
Instead of showing this by using the explicit formulae
\Ref{BeginOfSupertransf}--\Ref{EndOfSupertransf},
we presents here more general procedure in the WZ gauge:
\begin{align*}
 \delta b
 &=\xi^{i\alpha}[\nabla_{i\alpha},B]_\pm
   +\Bxi^{i\Dalpha}[\Bnabla_{i\Dalpha},B]_\pm \Bigr|,\\
 \delta_1\delta_2 b
% &=(\xi_2)^{i\alpha}\Bigl(
%   [\delta_1 \nabla_{i\alpha},B]_\pm
%   +[\nabla_{i\alpha},\delta_1 B]_\pm\Bigr)
%  +(\Bxi_2)^{i\Dalpha}\Bigl(
%   [\delta_1 \Bnabla_{i\Dalpha},B]_\pm
%   +[\Bnabla_{i\Dalpha},\delta_1 B]_\pm\Bigr)\Bigr|\\
% &=-(\xi_2)^{i\alpha}(\xi_1)^{j\beta}\Bigl(
%    [\nabla_{i\alpha},[Q_{j\beta},B]_\pm]_\mp
%    -[\{Q_{j\beta},\nabla_{i\alpha}\},B]\Bigr)\\
% &\phantom{=\ }
%   -(\xi_2)^{i\alpha}(\Bxi_1)^{j\Dbeta}\Bigl(
%    [\nabla_{i\alpha},[\OLL{Q}_{j\Dbeta},B]_\pm]_\mp
%    -[\{\OLL{Q}_{j\Dbeta},\nabla_{i\alpha}\},B]\Bigr)\\
% &\phantom{=\ }
%   -(\Bxi_2)^{i\Dalpha}(\xi_1)^{j\beta}\Bigl(
%    [\Bnabla_{i\Dalpha},[Q_{j\beta},B]_\pm]_\mp
%    -[\{Q_{j\beta},\Bnabla_{i\Dalpha}\},B]\Bigr)\\
% &\phantom{=\ }
%    -(\Bxi_2)^{i\Dalpha}(\Bxi_1)^{j\Dbeta}\Bigl(
%    [\Bnabla_{i\Dalpha},[\OLL{Q}_{j\Dbeta},B]_\pm]_\mp
%    -[\{\OLL{Q}_{j\Dbeta},\Bnabla_{i\Dalpha}\},B]\Bigr)\Bigr|\\
% &=-(\xi_1)^{j\beta}(\xi_2)^{i\alpha}
%    [Q_{j\beta},[\nabla_{i\alpha},B]_\pm]_\mp%\\
%% &\phantom{=\ }
%   -(\Bxi_1)^{j\Dbeta}(\xi_2)^{i\alpha}
%    [\OLL{Q}_{j\Dbeta},[\nabla_{i\alpha},B]_\pm]_\mp\\
% &\phantom{=\ }
%   -(\xi_1)^{j\beta}(\Bxi_2)^{i\Dalpha}
%    [Q_{j\beta},[\Bnabla_{i\Dalpha},B]_\pm]_\mp%\\
%% &\phantom{=\ }
%   -(\Bxi_1)^{j\Dbeta}(\Bxi_2)^{i\Dalpha}
%    [\OLL{Q}_{j\Dbeta},[\Bnabla_{i\Dalpha},B]_\pm]_\mp \Bigr|\\
 &=-(\xi_1)^{j\beta}(\xi_2)^{i\alpha}
    [\nabla_{j\beta},[\nabla_{i\alpha},B]_\pm]_\mp%\\
% &\phantom{=\ }
   -(\Bxi_1)^{j\Dbeta}(\xi_2)^{i\alpha}
    [\OLL{\nabla}_{j\Dbeta},[\nabla_{i\alpha},B]_\pm]_\mp\\
 &\phantom{=\ }
   -(\xi_1)^{j\beta}(\Bxi_2)^{i\Dalpha}
    [\nabla_{j\beta},[\Bnabla_{i\Dalpha},B]_\pm]_\mp%\\
% &\phantom{=\ }
   -(\Bxi_1)^{j\Dbeta}(\Bxi_2)^{i\Dalpha}
    [\OLL{\nabla}_{j\Dbeta},[\Bnabla_{i\Dalpha},B]_\pm]_\mp \Bigr|
 \qquad\text{(WZ gauge)},
\end{align*}
so that
\begin{align*}
 (\delta_1\delta_2-\delta_2\delta_1)b
% &=-(\xi_1)^{j\beta}(\xi_2)^{i\alpha}\Bigl(
%   [\nabla_{j\beta},[\nabla_{i\alpha},B]_\pm]_\mp+
%   [\nabla_{i\alpha},[\nabla_{j\beta},B]_\pm]_\mp\Bigr)\\
% &\phantom{=\ }
%   -(\Bxi_1)^{j\Dbeta}(\xi_2)^{i\alpha}\Bigl(
%   [\Bnabla_{j\Dbeta},[\nabla_{i\alpha},B]_\pm]_\mp+
%   [\nabla_{i\alpha},[\Bnabla_{j\Dbeta},B]_\pm]_\mp\Bigr)\\
% &\phantom{=\ }
%   -(\xi_1)^{j\beta}(\Bxi_2)^{i\Dalpha}\Bigl(
%   [\nabla_{j\beta},[\Bnabla_{i\Dalpha},B]_\pm]_\mp+
%   [\Bnabla_{i\Dalpha},[\nabla_{j\beta},B]_\pm]_\mp\Bigr)\\
% &\phantom{=\ }
%   -(\Bxi_1)^{j\Dbeta}(\Bxi_2)^{i\Dalpha}\Bigl(
%   [\Bnabla_{j\Dbeta},[\Bnabla_{i\Dalpha},B]_\pm]_\mp+
%   [\Bnabla_{i\Dalpha},[\Bnabla_{j\Dbeta},B]_\pm]_\mp\Bigr)\Bigr|\\
% &=+(\xi_1)^{j\beta}(\xi_2)^{i\alpha}
%   [B,\{\nabla_{j\beta},\nabla_{i\alpha}\}]%\\
%% &\phantom{=\ }
%   +(\Bxi_1)^{j\Dbeta}(\xi_2)^{i\alpha}
%   [B,\{\Bnabla_{j\Dbeta},\nabla_{i\alpha}\}]\\
% &\phantom{=\ }
%   +(\xi_1)^{j\beta}(\Bxi_2)^{i\alpha}
%   [B,\{\nabla_{j\beta},\Bnabla_{i\Dalpha}\}]%\\
%% &\phantom{=\ }
%   +(\Bxi_1)^{j\Dbeta}(\Bxi_2)^{i\alpha}
%   [B,\{\Bnabla_{j\Dbeta},\Bnabla_{i\Dalpha}\}]\Bigr|\\
% &=-(\xi_1)^{j\beta}(\xi_2)^{i\alpha}
%   [i\Omega_{ji}C_{\beta\alpha}\nabla_z-iC_{\beta\alpha}W_{ji},B]%\\
%% &\phantom{=\ }
%   -(\Bxi_1)^{j\Dbeta}(\xi_2)^{i\alpha}
%   [i\Omega_{ji}(\sigma^\mu)_{\alpha\Dbeta}\nabla_\mu,B]\\
% &\phantom{=\ }
%   -(\xi_1)^{j\beta}(\Bxi_2)^{i\alpha}
%   [i\Omega_{ji}(\sigma^\mu)_{\beta\Dalpha}\nabla_\mu,B]%\\
%% &\phantom{=\ }
%   -(\Bxi_1)^{j\Dbeta}(\Bxi_2)^{i\alpha}
%   [i\Omega_{ji}C_{\Dbeta\Dalpha}\nabla_z+iC_{\Dbeta\Dalpha}W_{ji},B]\Bigr|\\
 &=-i\Bigl(
  (\xi_1)_i{}^{\alpha}(\sigma^\mu)_{\alpha\Dalpha}(\Bxi_2)^{i\Dalpha}
  -(\xi_2)_i{}^{\alpha}(\sigma^\mu)_{\alpha\Dalpha}(\Bxi_1)^{i\Dalpha}\Bigr)   
   [\nabla_\mu,B] \\
 &\phantom{=\ }
   -i\Bigl(
   (\xi_1)_{i\alpha}(\xi_2)^{i\alpha}
   +(\xi_1)_{i\Dalpha}(\xi_2)^{i\Dalpha}\Bigr)
   [\nabla_z,B]
   -i\Bigl(
   (\xi_1)^i_{\alpha}(\xi_2)^{j\alpha}
   -(\xi_1)^i_{\Dalpha}(\xi_2)^{j\Dalpha}\Bigr)
   [W_{ij},B]   \Bigr|.
% &=-i\Bigl(
%   (\xi_1)_{i\alpha}(\xi_2)^{i\alpha}
%   +(\xi_1)_{i\Dalpha}(\xi_2)^{i\Dalpha}\Bigr)
%   [D_z,B]\\
% &\phantom{=\ }
%   -i\Bigl(
%  (\xi_1)_i{}^{\alpha}(\sigma^\mu)_{\alpha\Dalpha}(\Bxi_2)^{i\Dalpha}
%  -(\xi_2)_i{}^{\alpha}(\sigma^\mu)_{\alpha\Dalpha}(\Bxi_1)^{i\Dalpha}\Bigr)
%   [D_\mu-i\Gamma_\mu,B]\\
% &\phantom{=\ }
%  -\biggl[
%  (\xi_1)_i{}^{\alpha}(\sigma^\mu)_{\alpha\Dalpha}(\Bxi_2)^{i\Dalpha}
%  -(\xi_2)_i{}^{\alpha}(\sigma^\mu)_{\alpha\Dalpha}(\Bxi_1)^{i\Dalpha}
%   \Gamma_\mu\\
% &\phantom{=-\biggl[]}
%  +i\biggl[\Bigl(
%   (\xi_1)^i_{\alpha}(\xi_2)^{j\alpha}
%   -(\xi_1)^i_{\Dalpha}(\xi_2)^{j\Dalpha}\Bigr)
%   W_{ij}, B\biggr]\Bigr|.
\end{align*}
Thus, for the component fields
$b=\phi_{ij},\ \lambda_{i\alpha},\ \Blambda_{i\Dalpha},\ H_{ij},\  V_\mu$,
we find
\begin{align}
 (\delta_1\delta_2-\delta_2\delta_1)b
 &=-i\Bigl(
  (\xi_1)_i{}^{\alpha}(\sigma^\mu)_{\alpha\Dalpha}(\Bxi_2)^{i\Dalpha}
  -(\xi_2)_i{}^{\alpha}(\sigma^\mu)_{\alpha\Dalpha}(\Bxi_1)^{i\Dalpha}\Bigr)   
   [\MC{D}_\mu,b]\notag\\
 &\phantom{=\ }
   -i\Bigl(
   (\xi_1)_{i\alpha}(\xi_2)^{i\alpha}
   +(\xi_1)_{i\Dalpha}(\xi_2)^{i\Dalpha}\Bigr)
   [\del_z,b]
  +i\biggl[\Bigl(
   (\xi_1)^i_{\alpha}(\xi_2)^{j\alpha}
   -(\xi_1)^i_{\Dalpha}(\xi_2)^{j\Dalpha}\Bigr)
   \phi_{ij}, b\biggr],
% &\phantom{=\ }
\end{align}
where $\MC{D}_\mu$ is the ordinary gauge covariant derivative
showing that the $\del_\mu$ in \eRef{closeness} is correctly
gauge covariantized. The last term is the gauge transformation
of the adjoint field $b$ with the gauge parameter being
$((\xi_1)^i_{\alpha}(\xi_2)^{j\alpha}
   -(\xi_1)^i_{\Dalpha}(\xi_2)^{j\Dalpha})\phi_{ij}$.
Therefore the algebra closes up to a gauge transformation.
This term appears due to the fact we have taken the WZ gauge.
On the other hand, for $b=A_\mu$, we should take $B=i\nabla_\mu$
then we find
\begin{align}
 (\delta_1\delta_2-\delta_2\delta_1)A_\mu
 &=-\biggl[\MC{D}_\mu,\Bigl(
   (\xi_1)^i_{\alpha}(\xi_2)^{j\alpha}
   -(\xi_1)^i_{\Dalpha}(\xi_2)^{j\Dalpha}\Bigr)
   \phi_{ij}\biggr]\notag\\
 &\phantom{=\ }
   -i\Bigl(
   (\xi_1)_{i\alpha}(\xi_2)^{i\alpha}
   +(\xi_1)_{i\Dalpha}(\xi_2)^{i\Dalpha}\Bigr)
   [\del_z,A_\mu]\notag\\
 &\phantom{=\ }
   -i\Bigl(
  (\xi_1)_i{}^{\alpha}(\sigma^\nu)_{\alpha\Dalpha}(\Bxi_2)^{i\Dalpha}
  -(\xi_2)_i{}^{\alpha}(\sigma^\nu)_{\alpha\Dalpha}(\Bxi_1)^{i\Dalpha}\Bigr)   
   F_{\nu\mu},
\end{align}
so that the last term is again a gauge transformation of
the gauge field $A_\mu$ w.r.t.\ the gauge parameter
$((\xi_1)^i_{\alpha}(\xi_2)^{j\alpha}
   -(\xi_1)^i_{\Dalpha}(\xi_2)^{j\Dalpha})\phi_{ij}$.
Thus for any component field in this model the superalgebra
closes off-shell up to a gauge transformation.

One may suspect that the above algebraic proof of the off-shell
closure does not suffice to show the explicit supertransformations
\Ref{BeginOfSupertransf}--\Ref{EndOfSupertransf} certainly close.
However, those transformation laws have been computed
purely algebraically using only the constraints and the Bianchi identities,
and the computation in the proof above has also been done in exactly
the same manner. Thus the closeness is certainly and manifestly assured.

\section{Conclusion and Discussion}
\label{sec-conclusion}
We have seen how a superspace formulation using superconnections and
supercurvatures would be set up for the $\GR{USp}(4)$ super Yang-Mills
theory in four dimensions. We introduced a central charge to
prohibit unpleasant fields with spins higher than one appearing
in the super Yang-Mills multiplet. In order to consider $N=4$ supersymmetry
with a central charge, we had to break the $R$-symmetry $\GR{SU}(4)$
into some automorphic subgroup. We chose $\GR{USp}(4)$ as
such automorphic subgroup. We then obtained almost uniquely
the appropriate constraints for supercurvatures by noticing
the trivial algebraic symmetries,
dropping the unnecessary supercurvatures which contain spins higher than
one, and imposing the reality conditions on the supercurvatures
to make the multiplet irreducible. These constraints have been consistently
solved with the restrictions by the Bianchi identities. Then we have found
that the theory we set up has one vector, four Weyl spinors,
five real scalars, five auxiliary scalars and one extra vector-like field,
which thus contains the same number of bosons and fermions, namely,
3+5+5+3=16 bosons and 4\CD 2\CD 2=16 fermions, as the off-shell
degrees of freedom. Supertransformations
as well as transformations associated with the central charge
can be computed using the Bianchi identities, and
has been automatically shown to be off-shell in section
\ref{sec-USp-model}.
Thus concerning to supertransformations, we have succeeded to develop
a superspace formulation for the $\GR{USp}(4)$ super Yang-Mills theory
in four dimensions.

There are, however, some problems remained.
First, we have to notice the vector-like field $V_\mu$ noted above.
By definition, this field has (mass) dimension two, the same as that
of the auxiliary fields. This means that in four dimensions
the field can not have a kinetic term as an elementary field,
or, in other words, terms only like $V_\mu V^\mu /2$ can be
allowed. One may then consider that this field should merely be an auxiliary
or nondynamical field. We expect it is not~\cite{Sohnius-Stelle-West-2},
in fact, the field is further constrained by \eRef{additional-2}, i.e.\ 
\[
 [\nabla^\mu,g_\mu]
  = -\{G_{i\alpha},G^{i\alpha}\}
    -\{\OLL{G}_{i\Dalpha},\OLL{G}^{i\Dalpha}\}
    +\frac{i}{4}[W^{ij},[\nabla_z,W_{ij}]],
\]
which is the same as \eRef{additional} in \cite{Sohnius-Stelle-West-2}.
If we restrict the gauge group to be Abelian, this equation leads to
\begin{equation}
 [\nabla^\mu,g_\mu]=0.
\end{equation}
Then as a solution we can take
\begin{equation}
 g_\mu=[\nabla^\nu,(\ast A)_{\mu\nu}],\quad
  (\ast A)_{\mu\nu}+(\ast A)_{\nu\mu}=0.
\end{equation}
(Here the Hodge star is taken to make an gauge invariant quantity.)
Thus the constraint \Ref{additional-2} assures, in a way, that the field
$V_\mu=i g_\mu|$ is not elementary and does contain contain
dynamical degrees as in
\(
 V_\mu V^\mu
  = [\nabla^\nu,(\ast A)_{\mu\nu}][\nabla^\rho,(\ast A)^\mu{}_\rho\}]
\)
in the Abelian case.
Similarly even in the non-Abelian cases,
the field $V_\mu$ should be interpreted as non-elementary and dynamical
degrees of freedom due to the constraint \eRef{additional-2}.
Here again we emphasize that \eRef{additional-2} is implemented
in our formulation and derived automatically from the Bianchi
identities.

Another crucial issue is whether we can construct the manifestly invariant
action of the $\GR{USp}(4)$ model within the framework of our
superspace formulation.
Since we have the component formulation \Ref{action},
we should in principle construct the action on the superspace.
In fact, as is mentioned in \cite{Sohnius-Stelle-West-2},
it can be shown, by explicitly evaluating the components,
that the action can be constructed in the form,
roughly in our notation, like
\[
 S\sim \tr\int d^4x\,\left.\left(
  \left\{\nabla,\left[\nabla, G G\right]\right\}
  +\left\{\Bnabla,\left[\Bnabla, \OLL{G}\OLL{G}\right]\right\}\right)\right|.
\]
However, because of the fact that $N=4$ supersymmetry contains
total of 16 supercharges, it may not possible to construct a manifestly
supersymmetric action on superspace in four-dimensional spacetime.
Moreover we work on the theory with a central charge, which
makes its superspace formulation to develop much harder than usual.
Nevertheless, we expect such superspace formulation could be
established since the off-shell component formulation as its
counterpart has been presented in a simple form.
We have to reveal these points in the work in progress.

\section*{Acknowledgment}
I would like to thank my supervisor Prof.\ Noboru Kawamoto
for his encouraging and thoughtful advice. I am grateful
to all staffs in our division for their pedagogical instruction.
I would also like to thank Junji Kato, Akiko Miyake, Kazuhiro Nagata
and Issaku Kanamori for fruitful discussions.

\appendix

\section{Notation and Formulae}
\subsection{Notation}
We denote the symmetric group of degree $p$%
\footnote{It is the group consists of all bijective maps on the set
$N_p:=\{1,\cdots,p\}$.
%, or, more generally, on a set $X$ such that $\card(X)=p$.
} by $\MF{S}_p$. The signature for $\sigma\in\MF{S}_p$ is defined as usual,
i.e.\ 
\begin{equation}
  \sgn(\sigma):=
   \left\{\begin{array}{cl}
          +1, & (\sigma:\text{even permutation})\\
          -1, & (\sigma:\text{odd permutation})
   \end{array}\right. .
\end{equation}
We use the following notation for symmetrization and antisymmetrization
of any kind of indices as
\begin{equation}
 A^{(i_1\cdots i_p)}
  :=\sum_{\sigma\in\MF{S}_p}A^{i_{\sigma(1)}\cdots i_{\sigma(p)}},\quad
 A^{[i_1\cdots i_p]}
  :=\sum_{\sigma\in\MF{S}_p}\sgn(\sigma)
    A^{i_{\sigma(1)}\cdots i_{\sigma(p)}}.
\end{equation}

The set of all $n$ by $n$ matrices whose matrix components belong to
the field $K$ is denoted by $M(n,K)$.
The $n\times n$ unit matrix on $K$ is denoted as $\bs{1}_n\in M(n,K)$
or simply as $\bs{1}\in M(n,K)$.
We denote the Pauli matrices by $\tau^i\in M(2,\mathbb{C})\ (i=1,2,3)$.
We have as usual
\begin{equation}
 \tau^i\tau^j=\delta^{ij}\bs{1}+i\sum_{k=1}^3\epsilon^{ijk}\tau^k,\qquad
 \tau^2\tau^i\tau^2=-(\tau^i)^\ast,
\end{equation}
where $\epsilon^{ijk}$ is the totally antisymmetric symbol
with $\epsilon^{123}=1$.

Through this article, we adopt the Einstein's convention,
unless otherwise noted, to contract two identical indices
with one from upper and the other from lower.

\subsection{Weyl-Spinor Representations in Four Dimensions}
\subsubsection{Vector Basis in $M(2,\mathbb{C})$ --- Minkowski Basis}
Let
\begin{equation}
 \sigma^\mu       :=(\bs{1},\tau^i),\quad
 \bar{\sigma}^\mu :=(\bs{1},-\tau^i),
\end{equation}
and
\begin{equation}
 C_{\downarrow}   :=\tau^2,\quad
 C^{\uparrow}     :=(C_{\downarrow}{}^T)^{-1}\equiv -\tau^2,\quad
 C_{\downarrow}{}^\ast :=(C_{\downarrow})^\ast\equiv -\tau^2,\quad
 C^{\uparrow}{}^\ast   :=(C^{\uparrow})^\ast
      \equiv((C_{\downarrow}{}^\ast)^T)^{-1}\equiv\tau^2.
\end{equation}
Note that the last four matrices are all antisymmetric so
that satisfy
\begin{equation}
\label{contragradient}
 C^{\uparrow}=-(C_{\downarrow})^{-1},\quad
 C^{\uparrow}{}^\ast = -(C_{\downarrow}{}^\ast)^{-1}.
\end{equation}

Two matrices $\sigma^\mu$ and $\bar{\sigma}^\mu$ satisfy the relations
\begin{equation}
\label{anticomm-pauli}
 \sigma^\mu\bar{\sigma}^\nu+\sigma^\nu\bar{\sigma}^\mu
   =2\eta^{\mu\nu},\quad
 \bar{\sigma}^\mu\sigma^\nu+\bar{\sigma}^\nu\sigma^\mu
   =2\eta^{\mu\nu},
\end{equation}
where $\eta^{\mu\nu}=(+,-,-,-)$ is the Minkowski metric in four dimensions,
and are related by a transformation
\begin{equation}
\label{sigma-sigmabar}
 (\bar{\sigma}^\mu)^T = C^{\uparrow}\sigma^\mu
   (C^{\uparrow}{}^\ast)^T,\quad
 \text{or}\quad
 (\sigma^\mu)^T       = C_{\downarrow}{}^\ast\bar{\sigma}^\mu
   (C_{\downarrow})^T.
\end{equation}
Notice also that
\begin{equation}
 (\sigma^\mu)^\dagger=\sigma^\mu,\quad
 (\bar{\sigma}^\mu)^\dagger=\bar{\sigma}^\mu.
\end{equation}

We prescribe that, according to the reason which will be understood
later, these matrices are labeled by row and column indices as
\begin{equation}
 (\sigma^\mu)_{\alpha\Dbeta},\quad
 (\bar{\sigma}^\mu)^{\Dalpha\beta}
\end{equation}
and
\begin{equation}
 (C_{\downarrow})_{\alpha\beta}=:C_{\alpha\beta},\quad
 (C^{\uparrow})^{\alpha\beta}  =:C^{\alpha\beta},\quad
 (C_{\downarrow}{}^\ast)_{\Dalpha\Dbeta}:= C_{\Dalpha\Dbeta},\quad
 (C^{\uparrow}{}^\ast)^{\Dalpha\Dbeta}  := C^{\Dalpha\Dbeta}.
\end{equation}
Then the above equations (\ref{contragradient}),
(\ref{anticomm-pauli}), and (\ref{sigma-sigmabar}) can be expressed by
\begin{gather}
 C_{\alpha\beta}C^{\gamma\beta}=\delta_{\alpha}{}^{\gamma},\quad
 C_{\Dalpha\Dbeta}C^{\Dgamma\Dbeta}=\delta_{\Dalpha}{}^{\Dbeta};\\
 (\sigma^\mu)_{\alpha\Dgamma}(\bar{\sigma}^\nu)^{\Dgamma\beta}
   +(\sigma^\nu)_{\alpha\Dgamma}(\bar{\sigma}^\mu)^{\Dgamma\beta}
     =2\eta^{\mu\nu}\delta_\alpha{}^\beta,\quad
 (\bar{\sigma}^\mu)^{\Dalpha\gamma}(\sigma^\nu)_{\gamma\Dbeta}
   +(\bar{\sigma}^\nu)^{\Dalpha\gamma}(\sigma^\mu)_{\gamma\Dbeta}
     =2\eta^{\mu\nu}\delta^{\Dalpha}{}_{\Dbeta};\\
 (\bar{\sigma}^\mu)^{\Dalpha\beta}
   =C^{\beta\delta}C^{\Dalpha\Dgamma}(\sigma^\mu)_{\delta\Dgamma},\quad
 (\sigma^\mu)_{\alpha\Dbeta}
   =C_{\Dbeta\Ddelta}C_{\alpha\gamma}(\bar{\sigma}^\mu)^{\Ddelta\gamma},
\end{gather}
respectively.

\paragraph{Orthonormality and Completeness}
Using equations (\ref{anticomm-pauli}) and (\ref{sigma-sigmabar}),
we obtain the orthonormality of the matrices
$\sigma^\mu,\ \bar{\sigma}^\mu$:
\begin{equation}
\label{orthonormal-sigma}
 \tr \sigma^\mu\bar{\sigma}^\nu = 2\eta^{\mu\nu}.
\end{equation}
Then we can easily show that four matrices
$\sigma^\mu\in M(2,\mathbb{C})$ are linearly independent,
and because of the fact that $\dim M(2,\mathbb{C})=4$,
they form a basis of the complex vector
space $M(2,\mathbb{C})$. Similarly, four matrices
$\bar{\sigma}^\mu\in M(2,\mathbb{C})$
form another basis of $M(2,\mathbb{C})$. Completeness of these basis
can be expressed as
\begin{equation}
 (\sigma^\mu)_{\alpha\Dgamma}(\bar{\sigma}_\mu)^{\Ddelta\beta}
  =2\delta_{\alpha}{}^{\beta}\delta^{\Ddelta}{}_{\Dgamma}.
\end{equation}
So called the Firtz transformations can be derived from the last
identity; for instance,
\begin{equation}
 (\sigma^\mu)_{\alpha\Dgamma}(\sigma_\mu)_{\beta\Ddelta}
  =2C_{\alpha\beta}C_{\Dgamma\Ddelta},\quad
 (\bar{\sigma}^\mu)^{\Dalpha\gamma}(\bar{\sigma}_\mu)^{\Dbeta\delta}
  =2C^{\Dalpha\Dbeta}C^{\gamma\delta}.
\end{equation}

\paragraph{Some Formulae}
From \eRef{anticomm-pauli} we obtain
\begin{align}
\label{three-sigma-symm}
 \sigma^\mu\bar{\sigma}^\rho\sigma^\nu
 +\sigma^\nu\bar{\sigma}^\rho\sigma^\mu
 &=2(\eta^{\mu\rho}\sigma^\nu+\eta^{\nu\rho}\sigma^\mu-\eta^{\mu\nu}\sigma^\rho),\\
 \bar{\sigma}^\mu\sigma^\rho\bar{\sigma}^\nu
 +\bar{\sigma}^\nu\sigma^\rho\bar{\sigma}^\mu
 &=2(\eta^{\mu\rho}\bar{\sigma}^\nu+\eta^{\nu\rho}\bar{\sigma}^\mu
    -\eta^{\mu\nu}\bar{\sigma}^\rho).
\end{align}
Total antisymmetricity w.r.t.\ $\mu,\ \nu,\ \rho,\ \sigma$ in
\begin{equation}
 \tr(\sigma^\mu\bar{\sigma}^\rho\sigma^\nu\bar{\sigma}^\sigma
     -\sigma^\nu\bar{\sigma}^\rho\sigma^\mu\bar{\sigma}^\sigma),\quad
 \tr(\bar{\sigma}^\mu\sigma^\rho\bar{\sigma}^\nu\sigma^\sigma
     -\bar{\sigma}^\nu\sigma^\rho\bar{\sigma}^\mu\sigma^\sigma)
\end{equation}
leads to the equations
\begin{align}
\label{three-sigma-asymm}
 \sigma^\mu\bar{\sigma}^\rho\sigma^\nu
 -\sigma^\nu\bar{\sigma}^\rho\sigma^\mu
  &=2i\VE^{\mu\nu\rho\sigma}\sigma_\sigma,\\
 \bar{\sigma}^\mu\sigma^\rho\bar{\sigma}^\nu
 -\bar{\sigma}^\nu\sigma^\rho\bar{\sigma}^\mu
  &=-2i\VE^{\mu\nu\rho\sigma}\bar{\sigma}_\sigma.
\end{align}

\subsubsection{(Anti-) Symmetric Basis in $M(2,\mathbb{C})$
 --- Minkowski Basis}
Let
\begin{equation}
 \sigma^{\mu\nu}:=\frac{1}{2}(\sigma^\mu\bar{\sigma}^\nu
                             -\sigma^\nu\bar{\sigma}^\mu),\quad
 \bar{\sigma}^{\mu\nu}
                :=\frac{1}{2}(\bar{\sigma}^\mu\sigma^\nu
                             -\bar{\sigma}^\nu\sigma^\mu).
\end{equation}
Clearly, these matrices have standard index structure
\begin{equation}
 (\sigma^{\mu\nu})_\alpha{}^\beta,\quad
 (\bar{\sigma}^{\mu\nu})^{\Dalpha}{}_{\Dbeta},
\end{equation}
and are traceless
\begin{equation}
 \tr\sigma^{\mu\nu}=(\sigma^{\mu\nu})_\alpha{}^\alpha=0,\quad
 \tr\bar{\sigma}^{\mu\nu}=(\bar{\sigma}^{\mu\nu})^{\Dalpha}{}_{\Dalpha}=0.
\end{equation}
Thus if we define matrices
\begin{alignat}{2}
 (\sigma^{\mu\nu})_{\alpha\beta}
  &:=(\sigma^{\mu\nu})_\alpha{}^\gamma C_{\gamma\beta},
 &\quad
 (\sigma^{\mu\nu})^{\alpha\beta}
  &:=C^{\alpha\gamma}(\sigma^{\mu\nu})_\gamma{}^\beta,\\
 (\bar{\sigma}^{\mu\nu})^{\Dalpha\Dbeta}
  &:=C^{\Dbeta\Dgamma}(\bar{\sigma}^{\mu\nu})^{\Dalpha}{}_{\Dgamma},
 &\quad
 (\bar{\sigma}^{\mu\nu})_{\Dalpha\Dbeta}
  &:=(\bar{\sigma}^{\mu\nu})^{\Dgamma}{}_{\Dbeta}C_{\Dgamma\Dalpha},
\end{alignat}
they are all symmetric w.r.t.\ $\alpha,\ \beta$ or $\Dalpha,\ \Dbeta$.
This fact can also be recognized by
\begin{equation}
 \sigma^{\mu\nu}C_{\downarrow}
 =\frac{1}{2}\left(\sigma^\mu C^{\uparrow\ast}(\sigma^\nu)^T
             +(\sigma^\mu C^{\uparrow\ast}(\sigma^\nu)^T)^T\right),\quad
 \text{etc.}
\end{equation}
Note also that
\begin{equation}
 (\sigma^{\mu\nu})^\dagger=-\bsigma^{\mu\nu},\quad
 (\bsigma^{\mu\nu})^\dagger=-\sigma^{\mu\nu}.
\end{equation}

\paragraph{(Anti-) Self-duality}
Since $\sigma^{\mu\nu}$ and $\bar{\sigma}^{\mu\nu}$ are traceless,
each of them represents $4-1=3$ d.o.f.\ in $M(2,\mathbb{C})$,
which corresponds to that
$\sigma^{\mu\nu}C_{\downarrow}$, $\bar{\sigma}^{\mu\nu}(C^{\uparrow})^T$,
etc.,\ are symmetric and have $2\CD 3/2=3$ d.o.f. These facts
then imply that $\sigma^{\mu\nu}$ and $\bar{\sigma}^{\mu\nu}$
are (anti-) self-dual ``tensors''. This is the case as we can show
using the formulae above that
\begin{equation}
 \tilde{\sigma}^{\mu\nu}
  :=\frac{1}{2}\VE^{\mu\nu\rho\sigma}\sigma_{\rho\sigma}
  =i\sigma^{\mu\nu},\quad
 \tilde{\bar{\sigma}}^{\mu\nu}
  :=\frac{1}{2}\VE^{\mu\nu\rho\sigma}\bar{\sigma}_{\rho\sigma}
  =-i\bar{\sigma}^{\mu\nu}.
\end{equation}

\paragraph{Orthonormality and Completeness}
Using eqs.~\Ref{three-sigma-symm}, \Ref{three-sigma-asymm}, and
\Ref{orthonormal-sigma}
we find that
\begin{equation}
 \tr \sigma^{\mu\nu}\sigma^{\rho\sigma}
   =-4\wp^+{}^{\mu\nu\rho\sigma},\quad
 \tr \bar{\sigma}^{\mu\nu}\bar{\sigma}^{\rho\sigma}
   =-4\wp^-{}^{\mu\nu\rho\sigma},
\end{equation}
where
\begin{equation}
 \wp^+{}^{\mu\nu\rho\sigma}
  :=\frac{1}{2}
    (\eta^{\mu\rho}\eta^{\nu\sigma}-\eta^{\mu\sigma}\eta^{\nu\rho}
     -i\VE^{\mu\nu\rho\sigma}),\quad
 \wp^-{}^{\mu\nu\rho\sigma}
  :=\frac{1}{2}
    (\eta^{\mu\rho}\eta^{\nu\sigma}-\eta^{\mu\sigma}\eta^{\nu\rho}
     +i\VE^{\mu\nu\rho\sigma}).
\end{equation}
Two symbols $\wp^+$ and $\wp^-$
can be interpreted as the projection operators in four dimensions into
the self-dual and anti-self-dual tensors, respectively,
in the vector representation, in fact,
\begin{gather}
 \frac{1}{2}\wp^\pm{}^{\mu\nu\rho\sigma}\wp^\pm{}_{\rho\sigma}{}^{\tau\lambda}
  =\wp^\pm{}^{\mu\nu\tau\lambda},\quad
 \frac{1}{2}\wp^\pm{}^{\mu\nu\rho\sigma}\wp^\mp{}_{\rho\sigma}{}^{\tau\lambda}
  =0,\\
 \wp^\pm{}^{\rho\sigma\mu\nu}=\wp^\pm{}^{\mu\nu\rho\sigma},\quad
 \wp^\pm{}^{\nu\mu\rho\sigma}=\wp^\pm{}^{\mu\nu\sigma\rho}
   =-\wp^\pm{}^{\mu\nu\rho\sigma}.
\end{gather}
Then we find that the sets of four matrices
\begin{equation}
 (C_{\alpha\beta},\ (\sigma^{\mu\nu})_{\alpha\beta}),\quad
 (C_{\Dalpha\Dbeta},\ (\bar{\sigma}^{\mu\nu})_{\Dalpha\Dbeta}),
\end{equation}
or their variant with some indices raised, are separately taken to be
bases in $M(2,\mathbb{C})$. Note here that
\begin{equation}
 \tr C_\downarrow C^\uparrow = -2,\quad
 \tr \sigma^{\mu\nu}C_\downarrow = 0,\quad \text{etc.}
\end{equation}

Completeness of these bases is expressed by identities as
\begin{gather}
 C_{\alpha\beta}C^{\gamma\delta}
 +\frac{1}{2}(\sigma^{\mu\nu})_{\alpha\beta}(\sigma_{\mu\nu})^{\gamma\delta}
  =2\delta_\alpha{}^\gamma \delta_\beta{}^\delta,\\
 C_{\Dalpha\Dbeta}C^{\Dgamma\Ddelta}
 +\frac{1}{2}(\bar{\sigma}^{\mu\nu})_{\Dalpha\Dbeta}
              (\bar{\sigma}_{\mu\nu})^{\Dgamma\Ddelta}
  =2\delta_{\Dalpha}{}^{\Dgamma} \delta_{\Dbeta}{}^{\Ddelta}.
\end{gather}
With the use of these identities, we can show, for instance,
the following Firtz transformations
\begin{equation}
 C_{\alpha\beta}C_{\gamma\delta}
  =C_{\alpha\gamma}C_{\beta\delta}-C_{\alpha\delta}C_{\beta\gamma},\quad
 \frac{1}{2}
 (\sigma^{\mu\nu})_{\alpha\beta}(\sigma_{\mu\nu})_{\gamma\delta}
  =C_{\alpha\gamma}C_{\beta\delta}+C_{\alpha\delta}C_{\beta\gamma},
 \ \text{etc.}
\end{equation}

\subsubsection{Weyl Spinors in Four-Dimensional Minkowski Space}
Weyl spinors are elements of representation spaces of
$\GR{SL}(2,\mathbb{C})$ corresponding to the two fundamental
representations $(\bs{2},\bs{1})$ and $(\bs{1},\bs{2})$.
We choose to label a spinor which transforms under $(\bs{2},\bs{1})$
as $\psi_\alpha$, and a the other as $\OLL{\psi}_\Dalpha$.
Their contragradient representations are thus represented by
$\psi^\alpha$ and $\OLL{\psi}^\Dalpha$. We assume
these spinors are Grassmann odd.
Matrices $C_\downarrow$ and $C_\downarrow{}^\ast$, or their contragradient
$C^\uparrow,\ C^\uparrow{}^\ast$, can be taken as the
$\GR{SL}(2,\mathbb{C})$ invariant metrics.
We choose $(\sigma^\mu)$ as a basis of the representation space for
$(\bs{2},\bs{1})\otimes(\bs{1},\bs{2})$, and $(\bsigma^\mu)$ as its
contragradient. This is why we have labeled as
$C_{\alpha\beta},\ C_{\Dalpha\Dbeta}$,
$C^{\alpha\beta},\ C^{\Dalpha\Dbeta}$ and
$(\sigma^\mu)_{\alpha\Dbeta},\ (\bsigma^\mu)^{\Dalpha\beta}$.
Transformations are then represented as follows
\begin{gather}
 \psi'_\alpha=\exp(+\frac{1}{4}\omega_{\mu\nu}\sigma^{\mu\nu})_\alpha{}^\beta
               \psi_\beta,\quad
 \OLL{\psi}'_\Dalpha=\exp(-\frac{1}{4}\omega_{\mu\nu}%
                    \bsigma^{\mu\nu})^\Dbeta{}_\Dalpha
               \OLL{\psi}_\Dbeta,\label{transf-Weyl}\\
 \psi'^\alpha=\exp(-\frac{1}{4}\omega_{\mu\nu}\sigma^{\mu\nu})_\beta{}^\alpha
               \psi^\beta,\quad
 \OLL{\psi}'^\Dalpha=\exp(+\frac{1}{4}\omega_{\mu\nu}%
                    \bsigma^{\mu\nu})^\Dalpha{}_\Dbeta
               \OLL{\psi}^\Dbeta.\label{transf-Weyl-2}
\end{gather}
Thus indices can be raised or lowered by $C$'s by the same rule as before,
namely,
\begin{equation}
 \psi^\alpha=C^{\alpha\beta}\psi_\beta,\quad
 \psi_\alpha=\psi^\beta C_{\beta\alpha}.
\end{equation}
Notice here that
\begin{equation}
 \psi_\alpha\chi^\alpha=-\psi^\alpha\chi_\alpha=+\chi_\alpha\psi^\alpha.
\end{equation}
We also find that
\begin{equation}
 \exp(-\frac{1}{4}\omega_{\mu\nu}%
                    \bsigma^{\mu\nu})^\Dbeta{}_\Dalpha
 =(\exp(+\frac{1}{4}\omega_{\mu\nu}\sigma^{\mu\nu})_\alpha{}^\beta)^\dagger,
 \quad\text{i.e.}\quad
 \OLL{\psi}_\Dalpha=(\psi_\alpha)^\dagger,\quad
 \OLL{\psi}^\Dalpha=(\psi^\alpha)^\dagger,
\end{equation}
i.e.,\ indices $\alpha$ are dotted to $\Dalpha$
under Hermitian conjugation and vice versa, which is consistent
to that we have labeled $(C^\ast)_{\Dalpha\Dbeta}$,
$(\sigma^\mu)^\dagger{}_{\alpha\Dbeta}=(\sigma^\mu)_{\alpha\Dbeta}$,
etc.

Through this article, Hermitian conjugation, as well as transposition,
always changes an order of, even of Grassmann, quantities. Thus for example,
\begin{equation}
 (\psi_\alpha\chi^\alpha)^\dagger
 =(\psi_\alpha C^{\alpha\beta}\chi_\beta)^\dagger
 =(\chi_\beta)^\dagger(C^{\Deta\Dalpha})(\psi_\alpha)^\dagger
 =\OLL{\chi}_\Dbeta\OLL{\psi}^\Dbeta.
\end{equation}
We define a fermionic derivative by
\begin{equation}
 \left\{
  \frac{\del}{\del\theta^\alpha}, \theta^\beta\right\}
  =\delta_\alpha{}^\beta,
 \quad
 \left\{
  \frac{\del}{\del\OLL{\theta}^\Dalpha}, \OLL{\theta}^\Dbeta\right\}
  =\delta_\Dalpha{}^\Dbeta,
\end{equation}
so that, by taking the Hermitian conjugation of this equation, we find that%
\footnote{Note in passing that, since $[\del_\mu,x^\nu]=\delta_\mu{}^\nu$,
we have $\del_\mu{}^\dagger=-\del_\mu$.}
\begin{equation}
 \left(\frac{\del}{\del\theta^\alpha}\right)^\dagger
 =\frac{\del}{\del(\theta^\alpha)^\dagger}
 =\frac{\del}{\del\OLL{\theta}^\Dalpha}.
\end{equation}
We also find that
\begin{equation}
 C^{\alpha\beta}\frac{\del}{\del\theta^\beta}
 =-\frac{\del}{\del\theta_\alpha},\quad\text{etc.}
\end{equation}

\subsubsection{Euclidean Basis and Weyl Spinors in Four-Dimensional
Euclidean Space}
Let
\begin{equation}
 \sigma^\mu:=(\bs{1},i\tau^i),\quad
 \bsigma^\mu:=(\bs{1},-i\tau^i).
\end{equation}
Then we have
\begin{equation}
 \sigma^\mu\bar{\sigma}^\nu+\sigma^\nu\bar{\sigma}^\mu
   =2\eta^{\mu\nu},\quad
 \bar{\sigma}^\mu\sigma^\nu+\bar{\sigma}^\nu\sigma^\mu
   =2\eta^{\mu\nu},
\end{equation}
where $\eta^{\mu\nu}=(+,+,+,+)$ is the four-dimensional Euclidean
metric. Most of the definitions and
formulas in the preceding sections also hold here by replacing
the Minkowski metric by the Euclidean metric, and also
$\VE^{\mu\nu\rho\sigma}$ by $i\VE^{\mu\mu\rho\sigma}$. There are,
however, some exceptions; note
\begin{equation}
 (\sigma^\mu)^\dagger=\bsigma^\mu,\quad
 (\bsigma^\mu)^\dagger=\sigma^\mu,
\end{equation}
so that
\begin{equation}
 (\sigma^{\mu\nu})^\dagger=-(\sigma^{\mu\nu}),\quad
 (\bsigma^{\mu\nu})^\dagger=-(\bsigma^{\mu\nu}).
\end{equation}
These equations and eqs.~\Ref{transf-Weyl} and \Ref{transf-Weyl-2}
implies that indices $\alpha,\ \Dalpha$ are raised or lowered
under Hermitian conjugations, namely,
\begin{equation}
 (\psi_\alpha)^\dagger=\psi^\alpha,\quad
 (\OLL{\psi}_\Dalpha)^\dagger=\OLL{\psi}^\Dalpha,\quad\text{etc.}
\end{equation}
\subsection{$\GR{SU}(N)$}
\subsubsection{Special Unitary Group $\GR{SU}(N)$}
The special unitary group $\GR{SU}(N)$ is the group composed of
all unitary transformations on a vector space
without the trivial phase transformation group $\GR{U}(1)$.
For convenience, it is represented by matrices
w.r.t.\ an orthonormalized basis, namely
\begin{equation}
\label{unitary-fundrep}
 \GR{SU}(N)=\{(M_i{}^j)\in\GR{GL}(N,\mathbb{C})|
               (M^\dagger)_i{}^k M_k{}^j=\delta_i{}^j,\ \det M=1\}.
\end{equation}

\subsubsection{Special Unitary Algebra $\AL{su}(N)$}
The special unitary algebra $\AL{su}(N)$ is composed of
elements $X\in\AL{gl}(N,\mathbb{C})$ which satisfies that
$\exp(itX)\in\GR{SU}(N)$, namely,
\begin{equation}
 \AL{su}(N)=\{(X_i{}^j)\in\AL{gl}(N,\mathbb{C})|
               (X^\dagger)_i{}^j=X_i{}^j,\ \tr X=0\}.
\end{equation}
Note here that $\AL{su}(N)$ should be a real vector space.
As a consequence we find that $\dim\GR{SU}(N)=\dim\AL{su}(N)=N^2-1$.

Letting
\begin{equation}
 \langle X,Y \rangle:=\tr(X^\dagger Y),\quad
 X,\ Y\in M(n,\mathbb{C}),
\end{equation}
we introduce a positive-definite Hermitian form on $M(N,\mathbb{C})$.
Then, by taking a suitable linear combination, we can obtain
an orthonormalized basis $(X^a)$ in $\AL{su}(N)$ such as
\begin{equation}
 \langle X^a,X^b \rangle = \frac{1}{2}\delta^{ab}.
\end{equation}
In addition, if we define $X^0:=\bs{1}/\sqrt{2N}\in M(N,\mathbb{C})$,
$N^2$ matrices $X^a\ (a=0,\cdots, N^2-1)$ satisfy this relation,
so that they can be interpreted as an orthonormalized basis
in $M(N,\mathbb{C})$. The corresponding completeness relation is
readily computed from this relation as
\begin{gather}
 \sum_{a=0}^{N^2-1}(X^a)_i{}^j (X^a)_k{}^l
  = \frac{1}{2}\delta_i{}^l \delta_k{}^j,\\
 \sum_{a=1}^{N^2-1}(X^a)_i{}^j (X^a)_k{}^l
  = \frac{1}{2N}(N\delta_i{}^l \delta_k{}^j-\delta_i{}^j \delta_k{}^l).
    \label{complete-su}
\end{gather}

Let $(X^a)$ be an orthonormalized basis in $M(N,\mathbb{C})$, and let
\begin{equation}
 F^{abc}:=2\langle X^a,X^bX^c \rangle
         =\langle X^a, \{X^b,X^c\} \rangle
          +\langle X^a,[X^b,X^c] \rangle.
\end{equation}
Note $F^{abc}=F^{bca}=F^{cab}$. Since
\begin{gather}
 \langle X^a, \{X^b,X^c\} \rangle^\ast
  =\langle \{X^b,X^c\},X^a \rangle=\langle X^a, \{X^b,X^c\} \rangle,\\
 \langle X^a,[X^b,X^c] \rangle^\ast
  =\langle [X^b,X^c],X^a \rangle  =-\langle X^a,[X^b,X^c] \rangle,
\end{gather}
they are identified as
\begin{equation}
 \langle X^a,\{X^b,X^c\}\rangle= d^{abc}:=\Re F^{abc},\quad
 -i\langle X^a,[X^b,X^c] \rangle=f^{abc}:=\Im F^{abc},
\end{equation}
where $d^{abc}$ is totally symmetric while $f^{abc}$ is totally
antisymmetric and in particular
$d^{ab0}=\delta^{ab}/\sqrt{2n},\ f^{ab0}=0$, so that
\begin{align}
 \{X^a,X^b\}&=\frac{1}{\sqrt{2N}}\delta^{ab}+\sum_{c=1}^{N^2-1}d^{abc}X^c,\\
  [X^a,X^b] &=i\sum_{c=1}^{N^2-1}f^{abc}X^c.
\end{align}

We now construct a specific orthonormalized basis in $\AL{su}(N)$
explicitly. Since $X^\dagger = X\ (X\in\AL{su}(N))$, we find
\begin{equation}
 (X_{\MR{R}})^T=X_{\MR{R}},\quad (X_{\MR{I}})^T+X_{\MR{I}}=0,\qquad
 \text{where}\quad
 X_{\MR{R}}:=\Re X,\quad X_{\MR{I}}:=\Im X.
\end{equation}
Thus as a basis, we can take the following:
\begin{gather}
 (X^m)_{kl}
   :=\frac{1}{\sqrt{2m(m+1)}}
       \left(\sum_{i=1}^{m}\delta^i{}_k\delta^i{}_l
         -m\delta^{m+1}{}_k\delta^{m+1}{}_l\right),\quad
       (m=1,\ \cdots,\ N-1), \label{SU-gen-1}\\
 (X_{\MR{s}}^{ij})_{kl}
   :=\frac{1}{2}
       (\delta^i{}_k \delta^j{}_l+\delta^j{}_k \delta^i{}_l),\quad
 (X_{\MR{a}}^{ij})_{kl}
   :=\frac{1}{2}i
       (\delta^i{}_k \delta^j{}_l-\delta^j{}_k \delta^i{}_l),\quad
       (1\leqslant i<j\leqslant N), \label{SU-gen-2}
\end{gather}  
where we have used fixed indices which do not obey the general
transformation laws. Note especially the Cartan subalgebra is generated
by $\{X^m\}$, so that weights are given as
\begin{align}
 \nu^m &=\left(0,\ \cdots,\ 0,\ -\frac{m-1}{\sqrt{2(m-1)m}},\ 
             \mathop{\frac{1}{\sqrt{2m(m+1)}}}^m,\ \cdots,\ 
             \frac{1}{\sqrt{2(N-1)N}}\right)\in\mathbb{R}^{N-1},\notag\\
       &\phantom{\left(0,\ \cdots,\ 0,\ -\frac{m-1}{\sqrt{2(m-1)m}},\ 
             \mathop{\frac{1}{\sqrt{2m(m+1)}}}^m,\ \cdots\right)}
             \quad (m=1,\ \cdots,\ N-1),\label{SU-W-1}\\
 \nu^N &=\left(0,\ \cdots,\ 0,\ 
               -\frac{N-1}{\sqrt{2(N-1)N}}\right)  \in\mathbb{R}^{N-1},
          \label{SU-W-2}
\end{align}
which satisfy that
\begin{equation}
 \|\nu^i\|^2=\frac{N-1}{2N},\quad
 \nu^i\CD\nu^j=-\frac{1}{2N},\quad(i\neq j),\quad
 \sum_{i=1}^{N}\nu^i=0.
\end{equation}
Correspondingly, simple roots are given as
\begin{gather}
 \alpha^i:=\nu^i-\nu^{i+1},\quad(i=1,\ \cdots,\ N-1),\\
 \text{with}\quad
 \|\alpha^i\|^2=1,\quad
 \alpha^i\CD\alpha^j
  =\left\{\begin{array}{cl}
           0, & (j\neq i,\ i\pm1)\\
           \DS -\frac{1}{2}, & (j=i\pm1)
          \end{array}\right.,
\end{gather}
thus the Dynkin diagram for $\AL{su}(N)$ is as follows:
\begin{equation}
 \begin{picture}(50,10)
  \put(0,5){\circle{10}}\put(0,-1){\makebox(0,0)[t]{\small$\alpha^1$}}
  \put(5,5){\line(1,0){10}}
  \dottedline{2}(15,5)(35,5)
  \put(35,5){\line(1,0){10}}
  \put(50,5){\circle{10}}\put(50,-1){\makebox(0,0)[t]{\small$\alpha^{N-1}$}}
 \end{picture}
\end{equation}

\subsubsection{Fundamental Representations}
\label{sec-unitary-fund}
The fundamental representations $(\rho,V)$ of $\GR{SU}(N)$ are given
by the $N$-dimensional defining representation \eRef{unitary-fundrep}
and, more generally, its antisymmetric tensor products
$(\rho\otimes\cdots\otimes\rho,\MC{A}(V\otimes\cdots\otimes V))$.
This can be seen by the fact that the weight
\begin{equation}
 \mu^j:=\sum_{i=1}^j\nu^i
\end{equation}
becomes fundamental.
Their differential representations
$(d\rho\oplus\cdots\oplus d\rho,\MC{A}(V\otimes\cdots\otimes V))$
become the fundamental representations of $\AL{su}(N)$.
We denote these representations by
$[j]:=\bs{\colvec{N\\ j}}$. 
Since
\begin{equation}
 \sum_{i=1}^{N}\nu^i=0,\quad\text{i.e.}\quad
 \sum_{i=1}^{j}\nu^i=-\sum_{i=j+1}^{N}\nu^i,
\end{equation}
we find that
\begin{equation}
 \OLL{[j]}=[N-j],\quad\text{especially}\quad
 \OL{\bs{N}}=[N-1].
\end{equation}
Adjoint representations $(\Ad,\AL{su}(N))$ of $\GR{SU}(N)$ and,
as its differential representations,
$(\ad,\AL{su}(N))$ of $\AL{su}(N)$, are $N^2-1$-dimensional and
given from the fundamental representations as in
\begin{equation}
 \bs{N}\otimes\OL{\bs{N}}=\bs{1}\oplus(\bs{N^2-1}).
\end{equation}

\subsection{$\GR{USp}(2n)$}
\label{sec-symplectic}
\subsubsection{Symplectic Form}
Let $V$ be a complex vector space with dimension $\dim V=2n$,
and $\omega$ be a antisymmetric nondegenerate bilinear form%
\footnote{
A bilinear form $\omega:V\times V\RA K$ is said to be
antisymmetric if it satisfies that
\[
 \Any v,\ w\in V,\quad\omega(v,w)+\omega(w,v)=0,
\]
i.e.\ $\DS\omega\in \bigwedge^2 V^\ast$. An antisymmetric nondegenerate
bilinear form $\omega$ exists only if the vector space $V$ is
even dimensional,
since an antisymmetric matrix $(\omega(e^i,e^j))\in M(\dim V,K)$,
where $(e^i)$ is a basis in $V$, cannot be nondegenerate (has at least
one zero eigenvalue) when $V$ is odd dimensional.}
on $V$.

The existence of a nondegenerate
bilinear form allows us to define a linear isomorphism $G:V\RA V^\ast$
such that
\begin{equation}
 \Any v\in V,\quad
 \tilde{v}:=G(v)\quad\text{s.t.}\quad
 \Any w\in V,\quad
 \tilde{v}(w)=\omega(v,w).
\end{equation}
Similarly the inverse $G^{-1}$ can be used to define an induced
bilinear form $\tilde{\omega}$ on $V^\ast$ such that
\begin{equation}
 \Any \phi\in V^\ast,\quad
 \tilde{\phi}:=G^{-1}(\phi)\quad\text{s.t.}\quad
 \Any \psi\in V^\ast,\quad
 \psi(\tilde{\phi})=\tilde{\omega}(\psi,\phi).
\end{equation}
These operations are represented by components w.r.t.\ a basis $(e^i)$
in $V$ and its dual $(e_i)$ in $V^\ast$ as before; letting
\begin{equation}
 \omega^{ij}:=\omega(e^i,e^j),\quad
 \omega_{ij}:=\tilde{\omega}(e_i,e_j),\quad
 v_i:=e_i(v),\quad
 \phi^i:=\phi(e^i),\quad
 v\in V,\ \phi\in V^\ast,
\end{equation}
we find that
\begin{equation}
 \omega_{ik}\omega^{jk}=\delta_i{}^j,\quad
 \tilde{v}^i=\omega^{ij}v_j,\quad
 \tilde{\phi}_i=\phi^j\omega_{ji}.
\end{equation}

\subsubsection{Unitary Symplectic Group $\GR{USp}(2n)$}
A transformation $f\in\End(V)$ such that
\begin{equation}
\label{symplectic}
 \omega(f(v),f(w))=\omega(v,w),\quad
 \text{for}\Any v,\ w\in V,
\end{equation}
is called a symplectic transformation. The nondegeneracy of
the symplectic form $\omega$ implies that $f\in\Aut(V)$. Then
the set of all symplectic transformations on $V$ forms a group,
called the symplectic group on $V$.
If $V$ is complex and finite dimensional, as is the case in our assumption,
one can impose that a symplectic transformation is simultaneously unitary
w.r.t.\ a Hermitian form on $V$. The collection of all
both unitary and symplectic transformations on $V$ also forms a group,
which is called the unitary symplectic group, or the complex symplectic
group, or simply, the symplectic group on the complex vector space $V$,
and denoted as $\GR{USp}(V)$ or simply, $\GR{Sp}(V)$.

Let $f\in\GR{USp}(V)$ and $(e^i)$ be a basis in $V$. The condition
\Ref{symplectic} is represented w.r.t.\ the basis as in
\begin{equation}
 M_k{}^i\omega^{kl}M_l{}^j=\omega^{ij},\quad
 \text{equivalently},\quad
 M_i{}^k\omega_{kl}M_j{}^l=\omega_{ij},
\end{equation}
where $f(e^i)=e^j M_j{}^i$.
Notice here that such transformation satisfies $(\det M)^2=1$.
Since $V$ is even dimensional, we find that $\det M=1$.
Then assume the basis $(e^i)$ is orthonormalized w.r.t.\ a Hermitian form
\begin{equation}
 g(e^i,e^j)=\delta^{\OL{i}j}.
\end{equation}
By a suitable orthogonal, but not symplectic in general,
transformation of this basis, we can obtain a basis $(e'^i)$ such that
\begin{equation}
 \omega(e'^i,e'^j)=\Omega^{ij}
 :=\left(\begin{array}{cc}
         0 & +1 \\
        -1 &  0
        \end{array}\right).
\end{equation}
We call this matrix representation of $\omega$ the standard form.
The corresponding contragradient $\tilde{\omega}$ is represented
\begin{equation}
 \tilde{\omega}(e'^i,e'^j)=\Omega_{ij}
 :=\left(\begin{array}{cc}
         0 & +1 \\
        -1 &  0
        \end{array}\right),
\end{equation}
i.e.\ by the same matrix.
Note that an orthogonal transformation is a unitary transformation,
which in particular preserve the Hermitian form,
so that $(e'^i)$ is also orthonormalized.
An orthonormalized basis which represents $\omega$ as
the standard form is said to be standard.

The unitary symplectic group is usually represented
w.r.t.\ a standard basis as in
\begin{equation}
 \GR{USp}(2n)=\{(M_i{}^j)\in\GR{GL}(2n,\mathbb{C})|
               M_k{}^i \Omega^{kl} M_l{}^j=\Omega^{ij},\ 
               (M^\dagger)_i{}^k M_k{}^j=\delta_i{}^j\}.
\end{equation}
If $M\in\GR{USp}(2n)$, it automatically satisfies $\det M=1$ as noted above.

In passing, we list some formulae in an irreducible matrix representation
\begin{equation}
 \omega_\downarrow+(\omega_\downarrow)^T=0,\quad
 \omega^\uparrow+(\omega^\uparrow)^T=0,\quad
 \omega_\downarrow=((\omega^\uparrow)^T)^{-1},\quad
 (\omega_\downarrow)^\dagger=c(\omega_\downarrow)^{-1}
  =-c\omega^\uparrow,
\end{equation}
where $0\neq c\in\mathbb{R}$ and $\omega_\downarrow=(\omega_{ij})$ and
$\omega^\uparrow=(\omega^{ij})$.
In particular, the last equation is derived by comparing
\begin{equation}
 M(\omega_\downarrow)^\dagger M^T=(\omega_\downarrow)^\dagger,\quad
 \text{and}\quad
 M(\omega_\downarrow)^{-1} M^T=(\omega_\downarrow)^{-1},
\end{equation}
and noting the uniqueness of such $\omega_\downarrow$ up to a constant
as shown by the Schur's lemma for the irreducible representation.
For the standard form, we find
\begin{equation}
 (\Omega_\downarrow)^\dagger=(\Omega_\downarrow)^{-1}=-\Omega^\uparrow.
\end{equation}

\subsubsection{Unitary Symplectic Algebra $\AL{usp}(2n)$}
The unitary symplectic algebra $\AL{usp}(2n)$ is the Lie algebra
of the Lie group $\GR{USp}(2n)$, namely, iff
$X\in\AL{usp}(2n)$ then $\exp(itX)\in\GR{USp}(2n)$, so that
\begin{equation}
 \AL{usp}(2n)=\{(X_i{}^j)\in\AL{gl}(2n,\mathbb{C})|
               X_k{}^i \Omega^{kj}+\Omega^{ik}X_k{}^j=0,\ 
               (X^\dagger)_i{}^j=X_i{}^j\}.
\end{equation}
Here again the condition $\tr X=0\ (X\in\AL{usp}(2n))$ is automatically
assured. If we write an element $X\in\AL{usp}(2n)$ in the form
\begin{equation}
 X=\begin{pmatrix}
     T  & S \\
     S' & T'
   \end{pmatrix},\qquad
  T,\ T',\ S,\ S'\in\AL{gl}(n,\mathbb{C}),
\end{equation}
we find that
\begin{equation}
 T^\dagger=T,\quad
 S^T=S,\quad
 T'=-T^T,\quad
 S'=S^\dagger.
\end{equation}
Thus as a basis in $\AL{usp}(2n)$, we can take the following
\begin{align}
 X^0&=\begin{pmatrix}
      \bs{1} & 0 \\
       0 &    -\bs{1}
     \end{pmatrix},\quad
 X^a=\begin{pmatrix}
      T^a & 0  \\
       0  & -(T^a)^\ast
     \end{pmatrix},
     \quad T^a\in\AL{su}(n),\quad(a=1,\ \cdots,\ n^2-1),\\
 X^a_{\MR{s}}
    &=\begin{pmatrix}
       0  & S_{\MR{R}} \\
      S_{\MR{R}} & 0
     \end{pmatrix},\quad
 X^a_{\MR{a}}
    =\begin{pmatrix}
       0  & -iS_{\MR{I}} \\
      iS_{\MR{I}} & 0
     \end{pmatrix},
     \quad S_{\MR{R,I}}\in\AL{gl}(n,\mathbb{R}),\quad
           S_{\MR{R,I}}^T=S_{\MR{R,I}},\notag \\
 &\phantom{%
    X^a_{\MR{s}}
    =\begin{pmatrix}
       0  & S_{\MR{R}} \\
      S_{\MR{R}} & 0
     \end{pmatrix},\quad
    X^a_{\MR{a}}
    =\begin{pmatrix}
       0  & -iS_{\MR{I}} \\
      iS_{\MR{I}} & 0
     \end{pmatrix},
     \quad S_{\MR{R,I}}SSS}
     \quad(a=1,\ \cdots,\ \frac{1}{2}n(n+1)),
\end{align}
and, as a consequence, we find that
$\dim \GR{USp}(2n)=\dim\AL{usp}(2n)=n(2n+1)$.
 
We consider the Cartan subalgebra to be generated by the following
$n$ elements:
\begin{equation}
 \tilde{X}^0:=\frac{1}{\sqrt{2n}}X^0,\quad
 \tilde{X}^m:=\begin{pmatrix}
                X^m  & 0\\
                0    & -X^m
              \end{pmatrix},\quad(m=1,\ \cdots,\ n-1),
\end{equation}
where $X^m$ is given by \Ref{SU-gen-1}.
Then, using the weights \Ref{SU-W-1}, \Ref{SU-W-2} and
$\nu^{n+1}:=e^{n+1}=(0,\ \cdots,\ 0,\ 1)\in\mathbb{R}^{n+1}$,
the corresponding weights are given
\begin{equation}
 \pm\left(\nu^i+\frac{1}{\sqrt{2n}}\nu^{n+1}\right),
\end{equation}
so that the roots are
\begin{equation}
 \nu^i-\nu^j,\ (i\neq j),\quad
 \pm\left(\nu^i+\nu^j+\sqrt{\frac{2}{n}}\nu^{n+1}\right),
\end{equation}
and thus simple roots are
\begin{equation}
 \alpha^i=\left\{
 \begin{array}{ll}
   \nu^i-\nu^{i+1}, & (i=1,\ \cdots,\ n-1),\\
   2\nu^n+\sqrt{\frac{2}{n}}\nu^{n+1}, & (i=n),
 \end{array}\right.
\end{equation}
which satisfy that
\begin{equation}
 \|\alpha^i\|^2=\left\{
 \begin{array}{ll}
   1, & (i=1,\ \cdots,\ n-1),\\
   2, & (i=n),
 \end{array}\right.\ 
 \alpha^i\CD\alpha^j=\left\{
 \begin{array}{cl}
  \DS -\frac{1}{2}, &(j=i+1,\ i=1,\ \cdots,\ n-1,\ \text{or}\ i\LR j),\\
   -1,           &(i=n-1,\ j=n,\ \text{or}\ i\LR j),\\
   0,            &(\text{otherwise}).
 \end{array}\right.
\end{equation}
The Dynkin diagram for $\AL{usp}(2n)$ is as follows:
\begin{equation}
 \begin{picture}(85,10)
  \put(0,5){\circle{10}}\put(0,-1){\makebox(0,0)[t]{\small$\alpha^1$}}
  \put(5,5){\line(1,0){10}}
  \dottedline{2}(15,5)(35,5)
  \put(35,5){\line(1,0){10}}
  \put(50,5){\circle{10}}%\put(50,-1){\makebox(0,0)[t]{\small$\alpha^{n-1}$}}
  \drawline(54.33,7.5)(65,7.5)
  \put(70,5){\circle{10}}
  \put(70,-1){\makebox(0,0)[t]{\small$\alpha^{n}$}}
  \drawline(54.33,2.5)(65,2.5)
 \end{picture}
\end{equation}
This Dynkin diagram looks like that of $\AL{so}(2n+1)$.
However, norms of these simple roots are not the same with each other, thus
the two diagrams are actually different, except the case for $n=2$,
so that $\GR{SO}(5)\cong\GR{USp}(4)$.

\section{Clifford Algebra and Spinors}
\label{sec-Clifford}
In this appendix, we briefly examine some fundamental properties
of Clifford algebra which plays a role in this article.
Notation and convention follows the preceding appendix.

\subsection{Clifford Algebra}
\subsubsection{Definition}
Let $\eta$ be a (non-degenerate) metric with (Sylvester's) signature
$(t,s)$ which is defined on a vector space $M(t,s)$ with dimension
$D=t+s$~\cite{Kugo-Townsend}:
\begin{equation}
 \eta_{\mu\nu}=(\underbrace{+,\cdots,+}_t,\ \underbrace{-,\cdots,-}_s).
\end{equation}

Clifford Algebra $\MC{C}(t,s)$ is defined to be an algebra which is generated
by $D$ elements $\{\gamma^{1},\ \cdots,\ \gamma^{D}\}$ which satisfy
the anticommutation relations%
\footnote{
Let $T(M(t,s))$ be the tensor algebra on $M(t,s)$ and $I_\eta$ be
the two-sided ideal
generated by the set
\[
  \{t\in T(M(t,s))|
    t=v\otimes v-\eta(v,v),\ v\in M(t,s)\}.
\]
Then Clifford algebra $\MC{C}(t,s)$ is defined as the quotient algebra
$\MC{C}(t,s)=T(M(t,s))/I_{\eta}$. In $\MC{C}(t,s)$ tensor product
of $\gamma,\ \gamma'\in\MC{C}(t,s)$ is simply denoted as
$\gamma\gamma'$ instead of $\gamma\otimes\gamma'$.
Clifford algebra is determined uniquely by the anticommutation
relation $\{\gamma,\gamma'\}=2\eta(\gamma,\gamma')$
up to an isomorphism due to the universality of
tensor algebra. This fact is said to be the universality of
Clifford algebra.}
\begin{equation}
 \label{Clifford-alg}
 \{\gamma^{\mu},\gamma^{\nu}\}=2\eta^{\mu\nu},
\end{equation}
where $\eta^{\mu\nu}$ is defined to be the inverse of the metric
$\eta_{\mu\nu}$:
\begin{equation}
 \eta^{\mu\nu}\eta_{\nu\rho}=\eta^{\mu}{}_{\rho}\equiv\delta^{\mu}{}_{\rho}.
\end{equation}

\subsubsection{Irreducible Representation}
In what follows, we will be only interested in a finite dimensional
irreducible linear representation $(\rho,V)$ of the Clifford algebra
$\MC{C}(t,s)$ on a finite dimensional complex vector space $V$.
We will therefore identify each element $\gamma\in\MC{C}(t,s)$ of the Clifford
algebra with its image $\rho(\gamma)\in\End(V)$, which is
a linear transformation on the vector space $V$, or a matrix on
$\mathbb{C}^{\dim V}$, without distinguishing them. Further, we will only
consider the case where the dimension $D$ is even.

A basis of the algebra $\MC{C}(t,s)$ with
even dimensions $D=t+s$ can be defined as
\begin{gather}
\label{basis}
 (\gamma^{\mu_1\cdots\mu_p})_{%
   1\leqslant\mu_1<\cdots<\mu_p\leqslant D,\ 0\leqslant p\leqslant D},\\
 \gamma^{\mu_1\cdots\mu_p}
  :=\left\{\begin{array}{ll}
    \DS 1 & (p=0),\\
    \DS\frac{1}{p!}\sum_{\sigma\in\MF{S}_p}\sgn(\sigma)
    \gamma^{\mu_{\sigma(1)}}\cdots\gamma^{\mu_{\sigma(p)}} & (p>0),
    \end{array}\right.
\end{gather}
where $\MF{S}_p$ denotes the symmetric group of degree $p$.
As a consequence, it follows that
$\dim\MC{C}(t,s)=2^D=2^{D/2}\times 2^{D/2}$ and
$\dim V=2^{D/2}$.

That the elements in \eRef{basis} are linearly independent can be shown
by using the orthogonality relations
\begin{equation}
\label{orthonormal}
 \tr \gamma^{\mu_1\cdots\mu_p}\gamma^{\nu_1\cdots\nu_q}
 =(-1)^{p(p-1)/2}\dim V\delta_{pq}
 \eta_{\mu_1\cdots\mu_p,\nu_1\cdots\nu_p}
\end{equation}
where
\begin{equation}
 \eta_{\mu_1\cdots\mu_p,\nu_1\cdots\nu_p}
  :=\sum_{\sigma\MF{S}_p}\sgn(\sigma)
     \eta_{\mu_1\nu_{\sigma(1)}}\cdots\eta_{\mu_p\nu_{\sigma(p)}},
\end{equation}
which is essentially a direct consequence of the fact
\begin{equation}
 \tr\gamma^{\mu_1\cdots\mu_p}=0,\quad
 (1\leqslant\mu_1<\cdots<\mu_p\leqslant D,\ 1\leqslant p\leqslant D).
\end{equation}
On the other hand, the completeness of the elements in \eRef{basis}
can be proved by the fact that the Clifford algebra $\MC{C}(t,s)$
is generated by the elements $\{\gamma^1,\ \cdots,\ \gamma^D\}$.
Note that by using the orthogonality \eRef{orthonormal}
the completeness relation can be given as the identity
\begin{equation}
\label{completeness-Cliff}
 \gamma=\sum_{p=0}^{D}\frac{1}{p!}(-1)^{p(p-1)/2}\frac{1}{\dim V}
        \gamma^{\mu_1\cdots\mu_p}
        \tr\gamma_{\mu_1\cdots\mu_p}\gamma,\quad
 \text{for}\Any\gamma\in\MC{C}(t,s).
\end{equation}

For later convenience, we define a special element
(note here that $D=t+s$ is even and that so is $t-s$)
\begin{equation}
 \Gamma^5:=i^{(t-s)/2}(-1)^s\gamma^{1}\cdots\gamma^{D},
\end{equation}
which satisfies that
\begin{equation}
 \{\Gamma^5,\gamma^\mu\}=0,\quad(\Gamma^5)^2=1.
\end{equation}
We can also show that
\begin{equation}
 \gamma^{\mu_1\cdots\mu_p}\Gamma^5
  =\frac{1}{(D-p)!}
    i^{(t-s)/2}(-1)^{p(p-1)/2}\VE^{\mu_1\cdots\mu_p}{}_{\mu_{p+1}\cdots\mu_D}
          \gamma^{\mu_{p+1}\cdots\mu_D},
\end{equation}
where the totally antisymmetric tensor $\VE^{\mu_1\cdots\mu_D}$
is defined by %in \eRef{VE-upper}.
\begin{equation}
 \VE^{\mu_1\cdots\mu_D}=\frac{1}{|\det g|}\epsilon^{\mu_1\cdots\mu_D},\qquad
 \epsilon^{1\cdots D}=1.
\end{equation}

\subsubsection{Conjugations}
We then consider the complex conjugation of the representation
$(\rho,V)$. Eq.\ (\ref{Clifford-alg}) implies that
\begin{equation}
 \{-(\gamma^\mu)^*,-(\gamma^\nu)^*\}=2\eta^{\mu\nu},
\end{equation}
i.e.\ the complex conjugate $(-\rho^*,V)$ of the representation could also
generate a representation of Clifford algebra on the same space $V$.
The universality of the Clifford algebra%
\footnote{Clifford algebra is uniquely determined,
up to an isomorphic algebra, by the fundamental
anticommutation relations. This fact is called the universality of
Clifford algebra, and is a consequence of the universality of
tensor algebra.}
then requires the existence of a unitary transformation $B$ s.t.\footnote{%
Since
$\eta\gamma^\mu=B^{-1}(\gamma^\mu)^\ast B=B^\ast(\gamma^\mu)^\ast(B^{-1})^\ast$
the uniqueness of $B$ requires $B^\ast B=\VE$. Also it should be
unitary, which can be similarly seen by the uniqueness of $B$
and by the (anti-) Hermiticity
of $\gamma^\mu$ shown in the following.
All these analysis can be similarly applied to the case for $C$ below.}
\begin{equation}
 B\gamma^\mu B^{-1}=\eta(\gamma^\mu)^*,\qquad
 B^\dagger B=1,\qquad B^* B=\VE,\qquad\eta=\pm 1,\quad \VE=\pm 1.
\end{equation}
Similarly, since
\begin{equation}
 \{(-\gamma^\mu)^{T},(-\gamma^\nu)^{T}\}=2\eta^{\mu\nu},
\end{equation}
the contragradient representation $(-\rho^{T},V)$ should
be mapped by a unitary transformation $C$ s.t.\ 
\begin{equation}
 C\gamma^\mu C^{-1}=\eta'(\gamma^\mu)^T,\qquad
 C^\dagger C=1,\qquad C^T=\VE'C,\qquad\eta'=\pm 1,\quad\VE'=\pm 1.
\end{equation}
We could also write
\begin{equation}
 C^* B\gamma^\mu B^{-1}C^{*\,-1}=\eta(C\gamma^\mu C^{-1})^*
                                =\eta\eta'(\gamma^\mu)^\dagger,
\end{equation}
i.e.\ 
\begin{equation}
\label{gamma0}
 \Gamma^0(\gamma^\mu)^\dagger(\Gamma^0)^{-1}=\kappa\gamma^\mu,\qquad
 (\Gamma^0)^\dagger \Gamma^0 =1,\qquad
  \Gamma^0:=\VE(-1)^{t(t-1)/2}B^{-1}C^{*\,-1},\quad
  \kappa:=\eta\eta',
\end{equation}
(The sign $\VE(-1)^{t(t-1)/2}$ which appears in the definition of
$\Gamma^0$ is just for the later convenience.)
which implies that
$(\gamma^\mu)^\dagger=\kappa'_{(\mu)}\gamma^\mu,\ \kappa'_{(\mu)}=\pm 1$.
This sign $\kappa'_{(\mu)}$ can not be chosen arbitrarily; for,
\begin{equation}
  \left\{\begin{array}{cl}
           (\gamma^{\MR{t}})^2=+\bs{1},\\
           (\gamma^{\MR{s}})^2=-\bs{1},
  \end{array}\right.\quad\text{i.e.}\quad
  \left\{\begin{array}{cl}
        (\gamma^{\MR{t}})^\dagger(\gamma^{\MR{t}})=+\kappa'_{\MR{t}}\bs{1},\\
        (\gamma^{\MR{s}})^\dagger(\gamma^{\MR{s}})=-\kappa'_{\MR{s}}\bs{1},
  \end{array}\right.
\end{equation}
and, for any matrix $M$, $M^\dagger M$ is positive semidefinite,
so $\kappa'_{\MR{t}}=+1$ and $\kappa'_{\MR{s}}=-1$, i.e.,\ 
\begin{equation}
  \left\{\begin{array}{cl}
        (\gamma^{\MR{t}})^\dagger=+(\gamma^{\MR{t}}),\\
        (\gamma^{\MR{s}})^\dagger=-(\gamma^{\MR{s}}).
  \end{array}\right.
\end{equation}
This leads to
\begin{equation}
 \left\{\begin{aligned}
  \Gamma^0 &= \gamma^1\cdots\gamma^t,\\
  \kappa   &= (-1)^{t+1},
 \end{aligned}\right.\qquad\text{or}\qquad
 \left\{\begin{aligned}
  \Gamma^0 &= \gamma^{t+1}\cdots\gamma^D,\\
  \kappa   &= (-1)^{s},
 \end{aligned}\right.
\end{equation}
and also
\begin{equation}
 \left\{\begin{aligned}
 \eta' &= (-1)^{t+1}\eta,\\
 \VE'  &= \VE\eta^t (-1)^{t(t-1)/2},
 \end{aligned}\right.\qquad\text{or}\qquad
 \left\{\begin{aligned}
 \eta' &= (-1)^{s}\eta,\\
 \VE'  &= \VE\eta^s (-1)^{s(s-1)/2}.
 \end{aligned}\right.
\end{equation}
These two choices are equivalent, so that,
in what follows, we choose
the former representation for $\Gamma^0$.
Further, $\VE$ and $\eta$ can be related as follows.
Notice that
\begin{equation}
 (C\gamma^{\mu_1\cdots\mu_p})^T
  =(-1)^{p(t+1)+p(p-1)/2+t(t-1)/2}\eta^{t+p}\VE
    (C\gamma^{\mu_1\cdots\mu_p}),
\end{equation}
hence $C\gamma^{\mu_1\cdots\mu_p}$ are either symmetric or
antisymmetric. The total number of antisymmetric elements is
\begin{equation}
 \#(\MR{AS})=\frac{1}{2}2^{D/2}(2^{D/2}-1),
 \label{numAS-1}
\end{equation}
which can be also calculated as
\begin{equation}
 \#(\MR{AS})=\sum_p\colvec{D \\ p}
 \label{numAS-2}
\end{equation}
where the summation is over the numbers $p$ such that
\begin{equation}
 (-1)^{p(t+1)+p(p-1)/2+t(t-1)/2}\eta^{t+p}\VE=-1,\quad\text{i.e.}\quad
 (-1)^{p(t+1)+p(p-1)/2}\eta^{p}=-(-1)^{t(t-1)/2}\eta^t\VE.
\end{equation}
Equating eqs.~\Ref{numAS-1} and \Ref{numAS-2}, we obtain
Letting $f(p)=(-1)^{p(p-1)/2+p(t+1)}\eta^p$, we note
\begin{equation*}
 \begin{array}[t]{cc}
  p \pmod{4}  &  f(p) \\ \hline
  0           &   1   \\
  1           & (-1)^{t+1}\eta \\
  2           &  -1   \\
  3           & -(-1)^{t+1}\eta
 \end{array}\qquad
 \begin{array}[t]{cc}
  (-1)^{t+1}\eta  &  f(p) \\ \hline
   1              & 1,1,-1,-1 \\
  -1              & 1,-1,-1,1
 \end{array}
\end{equation*}
and, labeling the cases by $((-1)^{t+1}\eta,-(-1)^{t(t-1)}\eta^t\VE)$,
we can compute $\#(\MR{AS})$ as follows
\begin{computation}
$(+,+)$
\begin{align*}
 \#(\MR{AS})
 &=\sum_{p=0}^D\colvec{D \\ p}\frac{1}{2}\left(
    1+\Re\sqrt{2}e^{-i\frac{\pi}{4}+i\frac{n\pi}{2}}\right)\\
 &=\frac{1}{2}2^{D/2}\left(
   2^{D/2}-\sqrt{2}\cos\frac{\pi}{4}(D+3)\right).
\end{align*}
\end{computation}

\begin{computation}
$(+,-)$
\begin{align*}
 \#(\MR{AS})
 &=\sum_{p=0}^D\colvec{D \\ p}\frac{1}{2}\left(
    1-\Re\sqrt{2}e^{-i\frac{\pi}{4}+i\frac{n\pi}{2}}\right)\\
 &=\frac{1}{2}2^{D/2}\left(
   2^{D/2}-(-1)\sqrt{2}\cos\frac{\pi}{4}(D+3)\right).
\end{align*}
\end{computation}

\begin{computation}
$(-,+)$
\begin{align*}
 \#(\MR{AS})
 &=\sum_{p=0}^D\colvec{D \\ p}\frac{1}{2}\left(
    1+\Re\sqrt{2}e^{i\frac{\pi}{4}+i\frac{n\pi}{2}}\right)\\
 &=\frac{1}{2}2^{D/2}\left(
   2^{D/2}-\sqrt{2}\cos\frac{\pi}{4}(-D+3)\right).
\end{align*}
\end{computation}

\begin{computation}
$(-,-)$
\begin{align*}
 \#(\MR{AS})
 &=\sum_{p=0}^D\colvec{D \\ p}\frac{1}{2}\left(
    1-\Re\sqrt{2}e^{i\frac{\pi}{4}+i\frac{n\pi}{2}}\right)\\
 &=\frac{1}{2}2^{D/2}\left(
   2^{D/2}-(-1)\sqrt{2}\cos\frac{\pi}{4}(-D+3)\right).
\end{align*}
\end{computation}
Thus we find that
\begin{equation}
 \#(\MR{AS})
 =\frac{1}{2}2^{D/2}\left(
   2^{D/2}-\left(-(-1)^{t(t-1)/2}\eta^t\VE\right)\sqrt{2}
   \cos\frac{\pi}{4}\left(\eta(-1)^{t+1}D+3\right)\right).
\end{equation}
Equating this to $\DS\frac{1}{2}2^{D/2}(2^{D/2}-1)$, we obtain
\begin{equation}
 \VE
  =-\sqrt{2}\eta^t(-1)^{t(t-1)/2}\cos\frac{1}{4}\pi\left(
     \eta(-1)^{t+1}D+3\right)\\
  =\cos\frac{1}{4}\pi(s-t)-\eta\sin\frac{1}{4}\pi(s-t).
\end{equation}

Conjugations of $\Gamma^5$ are computed as
\begin{gather}
 B\Gamma^5 B^{-1}=(-1)^{(t-s)/2}(\Gamma^5)^\ast,\quad
 C\Gamma^5 C^{-1}=(-1)^{D/2}(\Gamma^5)^T,\\
 \Gamma^0(\Gamma^5)^\dagger(\Gamma^0)^{-1}
                 =(-1)^t\Gamma^5,\quad
 (\Gamma^5)^\dagger=\Gamma^5.
\end{gather}
Note also that
\begin{equation}
 B\Gamma^0 B^{-1}=\eta^t(\Gamma^0)^\ast,\quad
 C\Gamma^0 C^{-1}=\eta^t(-1)^{t(t-1)/2}(\Gamma^0)^T,\quad
 (\Gamma^0)^2=(-1)^{t(t-1)/2}\bs{1}.
\end{equation}

\subsubsection{The $\GR{SO}(t,s)$ Subalgebra, $\GR{Spin}(t,s)$}
Let
\begin{equation}
 \Sigma^{\mu\nu}:=\frac{i}{4}[\gamma^{\mu},\gamma^{\nu}]
                \equiv\frac{i}{2}\gamma^{\mu\nu}.
\end{equation}
Then it can be shown that $\{\Sigma^{\mu\nu}\}_{1\leqslant\mu<\nu\leqslant D}$
obeys the $\AL{so}(t,s)$ commutation relations
\begin{equation}
 [\Sigma^{\mu\nu},\Sigma^{\rho\sigma}]
 =i(\eta^{\nu\rho}\Sigma^{\mu\sigma}-\eta^{\nu\sigma}\Sigma^{\mu\rho}
    -\eta^{\mu\rho}\Sigma^{\nu\sigma}+\eta^{\mu\sigma}\Sigma^{\nu\rho}),
\end{equation}
and thus generates the $\AL{so}(t,s)$ subalgebra in $\MC{C}(t,s)$.
In fact, this subalgebra generates a group,
\begin{equation}
 \GR{Spin}(t,s):=
  \left\{\gamma\in\MC{C}(t,s)|
     \gamma=\exp\left(-i\frac{1}{2}\omega_{\mu\nu}\Sigma^{\mu\nu}\right)
  \right\},
\end{equation}
with
\begin{equation}
 \gamma\gamma^\mu\gamma^{-1}
  =\exp(\omega)^\mu{}_\nu \gamma^\nu.
\end{equation}
To show that $\GR{Spin}(t,s)$ is actually the double cover of $\GR{SO}(t,s)$,
let us consider the adjoint
representation on the vector space
\begin{equation}
 W=\langle\{\gamma^1,\ \cdots,\ \gamma^D\}\rangle,
\end{equation}
as in
\begin{equation}
 \Gamma\in\GR{Spin}(t,s),\ \gamma\in W,\qquad
 \Ad(\Gamma)(\gamma)=\Gamma\gamma\Gamma^{-1}.
\end{equation}
Let $\Ad(\Gamma)=1$. Then decompose $\Gamma=\Gamma_0+\gamma^p\Gamma'$,
where $\gamma_0$ and $\gamma'$ do not contain $\Gamma^p$.
Note here that $\Gamma$ contains terms with only even number of
$\gamma^p$'s by definition. We thus find
\begin{equation}
 \gamma^p=\Gamma\gamma^p\Gamma^{-1},\quad\text{i.e.}\quad
 \gamma^p(\Gamma_0+\gamma^p\Gamma')
   =(\Gamma_0+\gamma^p\Gamma')\gamma^p,
\end{equation}
so that
\begin{equation}
 (\gamma^p)^2\Gamma'=\gamma^p\Gamma'\gamma^p=-(\gamma^p)^2\Gamma',\quad
 \text{i.s.}\quad
 \Gamma'=0.
\end{equation}
Thus $\Gamma$ does not contain either of $\gamma^p$, so that
$\Gamma\in\mathbb{C}$, hence $\Gamma=\pm 1$.

Since (for even $D$)
\begin{equation}
 [\Gamma^5,\Sigma^{\mu\nu}]=0,
\end{equation}
the representation of $\GR{Spin}(t,s)$ on $V$
is reducible, although the representation of $\MC{C}(t,s)$ as a whole
is irreducible as noted before.
This reducible representation can be decomposed into the two irreducible
representations on the two eigenspaces of $\Gamma^5$ with eigenvalues,
$\pm 1$,  called the chirality of the representations.
Projection operators onto these two eigenspace are given as
\begin{equation}
 P^{\pm}=\frac{1\pm \Gamma^5}{2},
\end{equation}
where
\begin{equation}
 P^\pm P^\pm=P^\pm,\quad
 P^\pm P^\mp=0,\quad
 (P^\pm)^\dagger=P^\pm,\quad
 P^+ +P^- = 1.
\end{equation}

Note that
\begin{equation}
 [\Sigma^{2a-1,2a},\Sigma^{2b-1,2b}]=0.
\end{equation}
Hence $\Sigma^{2a-1,2a}$ can form a Cartan subalgebra
and can be simultaneously diagonalized.
Let
\begin{equation}
 S^a:=(-i)^{\delta_{a,(t+1)/2}+\theta(a-(t+2)/2)}\Sigma^{2a-1,2a}
     =\Gamma^{a+}\Gamma^{a-}-\frac{1}{2}.
\end{equation}
Then we find that $S^{a}$ takes half-integer values and
can be identified as the spin operators.

\subsection{Spinors}
\subsubsection{$\GR{Spin}(t,s)$ Spinors}
We have seen that the Clifford algebra $\MC{C}(t,s)$ is irreducibly
represented on $V=\mathbb{C}^{2^{D/2}}$.
This spinor representation also represents $\GR{Spin}(t,s)$,
the double cover of $\GR{SO}(t,s)$, which is not irreducible,
and decomposes into two irreducible representations on the eigenspaces
with chirality $\pm 1$. We call these
two irreducible representations of $\GR{Spin}(t,s)$ the Weyl
representations, and elements of the corresponding representation
spaces Weyl spinors. Weyl spinors are thus given from
a general $2^{D/2}$-component spinor $\psi\in V$ as
\begin{equation}
 \psi^\pm =P^\pm \psi,\qquad
 P^\pm\psi^\pm=\pm\psi^\pm,
\end{equation}
and can be considered as $2^{D/2-1}$-component objects.
On the other hand, a $2^{D/2}$-component spinor $\psi\in V$,
sometimes called a Dirac spinor, can be constructed from two Weyl spinors
\begin{gather}
 \psi = (P^+ +P^-)\psi =\psi^+ +\psi^-, \\
 \bs{2^{D/2}}= \bs{2^{D/2-1}}\oplus \bs{2^{D/2-1}}.
\end{gather}

We choose to label a $2^{D/2}$-component spinor as $\psi_\alpha$,
so that $\gamma\in\MC{C}(t,s)$, including $\Gamma^0$ and
$\Gamma^5$, is labeled as
\begin{equation}
 \gamma_\alpha{}^\beta,\quad
 (\gamma^{-1})_\alpha{}^\beta,\quad
 (\gamma^\dagger)_\alpha{}^\beta,\quad
 (\gamma^T)^\alpha{}_\beta=\gamma_\beta{}^\alpha,\quad
 (\gamma^\ast)^\alpha{}_\beta=(\gamma_\alpha{}^\beta)^\ast,
\end{equation}
and conjugation matrices $\MF{C}=C,\ B$ are labeled as
\begin{equation}
  \MF{C}^{\alpha\beta},\quad
 (\MF{C}^{-1})_{\alpha\beta},\quad
 (\MF{C}^T)^{\alpha\beta}=\MF{C}^{\beta\alpha},\quad
 (\MF{C}^\ast)_{\alpha\beta}=(\MF{C}^{\alpha\beta})^\ast,\quad
 (\MF{C}^\dagger)_{\alpha\beta}=(\MF{C}^{\beta\alpha})^\ast.
\end{equation}
Note that the indices are consistently set in
\begin{equation}
 (\Gamma^0)_\alpha{}^\beta
  =(B^{-1})_{\alpha\gamma}(C^{\ast -1})^{\gamma\beta}.
\end{equation}
Then a spinor transforms under $\GR{Spin}(t,s)$ as
\begin{equation}
 \psi'_\alpha
 =\exp\left(-\frac{i}{2}\omega_{\mu\nu}\Sigma^{\mu\nu}\right){}_\alpha{}^\beta
  \psi_\beta,\quad
 \text{or}\quad
 \delta\psi_\alpha
 =\frac{1}{4}\omega_{\mu\nu}(\gamma^{\mu\nu})_\alpha{}^\beta\psi_\beta.
\end{equation}
Upper indices are thus label the contragradient representations, as in
\begin{equation}
 \psi'^\alpha
 =\exp\left(+\frac{i}{2}\omega_{\mu\nu}\Sigma^{\mu\nu}\right){}_\beta{}^\alpha
  \psi^\beta,\quad
 \text{or}\quad
 \delta\psi^\alpha
 =-\frac{1}{4}\omega_{\mu\nu}(\gamma^{\mu\nu})_\beta{}^\alpha\psi^\beta.
\end{equation}

\subsubsection{Conjugations}
Conjugations discussed above are used to define some
conjugates of spinors.

First, we define the Dirac conjugate of $\psi\in V$ as
\begin{equation}
 \OLL{\psi}^\alpha
  :=\sum_\beta (\psi_\beta)^\dagger((\Gamma^0)^{-1})_\beta{}^\alpha.
\end{equation}
Note here that
\begin{equation}
 \Gamma^0 (\Sigma^{\mu\nu})^\dagger (\Gamma^0)^{-1}
 = \Sigma^{\mu\nu}.
\end{equation}
The conjugate spinor%
\footnote{In what follows we will omit the summation symbol for the
Dirac conjugation.}
does transform contragradiently as indicated by the upper indices:
\begin{equation}
 \OLL{\psi}'^\alpha
  =(\psi_\gamma)^\dagger
   \exp\left(+\frac{i}{2}\omega^{\mu\nu}(\Sigma_{\mu\nu})^\dagger\right){}%
     _\gamma{}^\beta(\Gamma^0)^{-1}_\beta{}^\alpha
  =\OLL{\psi}^\beta
   \exp\left(+\frac{i}{2}\omega^{\mu\nu}\Sigma_{\mu\nu}\right)_\beta{}^\alpha.
\end{equation}
For an upper-index spinor $\psi^\alpha$, the Dirac conjugate can be
defined as
\begin{equation}
 \OLL{\psi}_\alpha:=(\Gamma^0)^{-1}{}_\alpha{}^\beta(\psi^\beta)^\dagger,
\end{equation}
so that successive Dirac conjugations reduces to the identity operation
\begin{equation}
 \OLL{\OLL{\psi}}
 =(\Gamma^0)^{-1}(\OLL{\psi})^\dagger
 =(\Gamma^0)^{-1}\Gamma^0\psi=\psi.
\end{equation}

The next one is the charge conjugation. We define it for a lower-index
spinor,
\begin{equation}
 \tilde{\psi}^\alpha
   :=(\psi^T)_\beta (C^T)^{\beta\alpha}
    =C^{\alpha\beta}\psi_\beta,
\end{equation}
and for an upper-index spinor,
\begin{equation}
 \tilde{\psi}_\alpha
   :=(C^{-1})_{\alpha\beta}(\psi^T)^\beta.
\end{equation}
Again we have defined these conjugations so that $\tilde{\tilde{\psi}}=\psi$.
Since
\begin{equation}
 C\Sigma^{\mu\nu}C^{-1}=-(\Sigma^{\mu\nu})^T,
\end{equation}
these conjugates transform correctly as indicated by indices;
for instance,
\begin{equation}
 \tilde{\psi}'_\alpha
 =(C^{-1})_{\alpha\beta}
  \exp\left(+\frac{i}{2}\omega^{\mu\nu}\Sigma_{\mu\nu}\right)^T{}%
   ^\beta{}_\gamma(\psi^T)^\gamma
 =\exp\left(-\frac{i}{2}\omega^{\mu\nu}\Sigma_{\mu\nu}\right){}%
   _\alpha{}^\beta\tilde{\psi}_\beta.
\end{equation}
Conventionally, the term charge conjugation may only be used
for the operation to define
\begin{equation}
 (\psi_c)_\alpha
 :=(C^{-1})_{\alpha\beta}(\OLL{\psi}^T)^\beta
  =\widetilde{\OLL{\psi}}_\alpha,\qquad\text{with}\qquad
 \psi_{cc}=\VE\psi.
\end{equation}
Here we also define the charge conjugate with upper index
\begin{equation}
 (\psi_c)^\alpha
 :=C^{\alpha\beta}\OLL{\psi}_\beta
  =\widetilde{\OLL{\psi}}^\alpha,\qquad\text{with}\qquad
 \psi_{cc}=\VE\psi. 
\end{equation}

The last one is called the $B$-conjugation, defined with the matrix $B$
as%
\footnote{Remark here we have used somewhat
misleading labeling for $\psi^\ast$'s;
$(\psi^\ast)^\alpha=(\psi_\alpha)^\ast$ does not necessarily transform
as an upper-index spinor.}
\begin{gather}
 (\psi_b)_\alpha
  :=(B^{-1})_{\alpha\beta}(\psi^\ast)^\beta,\qquad
 (\psi^\ast)^\alpha:=(\psi_\alpha)^\ast,\\
 (\psi_b)^\alpha
  :=\eta^t B^{\alpha\beta}(\psi^\ast)_\beta,\qquad
 (\psi^\ast)_\alpha:=(\psi^\alpha)^\ast,
\end{gather}
with $\psi_{bb}=\VE\psi$. Notice that the $B$-conjugation is actually
the same operation as the charge conjugation%
\footnote{This is due to the fact that there are essentially
two kinds of conjugations in a complex vector space, one is the ordinary
contragradient, and the other is the complex conjugation. },
i.e.,\ $\psi_b=\psi_c$.

\subsubsection{Majorana Spinors}
Generally a $2^{D/2}$-component Dirac spinor is, though irreducible
w.r.t.\ the whole representation of the Clifford algebra,
reducible w.r.t.\ the representation of the $\GR{Spin}(t,s)$ subalgebra.
Thus, as a representation of $\GR{Spin}(t,s)$,
it is appropriate to use the Weyl representations with
$2^{D/2-1}$-component Weyl spinors, which is
irreducible w.r.t.\ $\GR{Spin}(t,s)$.
In such cases, it is sometimes more convenient to treats a pair of
a Weyl spinor and its partner with opposite chirality
as one $2^{D/2}$-component spinor, which is called a Majorana spinor.
Note that d.o.f.\ of the two Weyl spinors, one of which is a partner to
the other, is $2\times 2^{D/2-1}=2^{D/2}$,
while that of a general complex $2^{D/2}$-component spinor is
$2\times 2^{D/2}$. Thus we need to impose a reality condition on
a Majorana spinor to reduce the d.o.f. Such conditions are called
the Majorana conditions.
Of course, Majorana conditions have to be $\GR{Spin}(t,s)$-covariant.

Let $(\psi_{i\alpha})\ (i=1,\ \cdots,\ N)$ be a multiplet of
$N$ Majorana spinors with an internal symmetry $G$ labeled by the indices
$i$. We adopt the following Majorana condition
on such multiplets:
\begin{equation}
\label{Majorana}
 \psi_{i\alpha}=M_{ij}(\psi_c)^{j}{}_\alpha,\qquad
  M_{ij}M^{ik}=\delta_j{}^k,
\end{equation}
where
\begin{equation}
  (\psi_c)^i{}_\alpha
  =(C^{-1})_{\alpha\beta}(\OLL{\psi}^T)^{i\beta}
%   \left((\psi_{j\gamma})^\dagger(\Gamma^0)^{-1}_\gamma{}^\beta\right)^T
  =(C^{-1})_{\alpha\beta}((\Gamma^0)^{-1\ T})^\beta{}_\gamma
   (\psi^\ast)^{j\gamma},\quad
  (\psi^\ast)^{i\alpha}:=(\psi_{i\alpha}),
\end{equation}
and $M_{ij}$ is a $G$ (-invariant) metric. Notice here that
the charge conjugation matrix $C$ is a $\GR{Spin}(t,s)$-invariant metric,
and the condition \Ref{Majorana} is $\GR{Spin}(t,s)$ and
$\GR{G}$-covariant.
Since the condition \Ref{Majorana} implies that
\begin{equation}
 \psi_{i\alpha}=\VE M_{ik}(M^\ast)^{kj}\psi_{j\alpha},
\end{equation}
so that 
\begin{equation}
\label{Majorana-2}
 M_{ik}(M^\ast)^{kj}=\VE,\qquad\text{i.e.,}\qquad
 M^\dagger=\VE (M^{-1})^T.
\end{equation}
Thus the Majorana condition \Ref{Majorana} can be possibly imposed
only if there exists a metric $M_{ij}$ which satisfy the condition
\Ref{Majorana-2}. Since such a metric does not necessarily exist for
any internal symmetry group %, labeled by the indices $i$,
the Majorana condition restricts the possible internal symmetries.
This fact is seen more specifically as follows. For simplicity,
we assume a unitary representation for the internal symmetry metric,
namely, $M^\dagger=M^{-1}$. Then the condition \Ref{Majorana-2} is written
as $M=\VE M^T$.
\begin{computation}
If $\VE=+1$, the metric has to be symmetric: $M=M^T$.
If further $M$ is supposed to be invariant w.r.t.\ the internal
symmetry, we may take $G$ to be (a subgroup of) $\GR{O}(N)$.
Majorana spinors which obey the Majorana condition with $\VE=+1$
and $G\cong\GR{O}(N)$ are thus called $\GR{O}(N)$-Majorana spinors.

Especially, if $N=1$, i.e.,\ if we consider singlet Majorana spinors with
no internal symmetry, $M=1$ so that only the Majorana condition
with $\VE=+1$ is possible. In other words, if $\VE=-1$,
we can not consider singlet Majorana spinor.
\end{computation}

\begin{computation}
If $\VE=-1$, the metric has to be antisymmetric: $M+M^T=0$.
If further $M$ is supposed to be invariant w.r.t.\ the internal
symmetry, we may take $\GR{G}$ to be (a subgroup of)%
\footnote{As a nontrivial candidate, we can take, for example,
$\GR{G}\cong\GR{Spin}(4)$.}
$\GR{USp}(2n)$ with $N=2n$.
Majorana spinors which obey the Majorana condition with $\VE=-1$
and $\GR{G}\cong\GR{USp}(2n)$ are thus called $\GR{USp}(2n)$-Majorana spinors.
For instance, if $N=2$, we can consider as the internal symmetry
$\GR{USp}(2)\cong\GR{SU}(2)$ and $\GR{SU}(2)$-Majorana condition
with the invariant metric $C_{ij}=(\tau^2)_{ij}$.
\end{computation}

In some cases, Weyl spinors could independently become Majorana
(including $G$-Majorana) spinors, and, if so,
are called Majorana-Weyl spinors%
\footnote{Thus, each Weyl component (or Weyl decomposition)
of a Majorana-Weyl spinor is itself real, or self-conjugate, while
Weyl components of a general Majorana spinor are conjugate
to each other. In this sense, the Weyl decomposition
of a Majorana-Weyl spinor should be merely denoted as in
$\bs{2^{D/2}}\RA\bs{2^{D/2-1}}+\bs{2'^{D/2-1}}$, whereas
that of a non-Weyl-Majorana spinor can be as in
$\bs{2^{D/2}}\RA\bs{2^{D/2-1}}+\bs{\bar{2}^{D/2-1}}$.}.
The condition is that
\begin{equation}
 (\psi^\pm)_\MR{c}=\psi^\pm,
\end{equation}
which is possible only if%
\begin{equation}
 P^{\pm(-1)^t}=CP^{\pm}C^{-1}=P^{\pm(-1)^{D/2}},\qquad\text{i.e.}\qquad
 P^{\pm\sigma}=\id,\qquad \sigma:=(-1)^{(s-t)/2},
\end{equation}
so that $\sigma=1$, or, $s-t\equiv 0\pmod{4}$.

\section{Solutions of the $\GR{USp}(4)$ SYM}
\label{sec-comp}
In this section, we follow the whole computations to solve the system
of constraints \Ref{BeginOfConstraints}--\Ref{EndOfConstraints},
i.e.\ 
\begin{xalignat*}{2}
 \{\nabla_{i\alpha},\nabla_{j\beta}\}
  & =i\Omega_{ij}C_{\alpha\beta}\nabla_z
     -iC_{\alpha\beta}W_{ij},
 &
 \{\OLL{\nabla}_{i\Dalpha},\OLL{\nabla}_{j\Dbeta}\}
  & =i\Omega_{ij}C_{\Dalpha\Dbeta}\nabla_z
     +iC_{\Dalpha\Dbeta}W_{ij},\\
 \{\nabla_{i\alpha},\OLL{\nabla}_{j\Dbeta}\}
  & =i\Omega_{ij}(\sigma^\mu)_{\alpha\Dbeta}\nabla_\mu,
 &
  & \\
 [\nabla_{i\alpha},\nabla_\mu]
  &=-iF_{i\alpha\mu},
 &
 [\OLL{\nabla}_{i\Dalpha},\nabla_\mu]
  &=+i\OLL{F}_{i\Dalpha\mu},\\
 [\nabla_{i\alpha},\nabla_z]
  &=-iG_{i\alpha},
 &
 [\OLL{\nabla}_{i\Dalpha},\nabla_z]
  &=+i\OLL{G}_{i\Dalpha},\\
 [\nabla_\mu,\nabla_z]
  &=-ig_\mu,
 &
 [\nabla_\mu,\nabla_\nu]
  &=-iF_{\mu\nu},\\
 \Omega^{ij}W_{ij}
  &=0,
 &
 (W_{ij})^*
  &=W^{ij},
\end{xalignat*}
for the $\GR{USp}(4)$ model. Specifically, these constraints are further
restricted by various Bianchi identities for the superconnections
$\nabla_I$, and we treat all such relations to compute various
derivatives of the superfields in our system. We will find that
any higher derivatives of a superfield can be expressed by some other
superfields and some lower derivatives of the superfields. Thus
we obtain necessary and sufficient superfields and/or their
derivatives to express all the other superfields and their derivatives,
which corresponds to the independent degrees of freedom of the system.
Since then all derivatives of all superfield are obtained,
we can expand those superfields componentwisely. Thus
we completely determine the explicit forms of the superfields,
which we say we solve the constraints.

Let us now carry out the program explicitly.
Most of the computations are based on Bianchi identities.
For one simply Hermitian conjugate to the other only the result
will be listed.

\allowdisplaybreaks
\paragraph{Three Fermionic Derivatives}
First we compute relations derived from various combinations
of three fermionic derivatives. In order to show the typical
manner of the computations, we list them in the full detail.
\begin{computation}
\label{G}
Computation of $G_{i\alpha}$
\begin{align*}
 G_{i\alpha}
 &=i[\nabla_{i\alpha},\nabla_z]
  =i[\nabla_{i\alpha},
     -\frac{i}{8}C^{\beta\gamma}\Omega^{jk}
       \{\nabla_{j\beta},\nabla_{k\gamma}\}]\\
 &=-\frac{1}{8}C^{\beta\gamma}\Omega^{jk}
    \Bigl([\nabla_{j\beta},\{\nabla_{k\gamma},\nabla_{i\alpha}\}]
          +[\nabla_{k\gamma},\{\nabla_{i\alpha},\nabla_{j\beta}\}]\Bigr)\\
 &=-\frac{1}{8}C^{\beta\gamma}\Omega^{jk}
    \Bigl([\nabla_{j\beta},i\Omega_{ki}C_{\gamma\alpha}\nabla_z
                           -iC_{\gamma\alpha}W_{ki}]
          +[\nabla_{k\gamma},i\Omega_{ij}C_{\alpha\beta}\nabla_z
                            -iC_{\alpha\beta}W_{ij}]\Bigr)\\
 &=-\frac{i}{4}\Bigl(
    [\nabla_{i\alpha},\nabla_z]+[\nabla_{j\alpha},W^j{}_i]\Bigr)
  =-\frac{i}{4}\Bigl(
     -iG_{i\alpha}+[\nabla_{j\alpha},W^j{}_i]\Bigr),
\end{align*}
so that
\begin{equation}
 [\nabla_{j\alpha},W^j{}_i]=5iG_{i\alpha}.
\end{equation}
\end{computation}

\begin{computation}[p]
Computation of $\OLL{G}_{i\Dalpha}$
\begin{equation}
 [\OLL{\nabla}_{j\Dalpha},W^j{}_i]=5i\OLL{G}_{i\Dalpha},
\end{equation}
consistent to \Ref{G}.
\end{computation}

\begin{computation}
\label{G-2}
Another computation of $G_{i\alpha}$
\begin{align*}
 G_{i\alpha}
 &=i[\nabla_{i\alpha},\nabla_z]
  =i[\nabla_{i\alpha},-\frac{i}{8}C^{\Dbeta\Dgamma}\Omega^{jk}
                      \{\OLL{\nabla}_{j\Dbeta},\OLL{\nabla}_{k\Dgamma}\}]\\
 &=-\frac{1}{8}C^{\Dbeta\Dgamma}\Omega^{jk}
    \Bigl([\Bnabla_{j\Dbeta},
            i\Omega_{ik}(\sigma^\mu)_{\alpha\Dgamma}\nabla_\mu]
          +[\Bnabla_{k\Dgamma},
            i\Omega_{ij}(\sigma^\mu)_{\alpha\Dbeta}\nabla_{\mu}]\Bigr)\\
 &=\frac{i}{4}(\sigma^\mu C)_\alpha{}^\Dbeta [\Bnabla_{i\Dbeta},\nabla_\mu]
  =-\frac{1}{4}(\sigma^\mu C)_\alpha{}^\Dbeta \OLL{F}_{i\Dbeta\mu},
\end{align*}
so that
\begin{equation}
 G_{i\alpha}=-\frac{1}{4}(\sigma^\mu C)_\alpha{}^\Dbeta
                  \OLL{F}_{i\Dbeta\mu}.
\end{equation}
\end{computation}

\begin{computation}[p]
Another computation of $\OLL{G}_{i\Dalpha}$
\begin{equation}
\label{BG-F}
 \OLL{G}_{i\Dalpha}=-\frac{1}{4}(C\sigma^\mu)^\beta{}_\Dalpha
                       F_{i\beta\mu},
\end{equation}
consistent to \Ref{G-2}.
\end{computation}

\begin{computation}
\label{nablaW}
Computation of the first derivative of $W_{jk}$
\begin{align*}
 [\nabla_{i\alpha},W_{jk}]
 &=[\nabla_{i\alpha},\frac{i}{2}C^{\beta\gamma}\Bigl(
         \{\nabla_{j\beta},\nabla_{k\gamma}\}
         -iC{\beta\gamma}\Omega_{jk}\nabla_z\Bigr)]\\
 &=-\frac{i}{2}C^{\beta\gamma}\Bigl(
   [\nabla_{j\beta},i\Omega_{ki}C_{\gamma\alpha}\nabla_z
                    -iC_{\gamma\alpha}W_{ki}]
   +[\nabla_{k\gamma},i\Omega_{ij}C_{\alpha\beta}\nabla_z
                      -iC_{\alpha\beta}W_{ij}]\Bigr)\\
 &\phantom{=-}
   +\Omega_{jk}[\nabla_{i\alpha},\nabla_z]\\
 &=-\frac{1}{2}\Bigl(
   \Omega_{ki}[\nabla_{j\alpha},\nabla_z]
   +\Omega_{ij}[\nabla_{k\alpha},\nabla_z]\Bigr)
   +\Omega_{jk}[\nabla_{i\alpha},\nabla_z]\\
 &\phantom{=-}
   +\frac{1}{2}\Bigl(
   [\nabla_{j\alpha},W_{ki}]
   +[\nabla_{k\alpha},W_{ij}]\Bigr),
\end{align*}
so that
\begin{equation*}
 [\nabla_{i\alpha},W_{ji}]
  =\frac{i}{2}\Omega_{i[j}G_{k]\alpha}
   -i\Omega_{jk}G_{i\alpha}
   +\frac{1}{2}[\nabla_{[j\alpha},W_{k]i}].
\end{equation*}
\end{computation}

\begin{computation}[p]
Computation of the first derivative of $W_{jk}$
\begin{equation*}
 [\Bnabla_{i\Dalpha},W_{jk}]
 =\frac{i}{2}\Omega_{i[j}\OLL{G}_{k]\Dalpha}
  -i\Omega_{jk}\OLL{G}_{i\Dalpha}
  +\frac{1}{2}[\Bnabla_{[j\Dalpha},W_{k]i}],
\end{equation*}
consistent to \Ref{nablaW}.
\end{computation}

\begin{computation}
\label{nablaW-2}
Another computation of the first derivative of $W_{jk}$
\begin{align*}
 [\nabla_{i\alpha},W_{jk}]
 &=[\nabla_{i\alpha},-\frac{i}{2}C^{\Dbeta\Dgamma}\Bigl(
         \{\Bnabla_{j\Dbeta},\Bnabla_{k\Dgamma}\}
         -iC_{\Dbeta\Dgamma}\Omega_{jk}\nabla_z\Bigr)]\\
 &=\frac{i}{2}C^{\Dbeta\Dgamma}\Bigl(
   [\Bnabla_{j\Dbeta},i\Omega_{ik}(\sigma^\mu)_{\alpha\Dgamma}\nabla_\mu]
   +[\Bnabla_{k\Dgamma},i\Omega_{ij}(\sigma^\mu)_{\alpha\Dbeta}\nabla_\mu]
   \Bigr)\\
 &\phantom{=-}
   -\Omega_{jk}[\nabla_{i\alpha},\nabla_z]\\
 &=-\frac{i}{2}
   \Omega_{i[j}(\sigma^\mu C)_\alpha{}^\Dbeta\OLL{F}_{k]\Dbeta\mu}
   +i\Omega_{jk}G_{i\alpha},
\end{align*}
so that
\begin{equation}
 [\nabla_{i\alpha},W_{jk}]
  =2i\Omega_{i[j}G_{k]\alpha}+i\Omega_{jk}G_{i\alpha}.
\end{equation}
This solves \Ref{nablaW} consistently.
\end{computation}

\begin{computation}[p]
Another computation of the first derivative of $W_{jk}$
\begin{equation}
 [\Bnabla_{i\Dalpha},W_{jk}]
 =2i\Omega_{i[j}\OLL{G}_{k]\Dalpha}+i\Omega_{jk}\OLL{G}_{i\Dalpha},
\end{equation}
consistent to \Ref{nablaW-2}.
\end{computation}

\begin{computation}
\label{F}
Computation of $F_{i\alpha\mu}$
\begin{align*}
 F_{i\alpha\mu}
 &=i[\nabla_{i\alpha},\nabla_\mu]
  =i[\nabla_{i\alpha},
     -\frac{i}{8}\Omega^{jk}(\bar{\sigma}_\mu)^{\Dgamma\beta}
       \{\nabla_{j\beta},\Bnabla_{k\Dgamma}\}]\\
 &=-\frac{1}{8}\Omega^{jk}(\bsigma_\mu)^{\Dgamma\beta}\Bigl(
    [\nabla_{j\beta},i\Omega_{ik}(\sigma^\nu)_{\alpha\Dgamma}\nabla_\nu]
    +[\Bnabla_{k\Dgamma},i\Omega_{ij}C_{\alpha\beta}\nabla_z
                         -iC_{\alpha\beta}W_{ij}]\Bigr)\\
 &=-\frac{1}{8}(\sigma^\mu\bsigma_\mu)_\alpha{}^\beta F_{i\beta\nu}
   +\frac{3}{4}(\bsigma_\mu C)^\Dgamma{}_\alpha \OLL{G}_{i\Dgamma}\\
 &=-\frac{1}{4}F_{i\alpha\mu}
   -\frac{1}{8}(\bsigma_\mu C)^\Dgamma{}_\alpha (C\sigma^\nu)^\beta{}_\Dgamma
               F_{i\beta\nu}
   +\frac{3}{4}(\bsigma_\mu C)^\Dgamma{}_\alpha \OLL{G}_{i\Dgamma},
\end{align*}
so using \eRef{BG-F}, we obtain
\begin{equation}
 F_{i\alpha\mu}=(\bsigma_\mu C)^\Dgamma{}_\alpha \OLL{G}_{i\Dgamma}.
\end{equation}
\end{computation}

\begin{computation}[p]
Computation of $\OLL{F}_{i\Dalpha\mu}$
\begin{equation}
 \OLL{F}_{i\Dalpha\mu}
 =(C\bar{\sigma}_\mu)_\Dalpha{}^\beta G_{i\beta},
\end{equation}
consistent to \Ref{F}.
\end{computation}

\paragraph{Four Fermionic Derivatives}
We then show relations constructed by four fermionic derivatives.
Those computations tend to very complicated. Since we have seen
in the preceding paragraph the typical rules for such computations,
we now list the process more simply in the following.
\begin{computation}
Computation of the first derivative of $G_{j\beta}$
\begin{align*}
 \{\nabla_{i\alpha},G_{j\beta}\}
 &=i\{\nabla_{i\alpha},[\nabla_{j\beta},\nabla_z]\}\\
 &=-\{\nabla_{j\beta},G_{i\alpha}\}
   +[\nabla_z,\Omega_{ij}C_{\alpha\beta}\nabla_z-C_{\alpha\beta}W_{ij}],
\end{align*}
so that
\begin{equation*}
 \{\nabla_{i\alpha},G_{j\beta}\}
 +\{\nabla_{j\beta},G_{i\alpha}\}
 =-C_{\alpha\beta}[\nabla_z,W_{ij}].
\end{equation*}
\end{computation}

\begin{computation}[p]
\label{BnablaBG}
Computation of the first derivative of $\OLL{G}_{j\Dbeta}$
\begin{equation*}
 \{\Bnabla_{i\Dalpha},\OLL{G}_{j\Dbeta}\}
 +\{\Bnabla_{j\Dbeta},\OLL{G}_{i\Dalpha}\}
 =-C_{\Dalpha\Dbeta}[\nabla_z,W_{ij}].
\end{equation*}
\end{computation}

\begin{computation}
\label{nablaG-2}
Another computation of the first derivative of $G_{i\beta}$
\begin{align*}
 \{\nabla_{i\alpha},G_{j\beta}\}
 &=-\frac{i}{5}\{\nabla_{i\alpha},[\nabla_{k\beta},W^k{}_j]\}\\
 &=\frac{1}{5}\Bigl(
    \{\nabla_{j\beta},G_{i\alpha}\}-2\{\nabla_{i\beta},G_{j\alpha}\}
     +2\Omega_{ij}\{\nabla_{k\beta},G^k{}_\alpha\}\\
 &\phantom{=-()}
     -C_{\alpha\beta}[\nabla_z,W_{ij}]
     +C_{\alpha\beta}[W^k{}_j,W_{ik}]\Bigr).
\end{align*}
Taking $(i,j)$, we find that
\begin{equation}
\label{nablaG-symm}
 \{\nabla_{(i\alpha},G_{j)\beta}\}
  =-\frac{1}{2}[W_{ik},W^k{}_j],
\end{equation}
and, taking $[i,j]$, that
\begin{equation*}
 \{\nabla_{[i\alpha},G_{j]\beta}\}
  =-C_{\alpha\beta}[\nabla_z,W_{ij}]
   +\frac{1}{2}\Omega_{ij}\{\nabla_{k\alpha},G^k{}_\beta\}.
\end{equation*}
\end{computation}

\begin{computation}[p]
Another computation of the first derivative of $\OLL{G}_{j\Dbeta}$
\begin{gather}
 \{\Bnabla_{(i\Dalpha},\OLL{G}_{j)\Dbeta}\}
  =\frac{1}{2}C_{\Dalpha\Dbeta}[W_{ik},W^k{}_j],\\
 \{\Bnabla_{[i\Dalpha},\OLL{G}_{j]\Dbeta}\}
  =-C_{\Dalpha\Dbeta}[\nabla_z,W_{ij}]
   +\frac{1}{2}\Omega_{ij}\{\Bnabla_{k\Dalpha},\OLL{G}^k{}_\Dbeta\}.
       \notag
\end{gather}
\end{computation}

\begin{computation}
Further computation of the first derivative of $G_{j\beta}$
\begin{align*}
 \{\nabla_{i\alpha},G_{j\beta}\}
 &=-\frac{1}{4}(\sigma^\mu C)_\beta{}^\Dbeta
    \{\nabla_{i\alpha},\OLL{F}_{j\Dbeta\mu}\}
  =\frac{i}{4}(\sigma^\mu C)_\beta{}^\Dbeta
    \{\nabla_{i\alpha},[\Bnabla_{j\Dbeta},\nabla_\mu]\}\\
 &=-\frac{1}{2}C_{\alpha\beta}
    \{\Bnabla_{j\Dalpha},\OLL{G}_i{}^\Dalpha\}
   -\frac{i}{4}\Omega_{ij}(\sigma^{\mu\nu})_{\alpha\beta}F_{\mu\nu}.
\end{align*}
We then find that using \Ref{BnablaBG}
\begin{equation*}
 \{\nabla_{i\alpha},G^i{}_\beta\}
  =\frac{1}{2}C_{\alpha\beta}
   \{\Bnabla_{i\Dalpha},\OLL{G}^{i\Dalpha}\}
   -i(\sigma^{\mu\nu})_{\alpha\beta}F_{\mu\nu}
  =-i(\sigma^{\mu\nu})_{\alpha\beta}F_{\mu\nu},
\end{equation*}
which, together with \Ref{nablaG-2}, leads to
\begin{equation}
\label{nablaG-antisymm}
 \{\nabla_{[i\alpha},G_{j]\beta}\}
  =-C_{\alpha\beta}[\nabla_z,W_{ij}]
   -\frac{i}{2}\Omega_{ij}(\sigma^{\mu\nu})_{\alpha\beta}F_{\mu\nu}.
\end{equation}
Thus, from eqs.~\Ref{nablaG-symm} and \Ref{nablaG-antisymm},
we obtain that
\begin{equation}
 \{\nabla_{i\alpha},G_{j\beta}\}
 =-\frac{i}{4}\Omega_{ij}(\sigma^{\mu\nu})_{\alpha\beta}F_{\mu\nu}
  -\frac{1}{2}C_{\alpha\beta}[\nabla_z,W_{ij}]
  -\frac{1}{4}C_{\alpha\beta}[W_{ik},W^k{}_j].
\end{equation}
\end{computation}

\begin{computation}[p]
Further computation of the first derivative of $\OLL{G}_{j\Dbeta}$
\begin{gather}
 \{\Bnabla_{[i\Dalpha},\OLL{G}_{j]\Dbeta}\}
 =-C_{\Dalpha\Dbeta}[\nabla_z,W_{ij}]
  -\frac{i}{2}\Omega_{ij}(\bsigma^{\mu\nu})_{\Dalpha\Dbeta}F_{\mu\nu},\\
 \{\Bnabla_{i\Dalpha},\OLL{G}_{j\Dbeta}\}
 =-\frac{i}{4}\Omega_{ij}(\bsigma^{\mu\nu})_{\Dalpha\Dbeta}F_{\mu\nu}
  -\frac{1}{2}C_{\Dalpha\Dbeta}[\nabla_z,W_{ij}]
  +\frac{1}{4}C_{\Dalpha\Dbeta}[W_{ik},W^k{}_j].
\end{gather}
\end{computation}

\begin{computation}
Computation of the first derivative of $G_{j\beta}$
\begin{align*}
 \{\Bnabla_{i\Dalpha},G_{j\beta}\}
 &=i\{\Bnabla_{i\Dalpha},[\nabla_{j\beta},\nabla_z]\}\\
 &=-i\Bigl(
   i\{\nabla_{j\beta},\OLL{G}_{i\Dalpha}\}
   +i\Omega_{ji}(\sigma^\mu)_{\beta\Dalpha}
        [\nabla_z,\nabla_\mu]\Bigr)
  =\{\nabla_{j\beta},\OLL{G}_{i\Dalpha}\}
   -i\Omega_{ij}(\sigma^\mu)_{\beta\Dalpha}g_\mu.
\end{align*}
\end{computation}

\begin{computation}[p]
\label{nablaBG}
Computation of the first derivative of $\OLL{G}_{j\Dbeta}$
\begin{align*}
 \{\nabla_{i\alpha},\OLL{G}_{j\Dbeta}\}
  =\{\Bnabla_{j\Dbeta},G_{i\alpha}\}
   -i\Omega_{ij}(\sigma^\mu)_{\alpha\Dbeta}g_\mu.
\end{align*}
\end{computation}

\begin{computation}
\label{BnablaG-2}
Another computation of the first derivative of $G_{j\beta}$
\begin{align*}
 \{\Bnabla_{i\Dalpha},G_{j\beta}\}
 &=-\frac{i}{5}
   \{\Bnabla_{i\Dalpha},[\nabla_{k\beta},W^k{}_j]\}\\
 &=\frac{1}{5}\Bigl(
   \{\nabla_{j\beta},\OLL{G}_{i\Dalpha}\}
   -2\{\nabla_{i\beta},\OLL{G}_{j\Dalpha}\}
   +2\Omega_{ij}\{\nabla_{k\beta},\OLL{G}^k{}_\Dalpha\}\Bigr)
   +\frac{1}{5}(\sigma^\mu)_{\beta\Dalpha}[\nabla_\mu,W_{ij}].
\end{align*}
Taking $(i,j)$ we find, with the use of \Ref{nablaBG}, that
\begin{equation}
\label{BnablaG-symm}
 \{\Bnabla_{(i\Dalpha},G_{j)\beta}\}=0,
\end{equation}
and, similarly, $[i,j]$,
\begin{equation*}
 \{\Bnabla_{[i\Dalpha},G_{j]\beta}\}
  =-5i\Omega_{ij}(\sigma^\mu)_{\beta\Dalpha}g_\mu
   -2\Omega_{ij}\{\Bnabla_{k\Dalpha},G^k{}_\beta\}
   +(\sigma^\mu)_{\beta\Dalpha}[\nabla_\mu,W_{ij}].
\end{equation*}
Multiplying the last equation by $\Omega^{ij}$, we have
\begin{equation*}
 \{\Bnabla_{k\Dalpha},G^k{}_\beta\}
  =-2i(\sigma^\mu)_{\beta\Dalpha}g_\mu,
\end{equation*}
which in turn leads us to
\begin{equation}
\label{BnablaG-antisymm}
 \{\Bnabla_{[i\Dalpha},G_{j]\beta}\}
  =-i\Omega_{ij}(\sigma^\mu)_{\beta\Dalpha}g_\mu
   +(\sigma^\mu)_{\beta\Dalpha}[\nabla_\mu,W_{ij}].
\end{equation}
Thus using eqs.~\Ref{BnablaG-symm} and \Ref{BnablaG-antisymm},
we obtain that
\begin{equation}
 \{\Bnabla_{i\Dalpha},G_{j\beta}\}
  =-\frac{1}{2}(\sigma^\mu)_{\beta\Dalpha}\Bigl(
   i\Omega_{ij}g_\mu -[\nabla_\mu,W_{ij}]\Bigr).
\end{equation}
\end{computation}

\begin{computation}[p]
\label{nablaBG-2}
Another computation of the first derivative of $\OLL{G}_{j\Dbeta}$
\begin{gather}
 \{\nabla_{(i\alpha},\OLL{G}_{j)\Dbeta}\}=0,\\
 \{\nabla_{[i\alpha},\OLL{G}_{j]\Dbeta}\}
  =-i\Omega_{ij}(\sigma^\mu)_{\alpha\Dbeta}g_\mu
   -(\sigma^\mu)_{\alpha\Dbeta}[\nabla_\mu,W_{ij}],\\
 \{\nabla_{i\alpha},\OLL{G}_{j\Dbeta}\}
  =-\frac{1}{2}(\sigma^\mu)_{\alpha\Dbeta}\Bigl(
   i\Omega_{ij}g_\mu +[\nabla_\mu,W_{ij}]\Bigr).
\end{gather}
\end{computation}

\begin{computation}
\label{g}
Computation of $g_\mu$
\begin{equation*}
 g_\mu
  =i[\nabla_\mu,\nabla_z]
  =\frac{1}{8}(\bsigma^\mu)^{\Dbeta\alpha}
   [\{\nabla_{i\alpha},\Bnabla^i{}_\Dbeta\},\nabla_z]
  =\frac{i}{8}(\bsigma^\mu)^{\Dbeta\alpha}\Bigl(
   \{\nabla_{i\alpha},\OLL{G}^i{}_\Dbeta\}
   +\{\Bnabla_{i\Dbeta},G^i_\alpha\}\Bigr).
\end{equation*}
\end{computation}

\begin{computation}
\label{g-2}
Another computation of $g_\mu$
\begin{align*}
 g_\mu
 &=i[\nabla_\mu,\nabla_z]
  =\frac{1}{8}\Omega^{ij}C^{\alpha\beta}
   [\nabla_\mu,\{\nabla_{i\alpha},\nabla_{j\beta}\}]\\
 &=\frac{i}{4}(\bsigma_\mu)^{\Dbeta\alpha}
   \{\nabla_{i\alpha},\OLL{G}^i{}_\Dbeta\}.
\end{align*}
\end{computation}

\begin{computation}[p]
\label{g-3}
Another computation of $g_\mu$
\begin{equation*}
 g_\mu=\frac{i}{4}(\bsigma_\mu)^{\Dbeta\alpha}
   \{\Bnabla_{i\Dbeta},G^i{}_\alpha\}.
\end{equation*}
These results (\Ref{g}, \Ref{g-2}, \Ref{g-3})
are trivially obtainable from \Ref{BnablaG-2} and \Ref{nablaBG-2}.
\end{computation}

\paragraph{Five Fermionic Derivatives}
We now go on to the computation of relations containing
five fermionic derivatives.
\begin{computation}
\label{nablag}
Computation of $[\nabla_{i\alpha},g_\mu]$
\begin{equation*}
 [\nabla_{i\alpha},g_\mu]
  =i[\nabla_{i\alpha},[\nabla_\mu,\nabla_z]]
  =[\nabla_\mu,G_{i\alpha}]
   -(\bsigma_\mu C)^\Dalpha{}_\alpha
    [\nabla_z,\OLL{G}_{i\Dalpha}].
\end{equation*}
\end{computation}

\begin{computation}[p]
Computation of $[\Bnabla_{i\Dalpha},g_\mu]$
\begin{equation*}
 [\Bnabla_{i\Dalpha},g_\mu]
  =-[\nabla_\mu,\OLL{G}_{i\Dalpha}]
   +(C\bsigma_\mu)_\Dalpha{}^\alpha
    [\nabla_z,G_{i\alpha}].
\end{equation*}
\end{computation}

\begin{computation}
\label{nablag-2}
Another computation $[\nabla_{i\alpha},g_\mu]$
\begin{equation*}
 [\nabla_{i\alpha},g_\mu]
  =\frac{i}{4}(\bsigma^\mu)^{\Dbeta\beta}
   [\nabla_{i\alpha},\{\nabla_{j\beta},\OLL{G}^j{}_\Dbeta\}],
\end{equation*}
where
\begin{align*}
 [\nabla_{i\alpha},\{\nabla_{j\beta},\OLL{G}^j{}_\Dbeta\}]
 &=[\nabla^j{}_\beta,
    -\frac{1}{2}(\sigma^\mu)_{\alpha\Dbeta}\Bigl(
     i\Omega_{ij}g_\mu+[\nabla_\mu,W_{ij}]\Bigr)]
    -[\OLL{G}^j{}_\Dbeta,i\Omega_{ij}C_{\alpha\beta}\nabla_z
                         -iC_{\alpha\beta}W_{ij}]\\
 &=\frac{i}{2}(\sigma^\mu)_{\alpha\Dbeta}
     [\nabla_{i\beta},g_\mu]
   -\frac{5}{2}i(\sigma^\mu)_{\alpha\Dbeta}[\nabla_\mu,G_{i\beta}]\\
 &\phantom{=-}
   +2iC_{\alpha\beta}[\OLL{G}^j{}_\Dbeta,W_{ij}]
   -iC_{\alpha\beta}[\nabla_z,\OLL{G}_{i\Dbeta}].
\end{align*}
Thus we find that
\begin{align*}
 [\nabla_{i\alpha},g_\mu]
 &=-\frac{1}{8}(\sigma^\nu\bsigma_\mu)_\alpha{}^\beta
     [\nabla_{i\beta},g_\nu]
   -\frac{5}{8}(\sigma^\nu\bsigma_\mu)_\alpha{}^\beta
     [\nabla_\nu,G_{i\beta}]\\
 &\phantom{=-}
   +\frac{1}{2}(\bsigma_\mu C)^\Dbeta{}_\alpha[\OLL{G}^j{}_\Dbeta,W_{ij}]
   -\frac{1}{4}(\bsigma_\mu C)^\Dbeta{}_\alpha[\nabla_z,\OLL{G}_{i\Dbeta}].
\end{align*}
\end{computation}

\begin{computation}[p]
Another computation of $[\Bnabla_{i\Dalpha},g_\mu]$
\begin{align*}
 [\Bnabla_{i\Dalpha},g_\mu]
 &=-\frac{1}{8}(\bsigma_\mu\sigma^\nu)^\Dbeta{}_\Dalpha
     [\Bnabla_{i\Dbeta},g_\nu]
   -\frac{5}{8}(\bsigma_\mu\sigma_\nu)^\Dbeta{}_\Dalpha
     [\nabla^\nu,\OLL{G}_{i\Dbeta}]\\
 &\phantom{=-}
   +\frac{1}{2}(C\bsigma_\mu)_\Dalpha{}^\beta[G^j{}_\beta,W_{ij}]
   +\frac{1}{4}(C\bsigma_\mu)_\Dalpha{}^\beta[\nabla_z,G_{i\beta}].
\end{align*}
\end{computation}

\begin{computation}
Further computation of $[\nabla_{i\alpha},g_\mu]$
\begin{align*}
 [\nabla_{i\alpha},g_\mu]
 &=\frac{i}{4}(\bsigma)^{\Dbeta\beta}
   [\nabla_{i\alpha},\{\Bnabla_{j\Dbeta},G^j{}_\beta\}].
\end{align*}
Since
\begin{align*}
 [\nabla_{i\alpha},\{\Bnabla_{j\Dbeta},G^j{}_\beta\}]
 &=\left[\Bnabla^j{}_\Dbeta,
    -\frac{i}{4}\Omega_{ij}(\sigma^{\mu\nu})_{\alpha\beta}F_{\mu\nu}
    -\frac{1}{2}C_{\alpha\beta}[\nabla_z,W_{ij}]
    -\frac{1}{4}C_{\alpha\beta}[W_{ik},W^k{}_j]\right]\\
 &\phantom{=-}
    +i\Omega_{ij}(\sigma^\nu)_{\alpha\Dbeta}[\nabla_\nu,G^j{}_\beta],\\
 &\phantom{\ }
  \left|\quad\left(\begin{aligned}
   (\sigma^{\mu\nu})_{\alpha\beta}
   [\Bnabla_{i\Dbeta},F_{\mu\nu}]
   &=i(\sigma^{\mu\nu})_{\alpha\beta}
      [\Bnabla_{i\Dbeta},[\nabla_\mu,\nabla_\nu]]\\
   &=-2\bigl(C_{\alpha\beta}(C\bsigma^\mu)^{\Dgamma\gamma}
             [\nabla_\mu,G_{i\gamma}]
             +2(\sigma^\mu)_{\beta\Dbeta}
              [\nabla_{\mu,G_{i\alpha}}]\bigr),\\
   [\Bnabla^j{}_\Dbeta,[\nabla_z,W_ij]]
   &=5i[\nabla_z,\OLL{G}_{i\Dbeta}]+i[\OLL{G}^j{}_\Dbeta,W_{ij}],\\
   [\Bnabla^j{}_\Dbeta,[W_{ik},W^k{}_j]]
   &=-5i[\OLL{G}^k{}_\Dbeta,W_{ik}]
     -[i\Omega_{ik}\OLL{G}_{j\Dbeta}+2i\Omega_{j[i}\OLL{G}_{k]\Dbeta},W^{kj}]\\
   &=-8i[\OLL{G}^k{}_\Dbeta,W_{ik}],
  \end{aligned}\right)\right.\\
 &=-\frac{i}{2}C_{\alpha\beta}(C\bsigma^\mu)_\Dbeta{}^\gamma
    [\nabla_\mu,G_{i\gamma}]
   -i(\sigma^\mu)_{\beta\Dbeta}
    [\nabla_\mu,G_{i\alpha}]
   -i(\sigma^\mu)_{\alpha\Dbeta}
    [\nabla_\mu,G_{i\beta}]\\
 &\phantom{=-}
   -\frac{5}{2}iC_{\alpha\beta}
    [\nabla_z,\OLL{G}_{i\Dbeta}]
   +\frac{3}{2}iC_{\alpha\beta}
    [\OLL{G}^j{}_\Dbeta,W_{ij}],
\end{align*}
we find that
\begin{align*}
 [\nabla_{i\alpha},g_\mu] 
 &=\frac{5}{8}\left(
   [\nabla_\mu,G_{i\alpha}]
   -(\bsigma_\mu C)^\Dbeta{}_\alpha
   [\nabla_z,\OLL{G}_{i\Dbeta}]\right)\\
 &\phantom{=-}
   -\frac{3}{8}\left(
   (\sigma_{\mu\nu})_\alpha{}^\beta
   [\nabla^\nu,G_{i\beta}]
   -(\bsigma_\mu C)^\Dbeta{}_\alpha
   [\OLL{G}^j{}_\Dbeta,W_{ij}]\right).\tag{$\ast$}
\end{align*}
Equating this to the result in \Ref{nablag-2}, we have
\begin{equation*}
 (\sigma^\nu\bsigma_\mu)_\alpha{}^\beta
 [\nabla_{i\beta},g_\nu]
 =-2(\sigma_{\mu\nu})_\alpha{}^\beta[\nabla^\nu,G_{i\beta}]
  +(\bsigma_\mu C)^\Dbeta{}_\alpha
  [\OLL{G}^j{}_\Dbeta,W_{ij}]
  +3(\bsigma_\mu C)^\Dbeta{}_\alpha
  [\nabla_z,\OLL{G}_{i\Dbeta}].
\end{equation*}
After multiply this by
$\DS\frac{1}{4}(\sigma^\mu\bsigma_\rho)_\delta{}^\alpha$,
rename $\rho$ as $\mu$ to give
\begin{align*}
 [\nabla_{i\alpha},g_\mu]
 &=\frac{1}{2}\left(
   (\sigma_{\mu\nu})_\alpha{}^\beta
   [\nabla^\nu,G_{i\beta}]
   -(\bsigma_\mu C)^\Dbeta{}_\alpha
   [\OLL{G}^j{}_\Dbeta,W_{ij}]\right)\\
 &\phantom{=-}
   +\frac{3}{2}\left(
   [\nabla_\mu,G_{i\alpha}]
   -(\bsigma_\mu C)^\Dbeta{}_\alpha
   [\nabla_z,\OLL{G}_{i\Dbeta}]\right).
\end{align*}
Equating this to $(\ast)$, we find that
\begin{equation*}
   [\nabla_\mu,G_{i\alpha}]
   -(\bsigma_\mu C)^\Dbeta{}_\alpha
   [\nabla_z,\OLL{G}_{i\Dbeta}]
  =-(\sigma_{\mu\nu})_\alpha{}^\beta
   [\nabla^\nu,G_{i\beta}]
   +(\bsigma_\mu C)^\Dbeta{}_\alpha
   [\OLL{G}^j{}_\Dbeta,W_{ij}].
\end{equation*}
Thus going back again to $(\ast)$, we obtain
\begin{align}
 [\nabla_{i\alpha},g_\mu]
 &=[\nabla_\mu,G_{i\alpha}]
   -(\bsigma_\mu C)^\Dbeta{}_\alpha
   [\nabla_z,\OLL{G}_{i\Dbeta}] \notag \\
 &=-(\sigma_{\mu\nu})_\alpha{}^\beta
   [\nabla^\nu,G_{i\beta}]
   +(\bsigma_\mu C)^\Dbeta{}_\alpha
   [\OLL{G}^j{}_\Dbeta,W_{ij}],
\end{align}
which is completely consistent to \Ref{nablag}.
This result also leads to
\begin{equation}
 [\nabla_z,\OLL{G}_{i\Dalpha}]
 =-(C\sigma^\mu)^\alpha{}_\Dalpha
   [\nabla_\mu,G_{i\alpha}]
  -[\OLL{G}^j{}_\Dalpha,W_{ij}].
\end{equation}
\end{computation}

\begin{computation}[p]
Further computation of $[\Bnabla_{i\Dalpha},g_\mu]$
\begin{gather}
 \begin{aligned}[b]
 [\Bnabla_{i\Dalpha},g_\mu]
 &=-\frac{5}{8}\left(
   [\nabla_\mu,\OLL{G}_{i\Dalpha}]
   -(C\bsigma_\mu)_\Dalpha{}^\beta
   [\nabla_z,G_{i\beta}]\right)\\
 &\phantom{=-}
   -\frac{3}{8}\left(
   (\bsigma_{\mu\nu})^\Dbeta{}_\Dalpha
   [\nabla^\nu,\OLL{G}_{i\Dbeta}]
   -(C\bsigma_\mu)_\Dalpha{}^\beta
   [G^j{}_\beta,W_{ij}]\right),
 \end{aligned} \tag{$\ast'$} \\
 \begin{aligned}[b]
 [\Bnabla_{i\Dalpha},g_\mu]
 &=-[\nabla_\mu,\OLL{G}_{i\Dalpha}]
   +(C\bsigma_\mu)_\Dalpha{}^\beta
   [\nabla_z,G_{i\beta}] \\
 &=-(\bsigma_{\mu\nu})^\Dbeta_\Dalpha
   [\nabla^\nu,\OLL{G}_{i\Dbeta}]
   +(C\bsigma_\mu)_\Dalpha{}^\beta
   [G^j{}_\beta,W_{ij}],
 \end{aligned}\\
 [\nabla_z,G_{i\alpha}]
 =-(\sigma^\mu C)_\alpha{}^\Dbeta
   [\nabla_\mu,\OLL{G}_{i\Dbeta}]
  +[G^j{}_\alpha,W_{ij}].
\end{gather}
\end{computation}

\begin{computation}
Computation of $[\nabla_{i\alpha},[\nabla_z,W_{jk}]]$
\begin{align}
 [\nabla_{i\alpha},[\nabla_z,W_{jk}]]
 &=i[\nabla_z,\Omega_{jk}G_{i\alpha}+2\Omega_{i[j}G_{k]\alpha}]
   -i[G_{i\alpha},W_{jk}]\notag\\
 &=-i(\sigma^\mu C)_\alpha{}^\Dalpha\Bigl(
    \Omega_{jk}[\nabla_\mu,\OLL{G}_{i\Dalpha}]
    +2\Omega_{i[j}[\nabla_\mu,\OLL{G}_{k]\Dalpha}]\Bigr)\notag\\
 &\phantom{=\,}
   +i\left(
    \Omega_{jk}[G^l{}_\alpha,W_{il}]
    +2\Omega_{i[j}[G^l{}_\alpha,W_{k]l}]\right)\notag\\
 &\phantom{=\,}
   -i[G_{i\alpha},W_{jk}].
\end{align}
\end{computation}

\begin{computation}
Computation of $[\Bnabla_{i\Dalpha},[\nabla_z,W_{jk}]]$
\begin{align*}
 [\Bnabla_{i\Dalpha},[\nabla_z,W_{jk}]]
 &=-i(C\sigma^\mu)^\alpha{}_\Dalpha\Bigl(
    \Omega_{jk}[\nabla_\mu,G_{i\alpha}]
    +2\Omega_{i[j}[\nabla_\mu,G_{k]\alpha}]\Bigr)\\
 &\phantom{=\,}
   -i\left(
    \Omega_{jk}[\OLL{G}^l{}_\Dalpha,W_{il}]
    +2\Omega_{i[j}[\OLL{G}^l{}_\Dalpha,W_{k]l}]\right)\\
 &\phantom{=\,}
   +i[\OLL{G}_{i\Dalpha},W_{jk}].
\end{align*}
\end{computation}

\paragraph{Six Fermionic Derivatives}
Finally we compute relations containing six fermionic derivatives.
\begin{computation}
\label{nablanablaW}
Computation of $[\nabla_z,[\nabla_z,W_{jk}]]$
\begin{align*}
 [\nabla_z,[\nabla_z,W_{jk}]]
 &=-\frac{i}{8}\left[
     \{\nabla_{i\alpha},\nabla^{i\alpha}\},
     [\nabla_z,W_{jk}]\right]
  =-\frac{i}{4}\left\{\nabla_{i\alpha},
     [\nabla_{i\alpha},[\nabla_z,W_{jk}]]\right\}\\
 &=-\frac{1}{4}\Bigl\{\nabla^{i\alpha},
     (\sigma^\mu C)_\alpha{}^\Dalpha\Bigl(
       \Omega_{jk}[\nabla_\mu,\OLL{G}_{i\Dalpha}]
       +2\Omega_{i[j}[\nabla_\mu,\OLL{G}_{k]\Dalpha}]\Bigr)\\
 &\phantom{=-\frac{1}{4}{\nabla_{i\alpha}\sigma C}_{\alpha\beta}}
      -\left(\Omega_{jk}[G^j{}_\alpha,W_{il}]
             +2\Omega_{i[j}[G^l{}_\alpha,W_{k]l}]\right)
      +[G_{i\alpha},W_{jk}]\Bigr\}\\
 &\phantom{\,}\left|\quad\left(
  \begin{aligned}
    \{\nabla^{i\alpha},[\nabla_\mu,\OLL{G}_{i\Dalpha}]\}
     &=2i(C\sigma^\nu)^\alpha{}_\Dalpha[\nabla_\mu,g_\nu]
       -i(\bsigma_\mu)^{\Dbeta\alpha}
        \{\OLL{G}_{i\Dalpha},\OLL{G}^{i}{}_\Dalpha\},\\
    \{\nabla^{i\alpha},[\nabla_\mu,\OLL{G}_{k]\Dalpha}]\}
     &=-\frac{1}{2}(C\sigma^\nu)^\alpha{}_\Dalpha\left(
         -i\delta_{k]}{}^i[\nabla_\mu,g_\nu]
         +[\nabla_\mu,[\nabla_\nu,W^i{}_{k]}]]\right)\\
     &\phantom{=-}
         -i(\bsigma_\mu)^{\Dbeta\alpha}
           \{\OLL{G}_{k]\Dalpha},\OLL{G}^i{}_\Dbeta\},\\
    \{\nabla^{i\alpha},[G^l{}_\alpha,W_{il}]\}
     &=5i\{G^l{}_\alpha,G_l{}^\alpha\}
        -[W_{il},[\nabla_z,W^{il}]],\\
    \{\nabla^{i\alpha},[G^l{}_\alpha,W_{k]l}]\}
     &=3i\{G_{k]\alpha},G^{i\alpha}\}
       -2i\delta^i{}_{k]}\{G_{l\alpha},G^{l\alpha}\}\\
     &\phantom{=-}
      -[W_{k]l},[\nabla_z,W^{il}]]
      -\frac{1}{2}[W_{k]l},[W^i{}_m,W^{ml}]],\\
    \{\nabla^{i\alpha},[G_{i\alpha},W_{jk}]\}
     &=-i\Omega_{jk}\{G_{i\alpha},G^{i\alpha}\}
       +4i\{G_{j\alpha},G_k{}^\alpha\},
  \end{aligned}\right)\right.\\
 &=[\nabla_\mu,[\nabla_\mu,W_{jk}]]
   +\frac{1}{4}[W_{[j|l},[W_{k]m},W^{ml}]]\\
 &\phantom{=\,}
   +i\left(\Omega_{jk}\{G_{i\alpha},G^{i\alpha}\}
   -4\{G_{j\alpha},G_k{}^\alpha\}\right) %\\
   -i\left(\Omega_{jk}\{\OLL{G}_{i\Dalpha},\OLL{G}^{i\Dalpha}\}
   -4\{\OLL{G}_{j\Dalpha},\OLL{G}_k{}^\Dalpha\}\right)\\
 &\phantom{=\,}
   -\frac{1}{4}\left(
   \Omega_{jk}[W_{il},[\nabla_z,W^{il}]]
    -2[W_{[j|l},[\nabla_z,W_{k]}{}^l]]\right).
\end{align*}
\end{computation}

\begin{computation}[p]
Another computation of $[\nabla_z,[\nabla_z,W_{jk}]]$
\begin{align*}
 [\nabla_z,[\nabla_z,W_{jk}]]
 &=-\frac{i}{8}\left[
     \{\Bnabla_{i\Dalpha},\Bnabla^{i\Dalpha}\},
     [\nabla_z,W_{jk}]\right]
  =-\frac{i}{4}\left\{\Bnabla_{i\Dalpha},
     [\Bnabla_{i\Dalpha},[\nabla_z,W_{jk}]]\right\}\\
 &=[\nabla_\mu,[\nabla_\mu,W_{jk}]]
   +\frac{1}{4}[W_{[j|l},[W_{k]m},W^{ml}]]\\
 &\phantom{=\,}
   +i\left(\Omega_{jk}\{G_{i\alpha},G^{i\alpha}\}
   -4\{G_{j\alpha},G_k{}^\alpha\}\right) %\\
   -i\left(\Omega_{jk}\{\OLL{G}_{i\Dalpha},\OLL{G}^{i\Dalpha}\}
   -4\{\OLL{G}_{j\Dalpha},\OLL{G}_k{}^\Dalpha\}\right)\\
 &\phantom{=\,}
   +\frac{1}{4}\left(
   \Omega_{jk}[W_{il},[\nabla_z,W^{il}]]
    -2[W_{[j|l},[\nabla_z,W_{k]}{}^l]]\right).
\end{align*}
Comparing this result with the one in \Ref{nablanablaW},
we find immediately that we have to set
\begin{equation*}
\Omega_{jk}[W_{il},[\nabla_z,W^{il}]]
    -2[W_{[j|l},[\nabla_z,W_{k]}{}^l]]=0.
\end{equation*}
Thus we obtain that
\begin{align}
 [\nabla_z,[\nabla_z,W_{jk}]]
 &=[\nabla_\mu,[\nabla_\mu,W_{jk}]]
   +\frac{1}{4}[W_{[j|l},[W_{k]m},W^{ml}]]\notag \\
 &\phantom{=\,}
   +i\left(\Omega_{jk}\{G_{i\alpha},G^{i\alpha}\}
   -4\{G_{j\alpha},G_k{}^\alpha\}\right)
   -i\left(\Omega_{jk}\{\OLL{G}_{i\Dalpha},\OLL{G}^{i\Dalpha}\}
   -4\{\OLL{G}_{j\Dalpha},\OLL{G}_k{}^\Dalpha\}\right).
\end{align}
\end{computation}

\begin{computation}
Computation of $[\nabla_z,g_\mu]$
\begin{align*}
 [\nabla_z,g_\mu]
 &=\frac{i}{4}(\bsigma^\mu)^{\Dalpha\alpha}
   [\nabla_z,\{\nabla_{i\alpha},\OLL{G}^i{}_\Dalpha\}]\\
 &=\frac{i}{4}(\bsigma^\mu)^{\Dalpha\alpha}\Bigl(
   \{\nabla_{i\alpha},
    -(C\sigma^\nu)^\beta{}_\Dbeta[\nabla_\nu,G^i{}_\beta]
     -[\OLL{G}^j{}_\Dalpha,W^i{}_j]\}
     +i\{\OLL{G}^j{}_\Dalpha,G_{i\alpha}\}\Bigr)\\
 &\phantom{\,}
  \left|\quad\left(
  \begin{aligned}
    \{\nabla_{i\alpha},[\nabla_\nu,G^i{}_\beta]\}
   &=-i(\sigma^{\rho\sigma})_{\alpha\beta}
      [\nabla_\nu,F_{\rho\sigma}]
     -i(\bsigma_\nu C)^\Dbeta{}_\alpha
      \{G^i{}_\beta,\OLL{G}_{i\Dbeta}\},\\
    \{\nabla_{i\alpha},[\OLL{G}^j{}_\Dalpha,W^i{}_j]\}
   &=-5i\{\OLL{G}^j{}_\Dalpha,G_{j\alpha}\}
     +\frac{1}{2}(\sigma^\nu)_{\alpha\Dalpha}
      [W^i{}_j,[\nabla_\nu,W_i{}^j]],
   \end{aligned}\right)\right.\\
 &=\wp^+{}_\mu{}^{\nu\rho\sigma}
   [\nabla_\nu,F_{\rho\sigma}]
   -2(\bsigma_\mu)^{\Dalpha\alpha}
   \{G_{i\alpha},\OLL{G}^i{}_\Dalpha\}
   +\frac{i}{4}
   [W^{ij},[\nabla_\nu,W_{ij}]]. \tag{$\ast$}
\end{align*}
Using the Bianchi identity w.r.t.\
$(\nabla_\nu,\nabla_\rho,\nabla_\sigma)$, i.e.,\ 
\begin{equation*}
 \VE^{\mu\nu\rho\sigma}
  [\nabla_\nu,[\nabla_\rho,\nabla_\sigma]]=0,
\end{equation*}
the first term in $(\ast)$ equals simply to $[\nabla^\nu,F_{\mu\nu}]$.
Thus we obtain
\begin{equation}
 [\nabla_z,g_\mu]
  =[\nabla^\nu,F_{\mu\nu}]
   -2(\bsigma_\mu)^{\Dalpha\alpha}
   \{G_{i\alpha},\OLL{G}^i{}_\Dalpha\}
   +\frac{i}{4}
   [W^{ij},[\nabla_\nu,W_{ij}]].
\end{equation}
\end{computation}

\begin{computation}[p]
Another computation of $[\nabla_z,g_\mu]$
\begin{align*}
 [\nabla_z,g_\mu]
 &=\wp^-{}_\mu{}^{\nu\rho\sigma}
   [\nabla_\nu,F_{\rho\sigma}]
   -2(\bsigma_\mu)^{\Dalpha\alpha}
   \{G_{i\alpha},\OLL{G}^i{}_\Dalpha\}
   +\frac{i}{4}
   [W^{ij},[\nabla_\nu,W_{ij}]] \tag{$\ast'$}\\
 &=[\nabla^\nu,F_{\mu\nu}]
   -2(\bsigma_\mu)^{\Dalpha\alpha}
   \{G_{i\alpha},\OLL{G}^i{}_\Dalpha\}
   +\frac{i}{4}
   [W^{ij},[\nabla_\nu,W_{ij}]].
\end{align*}
\end{computation}

\begin{computation}
Computation of $[\nabla^\mu,g_\mu]$
\begin{align*}
 [\nabla^\mu,g_\mu]
 &=-\frac{i}{8}(\bsigma^\mu)^{\Dalpha\alpha}
   [\{\nabla_{i\alpha},\Bnabla^i{}_\Dalpha\},g_\mu]\\
 &=-\frac{i}{8}(\bsigma^\mu)^{\Dalpha\alpha}\Bigl(
   \{-(\bsigma_{\mu\nu})^\Dbeta{}_\Dalpha
       [\nabla^\nu,\OLL{G}^i{}_\Dbeta]
     +(C\bsigma_\mu)_\Dalpha{}^\beta
       [G^j{}_\beta,W^i{}_j], \nabla_{i\alpha}\}\\
 &\phantom{=-\frac{i}{8}(\bsigma^\mu)^{\Dalpha\alpha}()}
   \{-(\sigma_{\mu\nu})_\alpha{}^\beta
       [\nabla^\nu,G_{i\beta}]
     +(\bsigma_\mu C)^\Dbeta{}_\alpha
       [\OLL{G}^j{}_\Dbeta,W_{ij}], \Bnabla^i{}_{\Dalpha}\}\Bigr)\\
 &\phantom{\,}
  \left|\quad\left(
  \begin{aligned}
   \{[\nabla^\nu,\OLL{G}^i{}_\Dbeta],\nabla_{i\alpha}\}
   &=-2i(\sigma^\rho)_{\alpha\Dbeta}
     [\nabla^\nu,g_\rho]
     -i(\bsigma^\nu C)^\Dgamma{}_\alpha
     \{\OLL{G}_{i\Dgamma},\OLL{G}^i{}_\Dbeta\},\\
   \{[G^j{}_\beta,W^i{}_j],\nabla_{i\alpha}\}
   &=-5i\{G_{j\alpha},G^j{}_\beta\}
     +\frac{1}{2}C_{\alpha\beta}
      [[\nabla_z,W_{ij}],W^{ij}],\\
   \{[\nabla^\nu,G_{i\beta}],\Bnabla^i{}_\Dalpha\}
   &=2i(\sigma^\rho)_{\beta\Dalpha}
     [\nabla^\nu,g_\rho]
     +i(C\bsigma^\nu)_\Dalpha{}^\gamma
     \{G^i{}_{\gamma},G_{i\beta}\},\\
   \{[\OLL{G}^j{}_\Dbeta,W_{ij}],\Bnabla^i{}_\Dalpha\}
   &=5i\{\OLL{G}_{j\Dalpha},\OLL{G}^j{}_\Dbeta\}
     -\frac{1}{2}C_{\Dalpha\Dbeta}
      [[\nabla_z,W^{ij}],W_{ij}],
  \end{aligned}\right)\right.\\
 &=-3[\nabla^\rho,g_\rho]
   -4\{\OLL{G}_{i\Dgamma},\OLL{G}^i\Dgamma\}
   -4\{G_{i\gamma},G^{i\gamma}\}
   +i[W^{ij},[\nabla_z,W_{ij}]],
\end{align*}
so that
\begin{equation}
 [\nabla^\mu,g_\mu]
   =-\{\OLL{G}_{i\Dgamma},\OLL{G}^i\Dgamma\}
    -\{G_{i\gamma},G^{i\gamma}\}
   +\frac{i}{4}[W^{ij},[\nabla_z,W_{ij}]].
\end{equation}
\end{computation}


\begin{thebibliography}{99}
%
\bibitem{JS}
 J. Saito,
 Soryushiron Kenkyu (Kyoto) 111 (2005) 117.
%
\bibitem{D'Adda-Kanamori-Kawamoto-Nagata-1}% Twisted Superspace on a Lattice
 A. D'Adda, I. Kanamori, N. Kawamoto and K. Nagata,
 Nucl.\ Phys.\ \textbf{B707} (2005) 100, hep-lat/0406029;
 Nucl.\ Phys.\ Proc.\ Suppl.\ \textbf{140} (2005) 754,
 hep-lat/0409092.
\bibitem{D'Adda-Kanamori-Kawamoto-Nagata-2}%
 A. D'Adda, I. Kanamori, N. Kawamoto and K. Nagata,
 hep-lat/0507029.
%
\bibitem{Witten-1}% Topological Quantum Field Theory
 E. Witten, Comm.\ Math.\ Phys.\ \textbf{117} (1988) 353.
%
\bibitem{Witten-2}% Topological Sigma Models
 E. Witten, Comm.\ Math.\ Phys.\ \textbf{118} (1988) 411.
%
\bibitem{Baulieu-Singer}
 L. Baulieu, I. M. Singer,\ %  Topological Yang-Mills Symmetry,
 Nucl.\ Phys.\ Proc.\ Suppl.\ \textbf{15B} (1988) 12.
%
\bibitem{Brooks-Montano-Sonnenschein}%
 R. Brooks, D. Montano and J. Sonnenschein,
 Phys.\ Lett.\ \textbf{B214} (1988) 12.
%
\bibitem{Labastida-Pernici}%
 J. M. F. Labastida and M. Pernici,
 Phys.\ Lett.\ \textbf{B212} (1988) 56;
 Phys.\ Lett.\ \textbf{B213} (1988) 319.
%
\bibitem{Birmingham-Rakowski-Thompson}%
 D. Birmingham, M. Rakowski and G. Thompson,
 Phys.\ Lett.\ \textbf{B212} (1988) 187;
 Phys.\ Lett.\ \textbf{B214} (1988) 381;
 Nucl.\ Phys.\ \textbf{B315} (1989) 577.
%
\bibitem{Sohnius-1}% Bianchi Identities for Supersymmetric Gauge Theories
 M. F. Sohnius,
 Nucl.\ Phys.\ \textbf{B136} (1978) 461.
%
\bibitem{Sohnius-Stelle-West-1}% Off-Mass-Shell Formulation of Extended
                             % Supersymmetric Gauge Theories
 M. F. Sohnius, K. S. Stelle and P. C. West,
 Phys.\ Lett.\ \textbf{92B} (1980) 123.
%
\bibitem{Sohnius-Stelle-West-2}% Dimensional Reduction by Legendre
                             % Transformation Generates Off-Shell
                             % Supersymmetric Yang-Mills Theories
 M. F. Sohnius, K. S. Stelle and P. C. West,
 Nucl.\ Phys.\ \textbf{B173} (1980) 127.
%
\bibitem{Fayet-1}% Spontaneous Generation of Massive Multiplets and
               % Central Charges in Extended Supersymmetric Theories
 P. Fayet,
 Nucl.\ Phys.\ \textbf{B149} (1979) 137.
%
\bibitem{Wess-Bagger}
 J. Wess and J. Bagger,
 {\itshape Supersymmetry and Supergravity},
 Princeton University Press (1992).
%
\bibitem{Gates-Grisaru-Rocek-Siegel}
 S. J. Gates Jr, M. T. Grisaru, M. Roc\v{e}k and W. Siegel,
 {\itshape Superspace, or One thousand and one lessons in supersymmetry},
 Addison-Wesley (1983), hep-th/0108200.
%
\bibitem{Sohnius-3}
 M. F. Sohnius,
 {\itshape Introducing Supersymmetry},
 Phys.\ Reports 128 (1985) 39.
%
\bibitem{Donaldson}%
 S. K. Donaldson, J. Diff.\ Geom.\ \textbf{18} (1983) 279.
%
\bibitem{Eguchi-Yang}% N=2 Superconformal Models as
                     % Topological Field Theories
 T. Eguchi and S.-K. Yang,
 Mod.\ Phys.\ Lett.\ \textbf{A5} (1990) 1693.
%
\bibitem{Labastida-Llatas}% Topological Matter in Two Dimensions
 J. M. F. Labastida and P. M. Llatas,
 Nucl.\ Phys.\ \textbf{B379} (1992) 220.
%
\bibitem{Alvarez-Labastida}% Topological Matter in Four Dimensions
 M. Alvarez and J. M. F. Labastida,
 Nucl.\ Phys.\ \textbf{B437} (1995) 356, hep-th/9404115;
 Phys.\ Lett.\ \textbf{B315} (1993) 251, hep-th/9305028.
%
\bibitem{Yamron}
 J. P. Yamron,
 Phys.\ Lett.\ \textbf{B213} (1988) 325.
%
\bibitem{Vafa-Witten}% Dual String Pairs With $N=1$ and $N=2$ Supersymmetry 
                    % in Four Dimensions
 C. Vafa and E. Witten,
 Nucl.\ Phys.\ \textbf{B431} (1994) 3, hep-th/9507050.
%
\bibitem{Marcus}
 N. Marcus,
 Nucl.\ Phys.\ \textbf{B452} (1995) 331, hep-th/9506002.
%
\bibitem{Blau-Thompson}% Aspects of $N_T\geq 2$ Topological Gauge Theories
                       % and D-Branes
 M. Blau and G. Thompson,
 Nucl.\ Phys.\ \textbf{B492} (1997) 545, hep-th/9612143.
%
\bibitem{Labastida-Lozano}% MATHAI-QUILLEN FORMULATION OF TWISTED N=4
                          % SUPERSYMMETRIC GAUGE THEORIES IN FOUR-DIMENSIONS.
 J. M. F. Labastida and C. Lozano,
 Nucl.\ Phys.\ \textbf{B502} (1997) 741, hep-th/9709192.
%
\bibitem{Kawamoto-Tsukioka}%
 N. Kawamoto and T. Tsukioka,
 Phys.\ Rev.\ \textbf{D61} (2000) 105009, hep-th/9905222.
%
\bibitem{Kato-Kawamoto-Uchida}% TWISTED SUPERSPACE FOR N=D=2 SUPER BF
                              % AND YANG-MILLS WITH DIRAC-KAHLER FERMION
                              % MECHANISM
 J. Kato, N. Kawamoto and Y. Uchida,
 Int.\ J. Mod.\ Phys.\ \textbf{A19} (2004) 2149, hep-th/0310242.
%
\bibitem{Kato-Kawamoto-Miyake}% N=4 Twisted Superspace from Dirac-K\"{a}hler
                              % Twist and Off-Shell SUSY Invariant Actions
                              % in Four Dimensions
 J. Kato, N. Kawamoto and A. Miyake,
 Nucl.\ Phys.\ \textbf{B721} (2005) 229, hep-th/0502119.
%
\bibitem{Kato-Miyake}
 J. Kato and A. Miyake, in preparation.
\bibitem{Wess-Zumino-1}
 J. Wess and B. Zumino, Nucl.\ Phys.\ \textbf{B78} (1974) 1.
%
\bibitem{Ferrara-Zumino}
 S. Ferrara and B. Zumino, Nucl.\ Phys.\ \textbf{B79} (1974) 413.
%
\bibitem{Salam-Strathdee}
 A. Salam and J. Strathdee, Phys.\ Lett.\ \textbf{51B} (1974) 353.
%
\bibitem{Fayet-2}
 P. Fayet,
 Nucl.\ Phys.\ \textbf{B113} (1976) 135.
% 
\bibitem{Gliozzi-Scherk-Olive}%
 F. Gliozzi, J. Scherk and D. Olive,
 Nucl.\ Phys.\ \textbf{B122} (1977) 253.
%
\bibitem{Ferrara}
 S. Ferrara,
 Lett.\ Nuovo.\ Cim.\ 13 (1975) 629.
%
\bibitem{DiVecchia-Ferrara}
 P. Di Vecchia and S. Ferrara,
 Nucl.\ Phys.\ \textbf{B130} (1977) 93.
%
\bibitem{Wess-Zumino-2}
 J. Wess and B. Zumino, Phys.\ Lett.\ \textbf{66B} (1977) 361.
%
\bibitem{Wess}
 J. Wess, in {\itshape Topics in Quantum Field Theory},
 J. A. de Azc\'{a}rraga, ed.,\ Salamanca (1977).
%
\bibitem{Grimm-Sohnius-Wess}% Extended Supersymmetry and Gauge Theories
 R. Grimm, M. F. Sohnius and J. Wess,
 Nucl.\ Phys.\ \textbf{B133} (1978) 275.
%
%\bibitem{Hamada-Takao}% Supersymmetric Yang-Mills Theory in Two Dimensional
%                      % Superspace and Super Kac-Moody Algebra
% K. Hamada and M. Takao,
% Phys.\ Lett.\ \textbf{B210} (1988) 120.
%
\bibitem{Brink-Schwarz-Scherk}% Supersymmetric Yang-Mills Theories
 L. Brink, J. H. Schwarz and J. Scherk,
 Nucl.\ Phys.\ \textbf{B121} (1977) 77.
%
\bibitem{Milewski}% Superfield Formulation of the $N=2$ Super
                  % Yang-Mills Model with Central Charge
 B. Milewski,
 Phys.\ Lett.\ \textbf{B112} (1982) 148.
%
\bibitem{Claus-deWit-Faux-Kleijn-Siebelink-Termonia}
     % The Vector-Tensor Supermultiplet with Gauged Central Charge
 P. Claus, B. de Wit, M. Faux, B. Kleijn, R. Siebelink and P. Termonia,
 Phys.\ Lett.\ \textbf{B373} (1996) 81, hep-th/9512143.
%
\bibitem{Gaida}% Extended Supersymmetry with Gauged Central Charge
 I. Gaida,
 Phys.\ Lett.\ \textbf{B373} (1996) 89, hep-th/9512165.
%
\bibitem{Dragon-Ivanov-Kuzenko-Sokatchev-Theis}%
 N. Dragon, E. Ivanov, S. M. Kuzenko, E. Sokatchev and U. Theis,
 Nucl.\ Phys.\ \textbf{B538} (1999) 441, hep-th/9805152.
%
\bibitem{Hindawi-Ovrut-Waldram}% Vector-Tensor Multiplet in $N=2$
                               % Superspace with Central Charge
 A. Hindawi, B. A. Ovrut and D. Waldram,
 Phys.\ Lett.\ \textbf{B392} (1997) 85, hep-th/9609016.
%
%\bibitem{Dragon-Theis}% Gauging the Central Charge
% N. Dragon and Ulrich Theis,
%
\bibitem{Sohnius-2}% Supersymmetry and Central Charges
 M. F. Sohnius,
 Nucl.\ Phys.\ \textbf{B138} (1978) 109.
%
\bibitem{Kugo-Townsend}
 T. Kugo and P. Townsend,
 Nucl. Phys. \textbf{B221} (1983) 357.% - 380.
%
\bibitem{Polchinski}
 J. Polchinski,
 {\itshape String Theory}, Cambridge University Press (1998).

\end{thebibliography}
\end{document}